# Enhanced compound-protein binding affinity prediction by representing protein multimodal information via a coevolutionary strategy

Binjie Guo[1,2,3,6], Hanyu Zheng[1,2,3,6], Haohan Jiang[1,2,3,6], Xiaodan Li[2,3], Naiyu Guan[2,3], Yanming Zuo[2,3], Yicheng Zhang[2,3], Hengfu Yang[5], Xuhua Wang[1,2,3,4,*]

**Affiliations:**

1. Department of Neurobiology and Department of Rehabilitation Medicine, First Affiliated Hospital, Zhejiang University School of Medicine, Hangzhou, Zhejiang Province 310058, China
2. Liangzhu Laboratory, MOE Frontier Science Center for Brain Science and Brain-machine Integration, State Key Laboratory of Brain-machine Intelligence, Zhejiang University, 1369 West Wenyi Road, Hangzhou 311121, China
3. NHC and CAMS Key Laboratory of Medical Neurobiology, Zhejiang University, Hangzhou 310058, China
4. Co-innovation Center of Neuroregeneration, Nantong University, Nantong, 226001 Jiangsu, China
5. School of Computer Science, Hunan First Normal University, Changsha, 410205 Hunan, China
6. These authors contributed equally.

* Correspondence: xhw@zju.edu.cn (X.W.)

## Abstract

Due to the lack of a method to efficiently represent the multimodal information of a protein, including its structure and sequence information, predicting compound-protein binding affinity (CPA) still suffers from low accuracy when applying machine learning methods. To overcome this limitation, in a novel end-to-end architecture (named FeatNN), we develop a coevolutionary strategy to jointly represent the structure and sequence features of proteins and ultimately optimize the mathematical models for predicting CPA. Furthermore, from the perspective of data-driven approach, we proposed a rational method that can utilize both high- and low-quality databases to optimize the accuracy and generalization ability of FeatNN in CPA prediction tasks. Notably, we visually interpret the feature interaction process between sequence and structure in the rationally designed architecture. As a result, FeatNN considerably outperforms the state-of-the-art (SOTA) baseline in virtual drug screening tasks, indicating the feasibility of this approach for practical use. FeatNN provides an outstanding method for higher CPA prediction accuracy and better generalization ability by efficiently representing multimodal information of proteins via a coevolutionary strategy.

**Keywords:** Compound-protein binding affinity prediction; coevolutionary strategy; protein multimodal information; Protein 3D structure; Computational model; Artificial intelligence.

## Introduction

Since it is time and resource consuming to experimentally assess compounds and target protein binding affinities during drug discovery and development, effective virtual screening approaches using computational methods could greatly accelerate the drug candidate identification process by learning the abstract binding information between drug and target and accurately predicting compound-protein binding affinities (CPA) [1, 2], especially in cases where great numbers of sources for compound and protein interaction data are available through open source databases. For instance, BindingDB [3] currently provides a comprehensive collection of experimentally measured binding affinity data including more than 1 million protein–ligand complexes in the Protein Data Bank (PDB) [4], which substantially increases the potential for *in silico* CPA prediction. However, even with these abundant data, accurately predicting CPA is still the fundamental challenge preventing this method from being used in practical drug candidate screening applications due to the lack of a method to efficiently extract features from the data. To increase the accuracy of CPA prediction, the development of computational methods has proceeded with a variety of protein information embedding and representation strategies [5-8]. Despite substantial advancements, these strategies have met challenges with respect to further increasing the accuracy of CPA prediction.

Initially, researchers tended to represent protein features only using the protein sequence information, namely, the target (protein) is regarded as a sequence of residues. In these models, a pairwise array with the residue features of the protein as its column (or row) and the SMILES sequence information of the compound as its row (or column) is often utilized as the attention matrix to learn the potential interaction between a protein and a compound [9]. Typically, these models rely on the sequence information of the compounds and proteins of interest to learn their interactions via pairwise matrices, with the aim of predicting the binding affinities between them [9-13]. For example, multilayer 1-dimensional convolutional neural networks (1D-CNNs) are utilized to extract the features from the residue sequences of proteins, and the obtained vectors are used to represent the features of proteins, predict the CPAs and intensively study the noncovalent interaction between the ligand and binding target [14-16]. However, in addition to a protein's sequence of residues, the 3D structure of a protein also contributes significantly to its features [17, 18]. Therefore, neglecting the 3D spatial structure information of the protein may prevent the full realization of the potential of computational modeling in CPA prediction.

In this scenario, the approaches of representing and embedding protein structure information have been tentatively proposed to improve the accuracy in CPA prediction. To do so, molecular docking simulation methods [19, 20] based on background molecular dynamics knowledge and structure-based machine learning methods [8, 21] have been proposed. Relying on the knowledge of biophysics, the docking method computationally simulates the potential binding sites and 3D structures of compound-protein complexes, so it heavily depends on high-quality 3D protein structure data during CPA prediction [22, 23]. Despite a few successful stories, this method is severely limited due to the scarcity of high-quality 3D structure data of proteins (the precise position of each atom in a protein) [24]. By contrast, machine learning algorithm-based approaches can use 3D protein structure data with either high or low resolutions (the positions of key atoms in a protein). These models are fed with the spatial 3D information of the proteins in order to attain a superior ability to predict CPA [25-27]. For instance, the structural features of proteins were extracted through 3D atomic representations in voxel space by applying 3D CNNs [8]. However, the performance of these models was not significantly improved by introducing the structural information of the proteins [6, 8]. We hypothesized that this was due to the lack of the comprehensive consideration of the multimodal information (both sequence and structure information) of the protein by these methods. To address this problem, we sought to develop a method that can rationally incorporate the multimodal information of protein into CPA prediction models in order to improve CPA prediction performance.

Inspired by the multi-feature fusion tactics via coevolution [28], we designed an end-to-end neural network architecture (Fig. 1), named the fast evolutional aggregating and thoroughgoing graph neural network (FeatNN). Through the coevolutionary strategy, FeatNN efficiently represented the multimodal information (containing both structure and sequence information) of proteins and thus overcame the multimodal protein information representation challenge. Upon the $IC_{50}$ and KIKD datasets generated from PDBbind [29], FeatNN outperforms the SOTA method (MONN) in CPA prediction tasks by 21.33% and 17.07% with respect to the $R^2$ metric, 6.16% and 2.98% in terms of the root mean square error (RMSE), and 7.00% and 5.45% in the Pearson coefficients, respectively (Fig. 2).

The major technical advances of FeatNN are listed as follows.

1) An Evo-Updating block is employed in the protein encoding module to interactively update the sequence and structure information of proteins so that the high-quality features of proteins are extracted and presented, enabling FeatNN to outperform the SOTA model by great margins exceeding 21.33% in $R^2$.

2) In FeatNN, the distance matrices of protein residues are discretized into one dimension, and the word embedding strategy is applied to encode protein structure information, so that the network could effectively represent the multimodal protein information and lower the computational cost simultaneously.

3) With respect to the extraction of compound features, a specific residual connection is applied to represent the molecular graph, in which the features of the initial nodes are added onto each layer of the GCN [30], such that the graph features representation limitation caused by the notorious oversmoothing problem in traditional deep GCNs is solved.

4) With the pretraining and fine-tuning strategy, the $R^2$ performance of the optimized model, $FeatNN^{optm}$, further increases by 3.29% on average compared to that of FeatNN.

5) FeatNN has excellent generalization in the affinity prediction task, which is vital and pivotal in the drug screening domain. Targeting severe acute respiratory syndrome coronavirus 2 (SARS-CoV-2) 3-chymotrypsin (3C)-like protease and Akt-1, the generalization of FeatNN vastly outperforms the SOTA baseline in the affinity value prediction task.

6) The prediction results of FeatNN with different conformations of the same protein are robust when 3D structure information is directly introduced in the model while neglecting the molecular dynamics of the protein.

## Materials

### Dataset Construction

Even though PDBbind [31], BindingDB [3] and Binding MOAD [32-34] databases (Supplementary Fig. 2 and Supplementary Table. 1) contain paired information of protein-ligand complexes with structural data and the corresponding binding affinities, it was necessary to eliminate some data to comply with the quality standards of our model and baselines. The exclusion criteria included protein PDB file defects, and sequence information inconsistency in UniProt and PDB. Based on these criteria, we constructed a benchmark dataset based on PDBbind (version 2020, the general set) [29] that contains 12,699 compound-protein pairs. Meanwhile, a refined dataset [31] with higher quality of structural information has also been constructed from PDBbind (version 2020, the refined set, see Supplementary Fig. 2f). Additionally, we generated another dataset based on BindingDB (version Feb 6, 2022; the general set) [3] that is rich in data on compound-protein paired complexes but poor in protein diversity. The complex structure information in such dataset is not strictly paired and remains low-quality, because not all complexes in BindingDB have strictly paired 3D structure conformations, and most of these complexes correspond to multiple protein conformations with different PDB entries. Therefore, we preferentially chose the ligand-free or high-resolution PDB file for these complexes without strict correspondence between protein and compound. This generated dataset contains more than 210 thousand compound-protein pairs (Supplementary Table 1). To test the generalization ability of the models, we constructed new datasets from the Binding MOAD (see in Supplementary Table 1) database and excluded the complexes that appear in the datasets (train, validation, and test datasets) constructed from PDBbind (Supplementary Fig. Table 1). An affinity value of a certain measurement type (i.e., $K_i$, $K_d$, or $IC_{50}$.) for each complex was provided, and “KIKD” was used to refer to the combination of $K_i$-measured data and $K_d$-measured data due to their high homogeneity. More details about the dataset construction process are available in the Supplementary Methods 3.3.

## Training Data Generation

The PDBbind-based (both the general and refined datasets) training dataset generation process included three key steps. 1) Before performing data cleaning, we first assessed whether the regression labels (CPA values) in both PDBbind and BindingDB followed normal distributions to avoid the potential prediction deviation problem (Supplementary Fig. 2); 2) We then clustered the input compound and protein information according to a certain threshold (0.3, 0.4, 0.5, and 0.6) [9] to avoid the potential data leakage problem that can occur due to data similarities. In this evaluation, we assessed the similarity of the proteins using their multi-sequence alignment (MSA) scores and calculated the similarity of the compounds based on their fingerprints. Then, the same kinds of compounds or proteins with a certain threshold were divided into the same dataset; the details of this process are provided in Supplementary Methods 3.4 and 3.5; 3) Finally, we used a 5-fold cross-validation strategy [35] to generate training datasets to alleviate the potential overfitting problem. Then, the dataset was randomly shuffled with a training-validation-testing splitting ratio of approximately 7:1:2. For the generation of the BindingDB-based training dataset, we directly shuffled and split the dataset with the same training-validation-testing splitting ratio. The datasets generated from Binding MOAD were only used for testing the models’ generalization ability and transferability.

## Baseline Methods

To assess the performance of FeatNN, we chose to represent the SOTA algorithm architecture with the multiobjective neural network (MONN) [9], the structure-aware interactive graph neural network (SIGN) [26] and chose two classic methods, the drug-target binding affinity graph neural network (GraphDTA) [36], the bidirectional attention neural network for compound-protein interaction (BACPI) [37] as our baseline models. We followed the same experimental settings as those used in in the original studies that reported these baseline models.

·MONN applies a GCN block [30] to extract compound features and a 1D-CNN block to extract protein features and then constructs a pairwise matrix from the features of compounds and proteins to describe noncovalent interactions and predict CPA.

·GraphDTA comprises four models: the graph attention network (GATNet), graph convolutional network (GCNNet), the combined GAT and GCN (GATGCN) and graph isomorphism network (GINConvNet), all of which utilize architectures with a GCN block and an attention mechanism to extract protein and compound features and finally predict CPA through several dense layers that aggregate the features of compounds and proteins.

·BACPI serves as a bidirectional attention neural network and uses a 1D-CNN block to extract protein features from residue sequences and a graph attention network to extract compound features. CPA is predicted through several dense layers; this is similar to the GraphDTA approach.

·SIGN is as a structure-based method that converts the protein-ligand complex into a complex interaction graph and extract its features from such graph. The training data for this model must strictly contain the pair data (both protein and compound) in a complex with high-quality structure information.

## Results

### The Design of FeatNN with Input Protein Sequence and Structure Information

Given that the structure-based models that only consider the structure information of a protein might not well represent the protein's multimodal information, namely the sequence and structure information, we hypothesized that introducing the multimodal information of protein with a rational strategy in the CPA prediction model may further improve its CPA prediction performance.

To test this hypothesis, in an end-to-end neural network architecture, we first developed a method to represent the protein structure information (including the Euclidean distances between the residues of proteins in 3D space, the dihedral angles (Φ and ψ) on the backbones of proteins. Then we co-evolutionally updated this structure information with the residue sequences information of proteins, with the aim to comprehensively and efficiently represent their multimodal information. The general workflow of this model, FeatNN, is depicted in Fig. 1. FeatNN was designed based on a dexterous architecture that can process amino acid sequences and atom sequence with any lengths; thus, the whole set of information about proteins and compounds can be characterized. More specifically, the compound information proceeds through the compound extractor module (Fig. 1a and Supplementary Fig. 13) that consists of a multihead vertex representation (Fig. 1a and Supplementary Fig. 14) and deep GCN blocks (Fig. 1a and Supplementary Fig. 9). Notably, the deep GCN block is applied to prevent the oversmoothing problem during training process [38] of the compound extractor (the oversmoothing problem is described in more detail in Supplementary Note 1.1). To allow the remote atoms to communicate with a certain node, a master node is employed to simultaneously capture both local and global features so that FeatNN can learn comprehensive compound features from both global and local views at the same time.

Meanwhile, for the representation of protein structure information, the distance matrix of protein residues is discretized into one dimension, and the strategy of word embedding is applied to encode structure information regarding the Euclidean distances between protein residues as a discrete distance matrix (DDM), which greatly reduces the computational cost of obtaining structure information while still allowing the model to effectively represent the structure information of proteins. After that, the protein features are generally learned by the protein extractor module (Fig. 1b and Supplementary Fig. 15). In the protein extractor module, a Prot-Aggregation block (Fig. 1b and Supplementary Fig. 17) first converts the residue sequence of the given protein, the DDM, and the torsion matrix into two variables: a new matrix representing the residue sequence of the protein and a new distance matrix encoded with the structure information of the protein. The two outputs generated from the Prot-Aggregation block are then fed into the Evo-Updating block (Fig. 1b and Supplementary Fig. 18), which serves as the vital component in the protein encoder module (Fig. 1b and Supplementary Fig. 16). In this way, the structure and sequence information are interactively aggregated through a coevolutionary strategy in the Evo-Updating block, which ensures that FeatNN can learn preeminent features from multimodal protein information.

Finally, the learned representations of compound features and protein features are input into the affinity learning module (Fig. 1c and Supplementary Fig. 20). The detailed designs of the compound extraction module, protein extraction module and affinity learning module are described in the Methods and Supplementary sections.

**FeatNN Outperformed the SOTA Model in CPA Prediction**

To assess the performance of FeatNN, seven kinds of models mentioned above were trained on the dataset generated from the general PDBbind set, and their CPA prediction performances were compared (Fig. 2 and Supplementary Fig. 3). In addition to our model (FeatNN), the baseline models were BACPI [37], SIGN [26], MONN [9] and four variants of GraphDTA (i.e., GATGCN, GCNNet, GATNet and GINConvNet)[36]. Because some compounds and proteins tend to be highly similar and homologous, we followed the clustering strategy (for details, see Supplementary Methods 3.4 and 3.5) proposed in previous studies to prevent information leakage from the test set data during the model training process [9, 39]. Four different clustering thresholds were used to split and cluster the similarity data into training, valid and test sets in the control group experiment. They were 0.3, 0.4, 0.5 and 0.6, indicating the minimum distance between each similar class. For example, a 0.3 clustering threshold meant that any compounds from two different sets (training, valid, or test set) were at least 30% different in terms of their respective structures. In terms of the compound-clustered test group, $FeatNN^{general}$ outperformed the SOTA $baseline^{general}$ (MONN) by 21.33% in the $R^2$ metric under $IC_{50}$ (Fig. 2b and Supplementary Table 3) and 17.07% under KIKD (Fig. 2a and Supplementary Table 3). In addition, the evaluation results of the protein-clustered test group can be found in Supplementary Fig. 3. $FeatNN^{general}$ also surpassed the baseline models in most cases (Supplementary Fig. 3 and Supplementary Table 3). However, as shown by Supplementary Fig. 3a, the SIGN model achieved the best performance in RMSE but the worst in Pearson and $R^2$ on the "KIKD" dataset constructed from the general set of PDBbind-v2020, possibly because the SIGN model efficiently learned the absolute error (RMSE) between the prediction affinity and the real ones, but unable to learn their correlation (Pearson, $R^2$). Even though the similarity of the data (protein or compound) in the same dataset (training, validation, or test datasets) decreases with increasing threshold, the CPA prediction correlation performances of $\text{FeatNN}^{\text{general}}$ remained consistent and it outperformed the baselines, indicating the robustness and outstanding performance of FeatNN in comparison with the baseline models. Furthermore, we trained $\text{FeatNN}^{\text{refine}}$ on the refined datasets of PDBbind [31] to assess whether a high-quality structural dataset can enhance its CPA prediction performances. Interestingly, we found that the Pearson performances of $\text{FeatNN}^{\text{refine}}$ and SOTA $\text{baseline}^{\text{refine}}$ were respectively elevated by 2.65% and 5.45% compared to the corresponding methods trained on general datasets of PDBBind with the compound-clustered method (with the threshold of 0.3, details in Supplementary Fig. 4a, Supplementary Fig. 5a, Supplementary Table. 4, Supplementary Table 5). However, $R^2$ and Pearson values of $\text{FeatNN}^{\text{refine}}$ and SOTA $\text{baseline}^{\text{refine}}$ were found to be somewhat lower when applying the protein-clustered method, indicating that the accuracy and generalization of models were affected, possibly due to the limited number of high-quality data in the refined dataset of PDBbind-v2020 (Supplementary Fig. 4b, Supplementary Fig. 5b, Supplementary Table. 4, Supplementary Table 5). According to the statistic result (Supplementary Table 1), we found the protein diversity is poor in the refined dataset. Such a negative effect is observed possibly because the diversification of protein data is crucial for the performance of a computational model in CPA prediction tasks [40].

### Performances of FeatNN on the BindingDB Dataset

Even though the PDBbind database has rich protein diversity, the amount of paired information in this database is limited (12,699 records). By contrast, the BindingDB database is much larger (218,615 records), but the quality of the structural data in this database is not very high, and it is also poor in protein diversity and provides limited structure information for the compound and protein complexes. To comprehensively evaluate the performances of FeatNN, we first tested FeatNN and baseline models on BindingDB with a large-scale compound-protein interaction dataset. To do so, on the dataset generated from BindingDB with 218,615 compound-protein pairs, FeatNN and the baseline models were evaluated with 153,031 training samples, 21861 validation samples and 43,723 test samples[3]. To conduct a fair comparison, we evaluated the CPA prediction performance of the models by averaging the prediction results obtained over approximately 10 independent training processes on the dataset generated from BindingDB database. In contrast to the computer vision and natural language processing fields, the data in the biotechnology field are more flexible. The diversity of data in different datasets and the composition of data pairs may greatly change the performance of the model. As shown in Table 1, FeatNN outperformed the SOTA baseline with the best RMSE (0.765), Pearson correlation coefficient (0.850) and $R^2$ value (0.719).

**Applying Pretraining Strategy Enhanced the Performances of FeatNN**

First, to assess the generalization ability of FeatNN (Details in Supplementary Methods 3.6), we set up an independent third database named Binding MOAD with high-quality paired information data (the details for the generation of this dataset are provided in Supplementary Table 1). As shown in Supplementary Fig. 6, we found that the generalization ability of FeatNN was strongly depended on the amount of paired information in the training datasets. When trained on the general PDBbind dataset, $\text{FeatNN}^{\text{general}}$ showed superior generalization performance, outperforming the SOTA $\text{baseline}^{\text{general}}$ by 4.57% and 5.72% for the evaluation of the Pearson coefficient tested on $IC_{50}$ and KIKD measurement datasets constructed from Binding MOAD (Supplementary Fig. 6, Supplementary Fig. 7, Supplementary Table 6, Supplementary Table 8,). However, when trained on the refined datasets of PDBbind even with higher data quality, the models (both $\text{FeatNN}^{\text{refine}}$ and the SOTA $\text{baseline}^{\text{refine}}$) trained on the refined dataset of PDBbind showed considerably lower generalization ability compared to the corresponding models ($\text{FeatNN}^{\text{general}}$ and the SOTA $\text{baseline}^{\text{general}}$) trained on the general PDBbind dataset (Supplementary Fig. 6, Supplementary Table 6), with decreases by 62.95% and 93.10% in $R^2$ evaluation for FeatNN and SOTA baseline, respectively, possibly due to the limited amount of paired information used in the training process.

To further enhance the performance of FeatNN, $FeatNN^{optm}$ was tentatively trained by applying a pretraining strategy [41] to warm FeatNN up on the dataset with relatively low-quality structure data generated from BindingDB (Fig. 3a, Supplementary Methods 3.7). Considering that CPA prediction on PDBbind and BindingDB served as the same type of task, the parameters of the compound extractor learned from the two datasets could be highly generalized and portable. To test this hypothesis, we attempted to assess whether the performance of FeatNN on the PDBbind dataset could be improved by this parameter transfer strategy. To do so, the compound extractor parameters learned from BindingDB were frozen at first. The next steps were to fine-tune the protein extractor and affinity learning module, take the 'knowledge' learned from BindingDB as the initial parameters of the protein extractor and affinity learning module. In this way, we fine-tuned these two modules on the datasets generated from PDBbind, that is, to conduct multiple rounds of training and thus obtain $FeatNN^{optm}$ (Fig. 3a). As a result, the RMSE, Pearson coefficient, and $R^2$ of $FeatNN^{optm}$ for the PDBBind test dataset were increased by 3.29%, 1.93% and 5.47% (Fig. 3b and Supplementary Table 7) respectively, suggesting the excellent transferability of FeatNN to different datasets. Interestingly, the generalization ability of $FeatNN^{optm}$ is further enhanced by 2.04% and 5.79% for Pearson and $R^2$ compared with FeatNN directly trained on the PDBbind (Supplementary Fig. 7, Supplementary Table 8).

**The Functionality-based Interpretation of the FeatNN Module**

To elucidate the function of each block in FeatNN, we sought to assess the performance of FeatNN by ablating the blocks (Supplementary Methods 3.8) that were specifically designed to elevate its performance (for details, see Methods). The results shown in Fig. 4 demonstrate that a variety of components contribute significantly to the accuracy of FeatNN in CPA prediction. For instance, the robustness and prediction accuracy of FeatNN declined by approximately 14.34% in terms of the RMSE, 11.60% in the Pearson coefficient and 31.25% in $R^2$ without Evo-Updating, emphasizing the significance of the coevolutionary strategy in the protein extractor. Strikingly, the prediction accuracy decreased by approximately 15.22% in the RMSE, 15.61% in the Pearson coefficient and 36.33% in $R^2$ without addressing the oversmoothing problem via the deep GCN block. In addition, the master node in the deep GCN block, which represented the global information of each compound and communicated with the remote graph node through the graph warp unit (Fig. 4 and Supplementary Table 9), also contributed significantly to the accuracy of CPA prediction, highlighting the importance of interactively updating the global and local features and the importance of addressing the oversmoothing problem when representing the information of compounds. More importantly, the performance of the FeatNN versions that only used protein sequence information or structure information (DDM and torsion matrix) declined markedly by approximately 36.52% and 69.34%, respectively, in $R^2$ compared with the intact FeatNN baseline (Fig. 4 and Supplementary Table 9), emphasizing the importance of introducing the coevolutionary strategy to jointly aggregate and update the sequence and structure information of proteins. We ablated the compound-protein interactive matrix in the affinity learning module, which could help FeatNN to represent and learn the interaction information between compound and protein, and found that the $R^2$ performance declined by 38.09% (Fig. 4), indicating the rationality of learning effective interaction features by compound-protein interactive matrix. In addition, we ablated the torsion-related architecture and found that the performances declined by 13.48% in $R^2$ (Fig. 4 and Supplementary Table 9), highlighting the necessity of introducing the torsion information into FeatNN.

**The Interpretation of Information Flows in FeatNN**

To understand how information flows in the deep GCN, Evo-Updating, and affinity learning module, we visualized the original features in the intermediate layers of FeatNN (Supplementary Fig. 8). Because it is difficult to show the information transformation process in the original features directly, we applied t-distributed stochastic neighbor embedding (t-SNE) [42], a compression algorithm for high-dimensional data, to obtain a limpid data distribution in two dimension view (Fig. 5). As shown in Fig. 5, the atom features became more aggregated as the GCN layers deepened. This phenomenon dynamically explained why the node information flows in the layers and aggregates the features of neighbor nodes through the message passing mechanism [43] in the deep GCN block (Fig. 5a). In the Evo-Updating block, embedded sequence features and structure features were obtained from the Prot-Aggregation block, and then the sequence features and structure features were partially updated on each other, and part of their own information was integrated into the Evo-Updating block (Fig. 5b). When the Evo-Updating layers deepened, the difference between the sequence features and structural features gradually lessened, and the layers fused more multimodal information into themselves. Additionally, we extract the compound and protein features, which are learned from the deep GCN block and Evo-Updating block, respectively, in each layer for dimension reduction analysis (Fig. 5c, 5d). The distributions of compound features learned in the deep GCN block of each layer are clearly illustrated (Fig. 5c, 5d). We found that the features aggregated by the first three layers of the block have a certain degree of similarity, whereas the distribution of compound features tends to be more distinguishable in deep layers of GCN block (Fig. 5c), which might enable FeatNN to learn the precise features of the compound and address the notorious oversmoothing problem (Fig. 4 and Supplementary Fig .11a-c). In the Evo-Updating block, we showed that the eigenspace distance between protein structural features and sequence features that are learned in the same layer remains adjacent (Fig. 5d). More interestingly, we found that both the sequence and structural features learned in the deep layer of the block are updated along the same direction (evolution) through this coevolutionary strategy, which efficiently represents the multimodal information of proteins and ultimately benefits the CPA prediction accuracy (Fig. 4).

**FeatNN Outperformed the SOTA Baseline in Virtual Drug Screening Tasks**

To verify the feasibility of the use of FeatNN in virtual drug screening tasks [37, 44], we initially selected “SARS-CoV-2 3C-like protease” as the drug target (receptor), which is a verified target for developing drugs to cure SARS-CoV-2 [45]. We unbiasedly selected 28 bioactive small molecules [45-59] (listed in Supplementary Tab. 10, note: these molecules related to the target did not exist in PDBbind nor BindingDB) from publication research and the DrugBank database. The process of receptor-based affinity value prediction by applying FeatNN is shown in Fig. 6a. In addition, we selected a ligand-free protein structure of SARS-CoV-2 3C-like protease with the identity number of 7CWC in PDB. Strikingly, we found that the Pearson coefficient reached a value of 0.612 (Fig. 6b) in a CPA prediction task. Compared with the SOTA baseline (MONN) that obtained a Pearson coefficient of 0.402 (Fig. 6c), this was suggestive of the outstanding performance of FeatNN in searching for potential drug candidates from a massive database.

In addition, to verify the robustness of FeatNN, we repeated the prediction task many times and analyzed the results statistically (Fig. 6b). Nonetheless, a concern remained regarding the multimodality-based model of FeatNN: the prediction results obtained with different 3D protein structure conformations might have been variable. To assess this possibility, we selected the ligand-free protein conformations from 3 PDB files (recorded with PDB-ids of 7CWC, 7CWB and 7BAJ in the PDB Database, Supplementary Fig. 9a) of SARS-CoV-2 3C-like proteases as receptors for CPA prediction with FeatNN (Fig. 6b and Supplementary Figs. 9b-c). Remarkably, the CPA prediction task among 28 validated compounds still achieved robustness and exhibited excellent results with Pearson coefficients of 0.606 and 0.607, indicating that the prediction results obtained with FeatNN do not exhibit unstable changes in different target conformations (Fig. 6b and Supplementary Figs. 9b-c). To verify the feasibility of the use of FeatNN on different targets, we additionally chose a target named Akt-1 (PDB-id: 3O96) that is a critical receptor for the transmission of growth-promoting signals and resisting cancer [51]. In this experiment, 10 previously reported drugs (Supplementary Table. 11) that target Akt-1 [60-69] were selected for this virtual screening task, and FeatNN showed a better Pearson performance of 0.735 in the CPA prediction task compared with the SOTA baseline. Using different Akt-1 conformations (PDB-ids of 6HHJ, 3MV5, 3CQW and the ligand-free conformation predicted by AlphaFold2 [28]), the Pearson performance also remained stable (Fig. 6d, Supplementary Fig. 10 and Supplementary Table 11), indicating the robustness and reliable prediction ability of FeatNN in various virtual screening tasks with different targets.

## Discussion

The FeatNN model proposed in this study introduced a coevolutionary strategy to effectively represent multimodal protein features. Through a t-SNE visualization analysis and a module ablation study, from the perspective of interpretation, we showed that the information between protein sequences and structure features was jointly updated and aggregated, which ultimately benefited the CPA prediction accuracy of our approach. In this study, we found that the Evo-Updating block and deep GCN block in FeatNN function as the key components for aggregating and updating the features of both proteins and compounds (Fig. 4), emphasizing the significance of applying the coevolutionary strategy in protein feature extraction. Altogether, FeatNN learns efficiently from a limited data resource but is still able to cope with the complexity of structure data and achieve outstanding performance.

Although it is theoretically appealing to introduce the structural information of proteins in a CPA prediction model, we overcame numerous obstacles in the development of FeatNN. First, we elegantly overcame the oversmoothing problem[38] by introducing a specific residual connection in each layer of the GCN, which could add part of the initial information of the molecular graph into the current layers [70, 71]; therefore, the extraction ability of the model with respect to compound features was enhanced when the layers deepened (Supplementary Fig. 11a-c). Second, in the deep GCN block, a master node was employed to learn the global features during the training process, thus facilitating communication among remote nodes. Third, the protein distance matrix was discretely encoded to overcome the overwhelming information problem of the traditional continuous distance matrix. As a result, FeatNN greatly outperformed the SOTA model in tasks involving generalization ability on an independent database and targeting the “SARS-CoV-2 3C-like protease” and “Akt-1” affinity value prediction, indicating that FeatNN can be a powerful tool for advancing the drug development process.

Nevertheless, due to the scarcity of precise noncovalent interaction binding site data between the ligand and the binding pocket, and the data imbalance problem in the distribution of the few positive and predominantly negative data of binding sites, FeatNN faces difficulties in interpreting the CPA prediction results at the interaction level at current stage. Traditional methods such as upsampling and gradient penalty still cannot address such a dilemma (imbalance problem) without enough data for binding interactions [72]. Possibly, docking simulation combined with AI may be able to interpret the results predicted by the AI models at the interaction level [20], which may be a new research direction in the further development of FeatNN in our future study. Moreover, 3D structural information is not only relevant to proteins but also to other compounds [25, 73]. In this study, we only introduced the protein structural information, and experiments to additionally introduce compound geometry information are ongoing [74]. Theoretically, the strategy developed for protein feature extraction in our model could also be utilized to extract the geometric information of compounds. It could be appealing to introduce both protein and compound structure features in our model to further enhance its performance, given that the application of only the protein structure features in this study has already achieved a remarkable result. Other protein properties, such as the residue types of binding ligands, secondary structures and physicochemical characteristics, are also very important features. Incorporating these features into our model might further improve its performance. However, the challenge is how to represent these features with a rational method or provide an interpretable architecture, which is left to be addressed in future studies.

## Limitations

1) The training of the deep learning model depends strongly on the training data. In practice, if compounds or proteins are encountered with fairly different similarities that are very different from the data in the training set, the confidence in the prediction results will be greatly reduced. 2) Furthermore, because the architecture of FeatNN highly depends on the 3D structure of the protein, some protein data cannot be characterized due to the residue continuity defect of PDB files, so they must be discarded. Therefore, the number of training data will be decreased, but this will not significantly affect the performance of FeatNN. 3) Even though FeatNN can achieve improved precision and generalization ability in CPA prediction while ignoring the information regarding the binding pose between the ligand and the binding pocket, it is difficult for FeatNN to interpret the CPA prediction results at the interaction level, because of the scarcity and data imbalance problems of precise noncovalent interaction data between the ligand and the binding pocket.

## Conclusion

The proposed FeatNN model introduces a torsion matrix and a distance matrix in its protein extractor module, and it utilizes the deep GCN block with the master node in the compound extractor module to predict the affinity of a given compound-protein pair. The experimental results of our study showed that FeatNN outperformed the SOTA baseline by a significant margin, and the accessibility of FeatNN applied in lead compound screening was also verified; this approach demonstrates great potential for reducing the considerable time and expense involved in drug candidate screening experiments, and provides an interpretable architecture based on biology databases.

## Key Points

- We apply both 3D protein structure and sequence information with a coevolutionary strategy.
- We addressed the oversmoothing problem in graph representation of compounds.
- FeatNN achieved highly enhanced affinity prediction on well-known databases compared with the state-of-the-art methods.
- Generalization ability and feasibility of FeatNN are superior to the SOTA baseline both on the datasets generated from the Binding MOAD database and the virtual screening task targeting the receptor of the SARS-CoV-2 3CL protease and Akt-1.

## Acknowledgments

This study was supported by the Scientific and Technological Innovation 2030 Program of China - major projects (2021ZD0200408 to X.W.), the National Natural Science Foundation of China (81971866 to X.W.), the Natural Science Foundation of Zhejiang Province (LR20H090002 to X.W.), the Leading Innovative and Entrepreneur Team Introduction Program of Zhejiang (2019R01007 to X.W.), and the Fundamental Research Funds for the Central Universities (K20210195 to X.W.). We would like to thank professor Jie Yang for his feedback and advice on writing this paper.

### Conflict of Interest

Zhejiang University has filed a patent application related to this work, with X.W., B.G., H.Z. and H.J. listed as inventors. X.W. is a co-founder and scientific advisor of WeQure AI Inc., an AI-powered drug discovery start-up. The other authors declare no competing interests.

### Author Contributions

B.G., H.Z. and H.J. contributed equally to this work. X.W., B.G. and H.Z. conceptualized and designed the study. B.G., H.Z. and H.J. conducted the experiments and collected the data. B.G., H.Z., H.J., X.W., H.Y., X.L., N.G. and Y. Z. analyzed and interpreted the data. B.G., H.Z. and X.W. drafted the paper. All authors critically revised the manuscript and approved the final version for submission.

### Data Availability

The data that support the findings of this study are included in the paper, and further data are available from the corresponding author upon reasonable request.

**Table 1** Performance evaluation of different prediction approaches on the dataset generated from BindingDB. We apply RMSE, Pearson and $R^2$ to evaluate the CPA prediction performances. The results of each group were counted with 10 independent experiments. The mean value (and SD) of each independent experimental group are shown in the table. Note: The SIGN is highly dependent on the structure information of the complex and binding pockets while most structure information recorded in BindingDB is redundant and low-quality (lack of the information of pocket and binding site to represent the complex graph as the input training data), it is difficult to process the data before training the SIGN. Therefore, we did not train the SIGN on BindingDB.

| **Model** | $R^2$ ↑ | **RMSE** ↓ | **Pearson** ↑ |
|---|---|---|---|
| **FeatNN** | 0.719 (0.003) | 0.765 (0.004) | 0.850 (0.001) |
| **MONN** | 0.706 (0.004) | 0.783 (0.005) | 0.844 (0.002) |
| **BACPI** | 0.577 (0.005) | 0.935 (0.006) | 0.769 (0.002) |
| **GATGCN** | 0.543 (0.015) | 0.992 (0.016) | 0.742 (0.012) |
| **GCNNet** | 0.510 (0.023) | 1.030 (0.023) | 0.717 (0.015) |
| **GINConvNet** | 0.451 (0.124) | 1.080 (0.119) | 0.669 (0.094) |
| **GATConvNet** | 0.327 (0.027) | 1.200 (0.024) | 0.585 (0.001) |

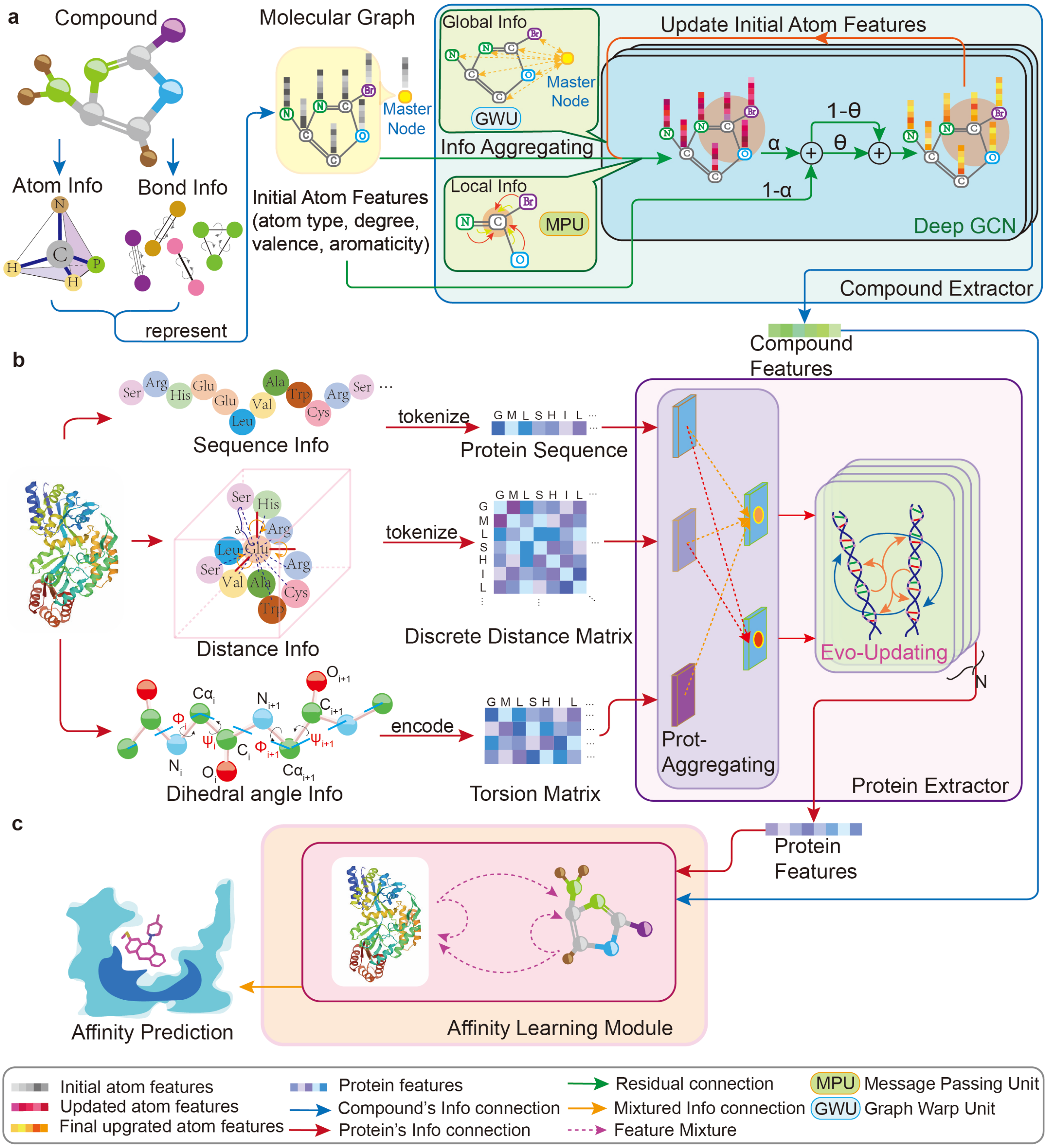

**Fig. 1 | Architecture overview of FeatNN. a.** The atom and bond information of a given compound is encoded into a molecular graph, which acts as the input for the compound extractor module to distill its features. The compound extractor includes a deep GCN block (Supplementary Fig. 12) and multihead attention blocks (Supplementary Fig. 14). **b.** The features of a protein are embedded with matrices and vectors as inputs to the Prot-Aggregation module (Supplementary Fig. 17), whose outputs are then fed to the Evo-Updating module (Supplementary Fig. 18), which co-evolutionarily updates the structure and sequence

features. Both the Prot-Aggregation module and the Evo-Updating module form the protein extractor block. **c.** The extracted atom and residue features are processed by the affinity learning module (Supplementary Fig. 20), which also enables FeatNN to learn the potential interaction features between the atoms of the compound and the residues of the protein. Additionally, the sets of information derived from the atom features and residue features are integrated through the affinity learning module to predict the CPA. The parameter settings of FeatNN are shown in Supplementary Table 2.

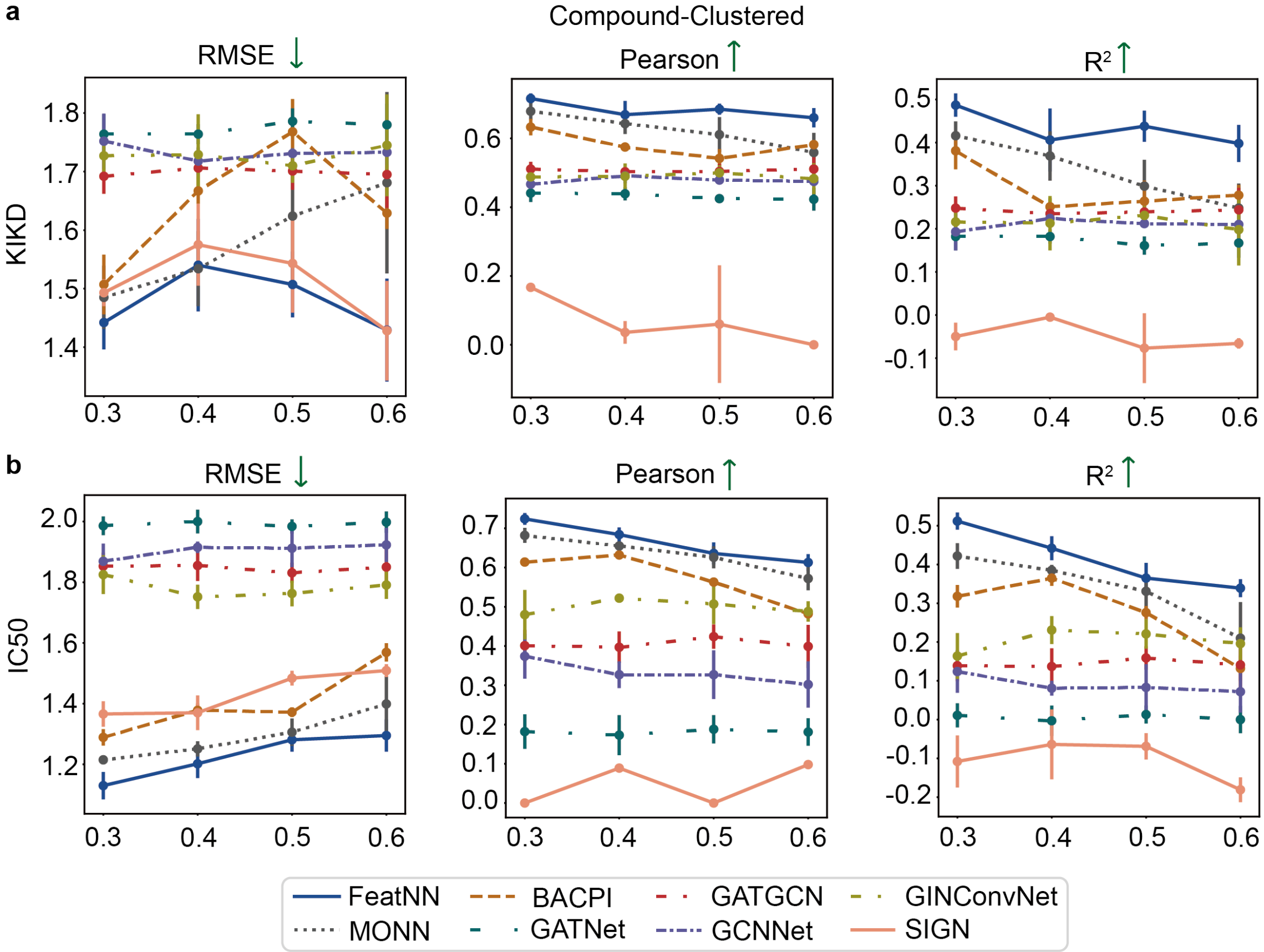

**Fig. 2 | Evaluation of FeatNN, BACPI, SIGN, GraphDTA (GATNet, GATGCN, GCNNet, GINConvNet) and MONN.** Performance evaluated on compound-clustered strategy datasets with similarity thresholds of 0.3, 0.4, 0.5 and 0.6 constructed from PDBbind with KIKD and $IC_{50}$ measurement, respectively. The benchmark dataset is generated from PDBbind (version 2020, the general set) and contains 12,699 compound-protein pairs. Performance results are plotted as the mean values and standard deviations (SD) by 5-fold cross-validation strategy with 10 independent experiments. Each point represents the independent experimental group mean with error bars indicating SD. We choose the three indicators (the RMSE, Pearson coefficient, and $R^2$) that can best evaluate the prediction performances of the methods in terms of the continuous values (CPA) they predicted. **a.** Performances evaluated on the dataset generated from PDBbind with KIKD measurement. **b.** Performances evaluated on the dataset generated from PDBbind with $IC_{50}$ measurement. Please note that the results of SIGN present here were different from the results reported by the original literature [26], possibly because we use PDBbind-v2020 as our benchmark database instead of PDBbind-v2016 used in their study. In

addition, considering the biology means behind the data, we split the dataset into two parts ("IC50" and "KIKD" [9]) instead of simply mixing the affinity measured with "$IC_{50}$", "$K_i$", and "$K_d$" together in their study. Moreover, we applied compound-cluster and protein-cluster strategies in our study to avoid data leakage caused by the biology-correlated knowledge (similarity structure or sequence in protein or compound). In most case, MONN achieved the best performances in baselines; therefore, we consider MONN as the SOTA baseline in our paper.

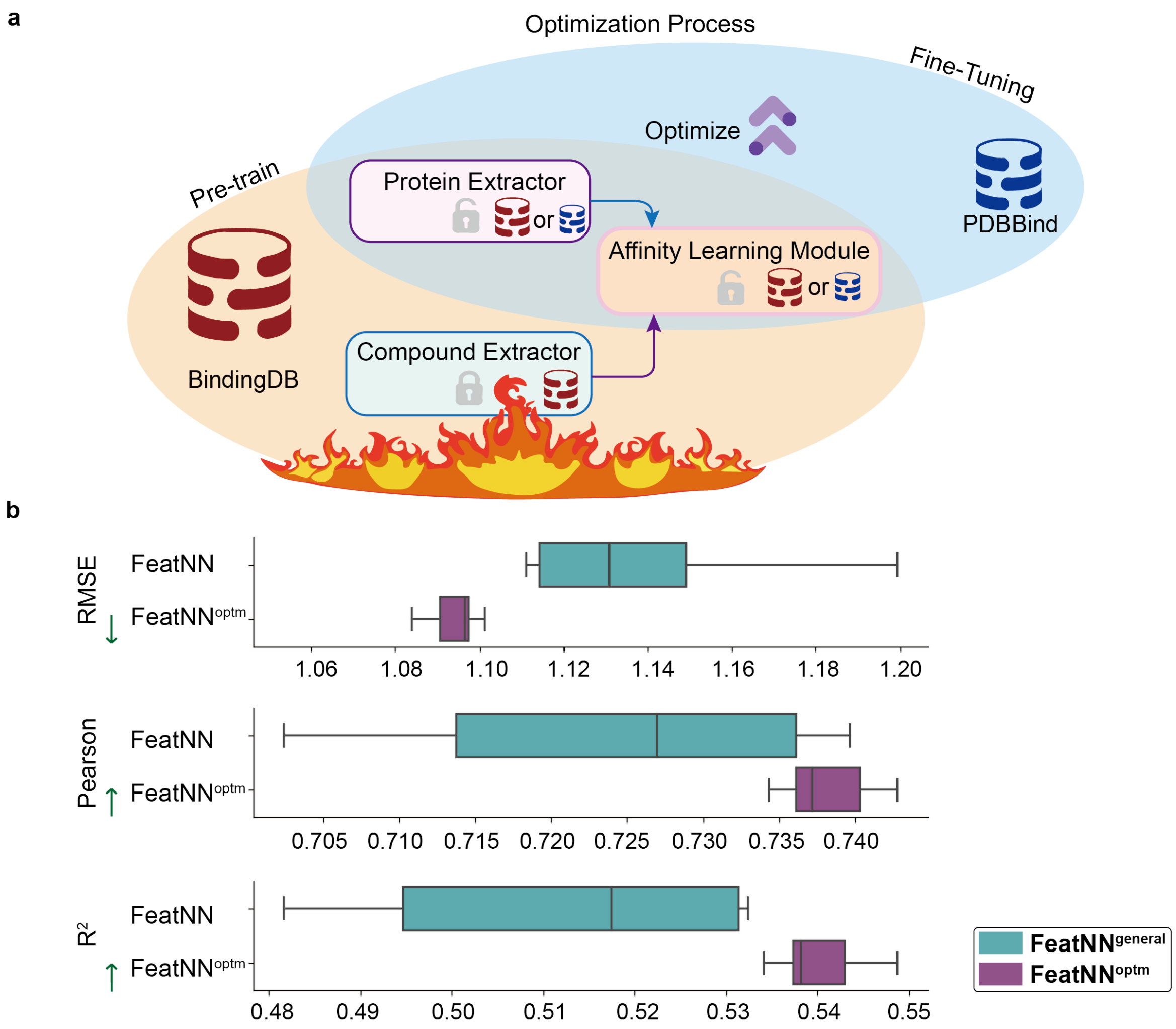

**Fig. 3 | The performance of FeatNN is greatly improved after optimization with fine-tuning strategy. a**. To optimize the performance of FeatNN, the parameters of the compound extractor obtained from the warm-up (pretraining) strategy on BindingDB are frozen, and then the protein extractor module and affinity learning module are fine-tuned on PDBbind to obtain $FeatNN^{optm}$. **b**. The RMSE, Pearson coefficient, and $R^2$ of FeatNN with the fine-tuning strategy ($FeatNN^{optm}$) were increased by 3.29%, 1.93% and 5.47% compared with that of the FeatNN version directly trained on PDBbind-v2020. FeatNN: original FeatNN trained on PDBbind. $FeatNN^{optm}$: FeatNN optimized with a fine-tuning strategy. The results of each group were counted with 10 independent experiments by 5-fold cross-validation strategy. The mean value, upper and lower quartiles, and SD of each independent experiment group are clearly shown

in Fig. 3b. Box plots; boxes depict the upper and lower quartiles of the data, and the vertical line in the box indicates the median of the statistical value of the group.

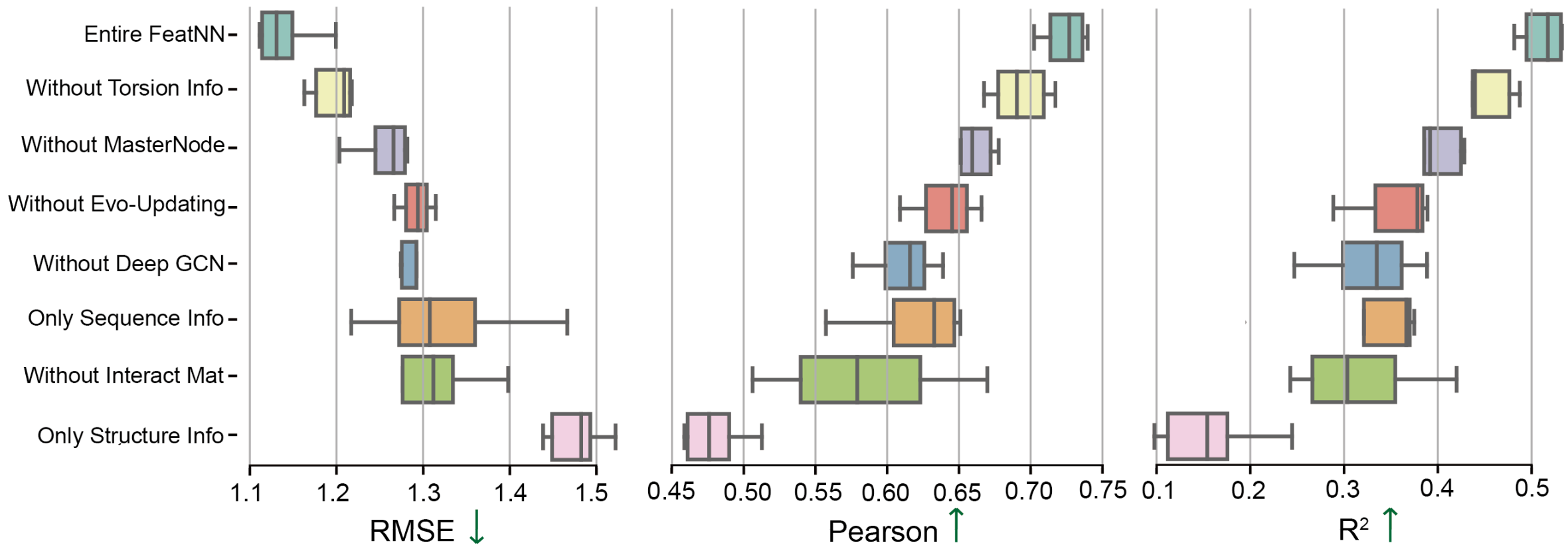

**Fig. 4 | Essential block ablation results of FeatNN.** Ablation results of FeatNN on the dataset generated from PDBbind, emphasizing the functionality of the essential blocks of FeatNN. The accuracy and robustness of FeatNN in terms of CPA prediction dramatically decline without the Evo-Updating block or torsion information, which functions as the core in protein feature extraction. Addressing the oversmoothing problem in the deep GCN block also remarkably increases the ability of the compound extractor to extract features from compounds, which in turn enhances the CPA prediction accuracy of the overall model. In addition, introducing the master node into the network to learn the global information of compounds is also important. The performances of the FeatNN version that only uses protein sequence information or structure information also remarkably decline compared with the entire FeatNN baseline, suggesting the importance of applying the coevolutionary strategy to interactively represent and update features of both sequence and 3D protein structure information. Furthermore, with ablation of the compound-protein interactive matrix, significant decline is observed in performances of the FeatNN, indicating the importance of learning the interaction features between protein and compound. The results of each group were counted with 10 independent experiments by 5-fold cross-validation strategy. The mean value, upper and lower quartiles, and SD of each independent experimental group are clearly depicted in Fig. 4. Box plots; boxes depict the upper and lower quartiles of the data, and the vertical line in the box indicates the median of the statistical value of the group. Abbreviations: Info: information.

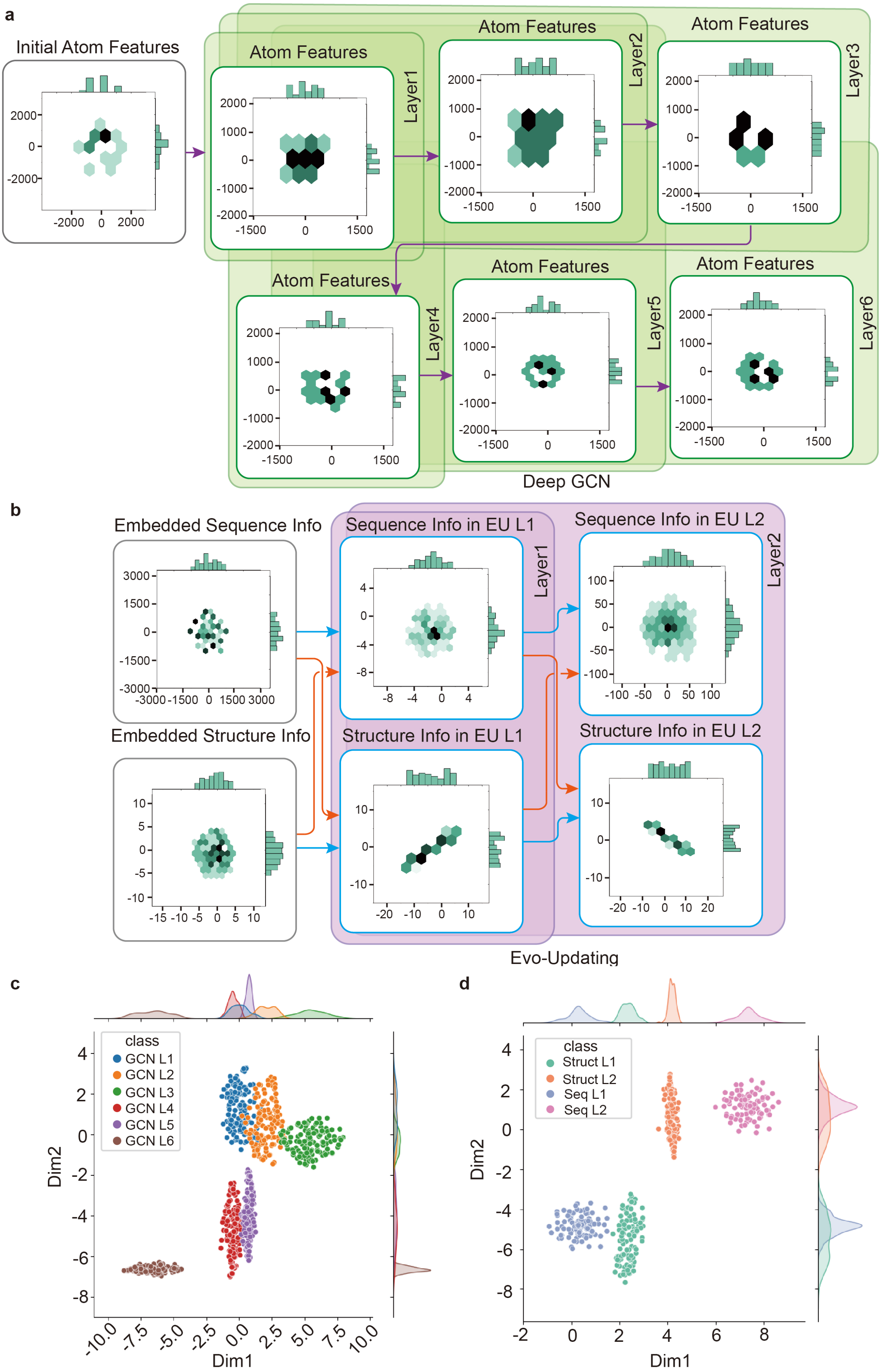

**Fig. 5 | Information flows in FeatNN's deep GCN and Evo-Updating blocks. a.** Visualization of the compound information aggregation process in the deep GCN block. **b**. Visualization of the coevolutionary process between the protein sequence and structure information in the Evo-Updating block. **c.** t-SNE dimensionality reduction analysis of deep GCN block (6 layers). **d.** t-SNE dimensionality reduction analysis of Evo-Updating block (2 layers). Abbreviations: EU L1 or L2: Evo-Updating Layer1 or Layer2. GCN L1 or L2: GCN block Layer1 or Layer2. Struct L1 or L2: Structure features in EU L1 or L2. Seq L1 or L2: Sequence features in EU L1 or L2. Embedded Sequence Info: sequence features obtained from the Prot-Aggregation block. Embedded Structure Info: structure features obtained from the Prot-Aggregation block. Initial atom features: atom features obtained from the graph embedding.

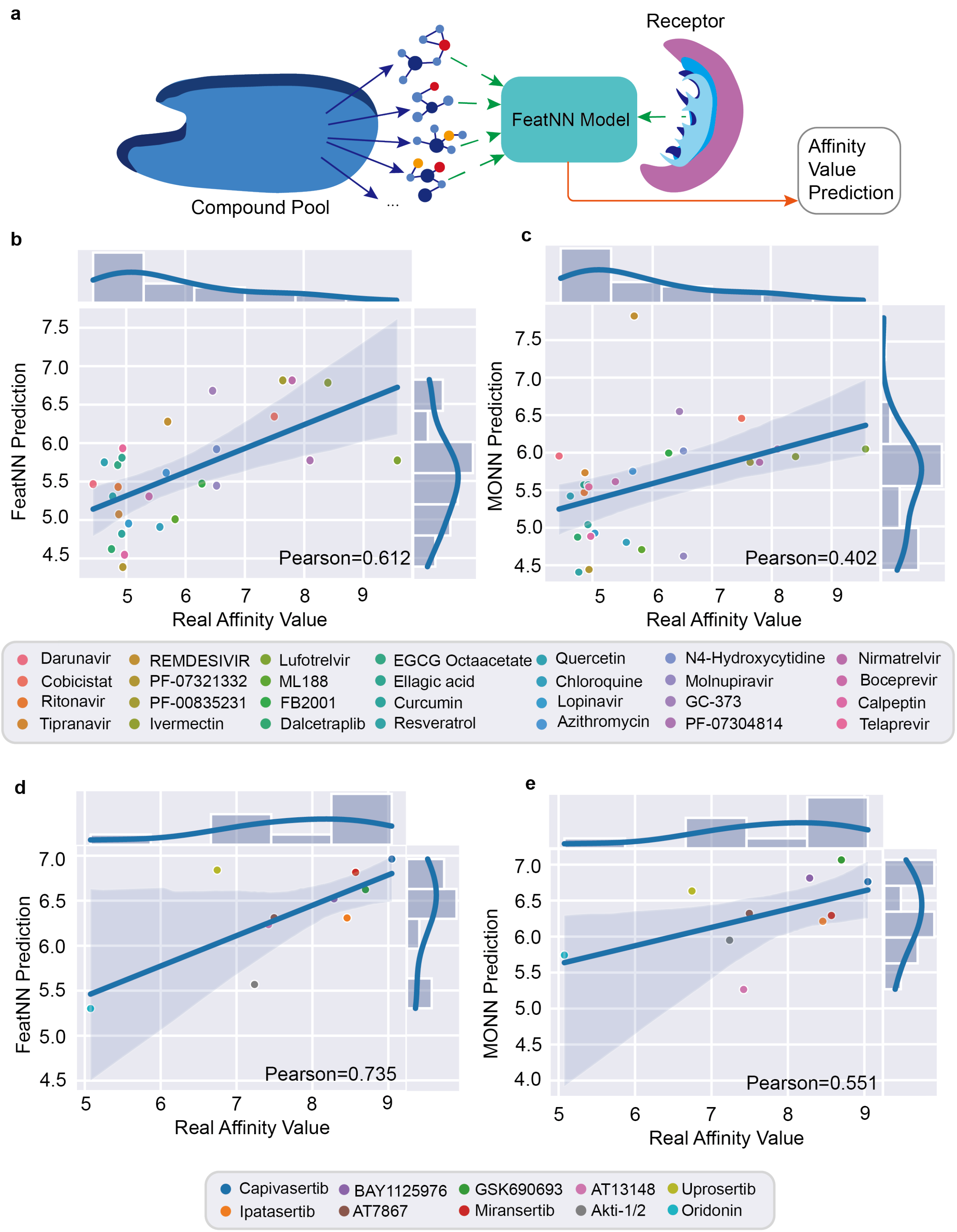

**Fig. 6 | Affinity prediction results of FeatNN and the SOTA baseline in practice. a.** Receptor-based virtual screening tasks: targeting both receptors of the SARS-CoV-2 3C-like protease and Akt-1, related bioactive compounds were unbiasedly selected (Supplementary Table 10 and 11) from published research and the DrugBank database to test the affinity

prediction precision and generalization ability of FeatNN. Targeting 3CL protease, **b**. the affinity prediction of 28 validated bioactive compounds by FeatNN result in a Pearson coefficient of 0.612. **c**. The affinity prediction of 28 validated bioactive compounds by MONN result in a Pearson coefficient of 0.402. Targeting Akt-1, **d.** the affinity prediction of 10 validated bioactive compounds by FeatNN results in a Pearson coefficient of 0.735. **e.** The affinity prediction of 10 validated bioactive compounds by MONN results in a Pearson coefficient of 0.551. Note: From the above experiments, it can be seen that MONN serves as the SOTA baseline in both datasets that generated from PDBbind and BindingDB databases, which is the reason that we only used MONN as a representative baseline model for testing. Both structure conformations of 3CL protease and Akt-1 are extracted from the PDB file with the PDB id of 7CWC and 3O96. Each point was obtained by the average of 15 independent experiments.

## References:

1. Chen X, Yan CC, Zhang X et al. Drug-target interaction prediction: databases, web servers and computational models, Brief Bioinform 2016;17:696-712.
2. Rester U. From virtuality to reality - Virtual screening in lead discovery and lead optimization: a medicinal chemistry perspective, Curr Opin Drug Discov Devel 2008;11:559-568.
3. Gilson MK, Liu T, Baitaluk M et al. BindingDB in 2015: A public database for medicinal chemistry, computational chemistry and systems pharmacology, Nucleic Acids Res 2016;44:D1045-1053.
4. Rose Y, Duarte JM, Lowe R et al. RCSB Protein Data Bank: Architectural Advances Towards Integrated Searching and Efficient Access to Macromolecular Structure Data from the PDB Archive, J Mol Biol 2021;433:166704.
5. Ozturk H, Ozgur A, Ozkirimli E. DeepDTA: deep drug-target binding affinity prediction, Bioinformatics 2018;34:i821-i829.
6. Zheng S, Li Y, Chen S et al. Predicting drug–protein interaction using quasi-visual question answering system, Nature Machine Intelligence 2020;2:134-140.
7. Qureshi R, Zhu M, Yan H. Visualization of Protein-Drug Interactions for the Analysis of Drug Resistance in Lung Cancer, IEEE J Biomed Health Inform 2021;25:1839-1848.
8. Jones D, Kim H, Zhang X et al. Improved Protein-Ligand Binding Affinity Prediction with Structure-Based Deep Fusion Inference, J Chem Inf Model 2021;61:1583-1592.
9. Li S, Wan F, Shu H et al. MONN: A Multi-objective Neural Network for Predicting Compound-Protein Interactions and Affinities, Cell Systems 2020;10:308-322.e311.
10. Ru X, Ye X, Sakurai T et al. NerLTR-DTA: Drug-target binding affinity prediction based on neighbor relationship and learning to rank, Bioinformatics 2022.
11. Bleakley K, Yamanishi Y. Supervised prediction of drug-target interactions using bipartite local models, Bioinformatics 2009;25:2397-2403.
12. Cao DS, Zhang LX, Tan GS et al. Computational Prediction of DrugTarget Interactions Using Chemical, Biological, and Network Features, Mol Inform 2014;33:669-681.
13. Ozturk H, Ozkirimli E, Ozgur A. A comparative study of SMILES-based compound similarity functions for drug-target interaction prediction, BMC Bioinformatics 2016;17:128.
14. Ragoza M, Hochuli J, Idrobo E et al. Protein-Ligand Scoring with Convolutional Neural Networks, J Chem Inf Model 2017;57:942-957.
15. Lee I, Keum J, Nam H. DeepConv-DTI: Prediction of drug-target interactions via deep learning with convolution on protein sequences, PLoS Comput Biol 2019;15:e1007129.
16. Rifaioglu AS, Nalbat E, Atalay V et al. DEEPScreen: high performance drug-target interaction prediction with convolutional neural networks using 2-D structural compound representations, Chem Sci 2020;11:2531-2557.
17. Klepeis JL, Lindorff-Larsen K, Dror RO et al. Long-timescale molecular dynamics simulations of protein structure and function, Curr Opin Struct Biol 2009;19:120-127.
18. Zhang J, Liang Y, Zhang Y. Atomic-level protein structure refinement using fragment-guided molecular dynamics conformation sampling, Structure 2011;19:1784-1795.

19. Ballester PJ, Mitchell JB. A machine learning approach to predicting protein-ligand binding affinity with applications to molecular docking, Bioinformatics 2010;26:1169-1175.
20. Bai Q, Liu S, Tian Y et al. Application advances of deep learning methods for de novo drug design and molecular dynamics simulation, WIREs Computational Molecular Science 2021;12.
21. Smith JS, Roitberg AE, Isayev O. Transforming Computational Drug Discovery with Machine Learning and AI, ACS Med Chem Lett 2018;9:1065-1069.
22. Kitchen DB, Decornez H, Furr JR et al. Docking and scoring in virtual screening for drug discovery: methods and applications, Nat Rev Drug Discov 2004;3:935-949.
23. Weiss DR, Karpiak J, Huang XP et al. Selectivity Challenges in Docking Screens for GPCR Targets and Antitargets, J Med Chem 2018;61:6830-6845.
24. Yamanishi Y, Araki M, Gutteridge A et al. Prediction of drug-target interaction networks from the integration of chemical and genomic spaces, Bioinformatics 2008;24:i232-240.
25. Fang X, Liu L, Lei J et al. Geometry-enhanced molecular representation learning for property prediction, Nature Machine Intelligence 2022;4:127-134.
26. Li S, Zhou J, Xu T et al. Structure-aware Interactive Graph Neural Networks for the Prediction of Protein-Ligand Binding Affinity. Proceedings of the 27th ACM SIGKDD Conference on Knowledge Discovery & Data Mining. 2021, 975-985.
27. Jiang D, Hsieh CY, Wu Z et al. InteractionGraphNet: A Novel and Efficient Deep Graph Representation Learning Framework for Accurate Protein-Ligand Interaction Predictions, J Med Chem 2021;64:18209-18232.
28. Jumper J, Evans R, Pritzel A et al. Highly accurate protein structure prediction with AlphaFold, Nature 2021;596:583-589.
29. Wang R, Fang X, Lu Y et al. The PDBbind Database: Methodologies and Updates, Journal of Medicinal Chemistry 2005;48:4111-4119.
30. Kipf TN, Welling M. Semi-Supervised Classification with Graph Convolutional Networks, CoRR 2016;abs/1609.02907.
31. Liu Z, Li Y, Han L et al. PDB-wide collection of binding data: current status of the PDBbind database, Bioinformatics 2015;31:405-412.
32. Ahmed A, Smith RD, Clark JJ et al. Recent improvements to Binding MOAD: a resource for protein-ligand binding affinities and structures, Nucleic Acids Res 2015;43:D465-469.
33. Smith RD, Clark JJ, Ahmed A et al. Updates to Binding MOAD (Mother of All Databases): Polypharmacology Tools and Their Utility in Drug Repurposing, J Mol Biol 2019;431:2423-2433.
34. Hu L, Benson ML, Smith RD et al. Binding MOAD (Mother Of All Databases), Proteins 2005;60:333-340.
35. Refaeilzadeh P TL, Liu H. Cross-validation, Encyclopedia of database systems 2009:532-538.
36. Nguyen T, Le H, Quinn TP et al. GraphDTA: predicting drug-target binding affinity with graph neural networks, Bioinformatics 2021;37:1140-1147.
37. Li M, Lu Z, Wu Y et al. BACPI: a bi-directional attention neural network for compound-protein interaction and binding affinity prediction, Bioinformatics 2022.

38. Ishiguro KaM, Shin-ichi and Koyama, Masanori. Graph Warp Module: an Auxiliary Module for Boosting the Power of Graph Neural Networks in Molecular Graph Analysis, arXiv 2019.
39. Mayr A, Klambauer G, Unterthiner T et al. Large-scale comparison of machine learning methods for drug target prediction on ChEMBL, Chem Sci 2018;9:5441-5451.
40. Freschlin CR, Fahlberg SA, Romero PA. Machine learning to navigate fitness landscapes for protein engineering, Curr Opin Biotechnol 2022;75:102713.
41. Pengfei Liu WY, Jinlan Fu, Zhengbao Jiang, Hiroaki Hayashi, Graham Neubig. Pre-train, Prompt, and Predict: A Systematic Survey of Prompting Methods in Natural Language Processing, arXiv 2021.
42. Hinton. LJPvdMaGE. Visualizing High-Dimensional Data Using t-SNE., Journal of Machine Learning Research 2008;9.
43. Stachenfeld KaG, Jonathan and Battaglia, Peter. Graph Networks with Spectral Message Passing, arXiv 2021.
44. Bai Q, Tan S, Xu T et al. MolAICal: a soft tool for 3D drug design of protein targets by artificial intelligence and classical algorithm, Brief Bioinform 2021;22.
45. Hillen HS, Kokic G, Farnung L et al. Structure of replicating SARS-CoV-2 polymerase, Nature 2020;584:154-156.
46. Li J, Lin C, Zhou X et al. 2021.
47. Mahdi M, Motyan JA, Szojka ZI et al. Analysis of the efficacy of HIV protease inhibitors against SARS-CoV-2's main protease, Virol J 2020;17:190.
48. Chen J, Xia L, Liu L et al. Antiviral Activity and Safety of Darunavir/Cobicistat for the Treatment of COVID-19, Open Forum Infect Dis 2020;7:ofaa241.
49. Ahmed MH, Hassan A. Dexamethasone for the Treatment of Coronavirus Disease (COVID-19): a Review, SN Compr Clin Med 2020:1-10.
50. Hoffman RL, Kania RS, Brothers MA et al. Discovery of Ketone-Based Covalent Inhibitors of Coronavirus 3CL Proteases for the Potential Therapeutic Treatment of COVID-19, J Med Chem 2020;63:12725-12747.
51. Hinz N, Jucker M. Distinct functions of AKT isoforms in breast cancer: a comprehensive review, Cell Commun Signal 2019;17:154.
52. Lopez-Medina E, Lopez P, Hurtado IC et al. Effect of Ivermectin on Time to Resolution of Symptoms Among Adults With Mild COVID-19: A Randomized Clinical Trial, JAMA 2021;325:1426-1435.
53. Shamsi A, Mohammad T, Anwar S et al. Glecaprevir and Maraviroc are high-affinity inhibitors of SARS-CoV-2 main protease: possible implication in COVID-19 therapy, Biosci Rep 2020;40.
54. Vankadara S, Wong YX, Liu B et al. A head-to-head comparison of the inhibitory activities of 15 peptidomimetic SARS-CoV-2 3CLpro inhibitors, Bioorg Med Chem Lett 2021;48:128263.
55. Mody V, Ho J, Wills S et al. Identification of 3-chymotrypsin like protease (3CLPro) inhibitors as potential anti-SARS-CoV-2 agents, Commun Biol 2021;4:93.
56. Xiang R, Yu Z, Wang Y et al. Recent advances in developing small-molecule inhibitors against SARS-CoV-2, Acta Pharm Sin B 2022;12:1591-1623.

57. Costanzo M, De Giglio MAR, Roviello GN. SARS-CoV-2: Recent Reports on Antiviral Therapies Based on Lopinavir/Ritonavir, Darunavir/Umifenovir, Hydroxychloroquine, Remdesivir, Favipiravir and other Drugs for the Treatment of the New Coronavirus, Curr Med Chem 2020;27:4536-4541.
58. Lo HS, Hui KPY, Lai HM et al. Simeprevir Potently Suppresses SARS-CoV-2 Replication and Synergizes with Remdesivir, ACS Cent Sci 2021;7:792-802.
59. Hosseini-Zare MS, Thilagavathi R, Selvam C. Targeting severe acute respiratory syndrome-coronavirus (SARS-CoV-1) with structurally diverse inhibitors: a comprehensive review, RSC Adv 2020;10:28287-28299.
60. Grimshaw KM, Hunter LJ, Yap TA et al. AT7867 is a potent and oral inhibitor of AKT and p70 S6 kinase that induces pharmacodynamic changes and inhibits human tumor xenograft growth, Mol Cancer Ther 2010;9:1100-1110.
61. Politz O, Siegel F, Barfacker L et al. BAY 1125976, a selective allosteric AKT1/2 inhibitor, exhibits high efficacy on AKT signaling-dependent tumor growth in mouse models, Int J Cancer 2017;140:449-459.
62. Rhodes N, Heerding DA, Duckett DR et al. Characterization of an Akt kinase inhibitor with potent pharmacodynamic and antitumor activity, Cancer Res 2008;68:2366-2374.
63. Wu WI, Voegtli WC, Sturgis HL et al. Crystal structure of human AKT1 with an allosteric inhibitor reveals a new mode of kinase inhibition, PLoS One 2010;5:e12913.
64. Andrikopoulou A, Chatzinikolaou S, Panourgias E et al. "The emerging role of capivasertib in breast cancer", Breast 2022;63:157-167.
65. McLeod R, Kumar R, Papadatos-Pastos D et al. First-in-Human Study of AT13148, a Dual ROCK-AKT Inhibitor in Patients with Solid Tumors, Clin Cancer Res 2020;26:4777-4784.
66. Nandan D, Zhang N, Yu Y et al. Miransertib (ARQ 092), an orally-available, selective Akt inhibitor is effective against Leishmania, PLoS One 2018;13:e0206920.
67. Weisner J, Landel I, Reintjes C et al. Preclinical Efficacy of Covalent-Allosteric AKT Inhibitor Borussertib in Combination with Trametinib in KRAS-Mutant Pancreatic and Colorectal Cancer, Cancer Res 2019;79:2367-2378.
68. Song M, Liu X, Liu K et al. Targeting AKT with Oridonin Inhibits Growth of Esophageal Squamous Cell Carcinoma In Vitro and Patient-Derived Xenografts In Vivo, Mol Cancer Ther 2018;17:1540-1553.
69. Iksen, Pothongsrisit S, Pongrakhananon V. Targeting the PI3K/AKT/mTOR Signaling Pathway in Lung Cancer: An Update Regarding Potential Drugs and Natural Products, Molecules 2021;26.
70. He K, Zhang X, Ren S et al. Deep Residual Learning for Image Recognition, CoRR 2015;abs/1512.03385.
71. Chen M, Wei Z, Huang Z et al. Simple and Deep Graph Convolutional Networks, arXiv [cs.LG] 2020.
72. Susan S, Kumar A. The balancing trick: Optimized sampling of imbalanced datasets—A brief survey of the recent State of the Art, Engineering Reports 2020;3.

73. Shin WH, Zhu X, Bures MG et al. Three-dimensional compound comparison methods and their application in drug discovery, Molecules 2015;20:12841-12862.
74. Hadfield TE, Deane CM. AI in 3D compound design, Curr Opin Struct Biol 2022;73:102326.

## Supplementary information for "Enhanced compound-protein binding affinity prediction by representing protein multimodal information via a coevolutionary strategy"

Binjie Guo[1,2,3,6], Hanyu Zheng[1,2,3,6], Haohan Jiang[1,2,3,6], Xiaodan Li[2,3], Naiyu Guan[2,3], Yanming Zuo[2,3], Yicheng Zhang[2,3], Hengfu Yang[5], Xuhua Wang[1,2,3,4,*]

Affiliations:

1. Department of Neurobiology and Department of Rehabilitation Medicine, First Affiliated Hospital, Zhejiang University School of Medicine, Hangzhou, Zhejiang Province 310003, China
2. Liangzhu Laboratory, MOE Frontier Science Center for Brain Science and Brain-machine Integration, State Key Laboratory of Brain-machine Intelligence, Zhejiang University, 1369 West Wenyi Road, Hangzhou 311121, China
3. NHC and CAMS Key Laboratory of Medical Neurobiology, Zhejiang University, Hangzhou 310058, China
4. Co-innovation Center of Neuroregeneration, Nantong University, Nantong, 226001 Jiangsu, PR China
5. School of Computer Science, Hunan First Normal University, Changsha, 410205 Hunan, PR China
6. These authors contributed equally

* Correspondence: xhw@zju.edu.cn (X.W.)

**Contents**

# 1. Supplementary Notes

## 1.1. The Oversmoothing Issue in GCNs

Deep graph convolutional networks (GCNs) have been very popular since 2017, when Kipf and Welling achieved great success by obtaining SOTA performance on a semisupervised classification task[1]. This method can also be used in biological research to represent compound features and optimize compound property predictions[2, 3]. However, this method always encounters an oversmoothing issue due to the limitation of depth[4]. In other words, the performance of the GCN becomes worse when the number of layers increases because the representations of the nodes in the GCN converge to approximately the same values. Applying the residual network (ResNet)[5] and appending residual connections in GCN models can hardly solve this problem, while oversmoothing in a GCN is a type of Laplacian smoothing. To circumvent this issue, inspired by GCNII[6], a specific residual connection with the initial features of each node in the molecular graph is applied to extract compound features in our work; this strategy increases the number of layers from 2 to 4, enabling the model to extract more information. We mathematically interpret the oversmoothing issue in a traditional GCN as follows.

First, we define a simple and connected undirected graph $G$ (Supplementary Fig. 1a) with $n$ nodes and $m$ edges. We use $A$ as the adjacency matrix and $D$ as the degree matrix of graph $G$, where $d(v_i)$ is the degree of node $v_i$. Let $\tilde{A}$ and $\widetilde{D}$ be the adjacency and degree matrices of graph G augmented with self-loops. The normalized graph Laplacian matrix is defined as $L = I - \tilde{P} = I - \widetilde{D}^{-1/2}\tilde{A}\widetilde{D}^{-1/2}$, and time proceeds in unit steps: $t = 1,2, \dots n$. At each time t, the walk stays at some node $v_i \in V$, and at time $t + 1$, based on the transition matrix $P$, as $P = AD^{-1}$, the walk randomly chooses one of $v_i$'s neighbors to move to (Supplementary Fig. 1b); this is described as a random walk. A lazy random walk is a modified version of the original random walk. In a lazy random walk, at time t, the walker stays at the current vertex with the probability of $\frac{1}{2}$ and takes a step as in the original random walk with the probability of $1/2$ (Supplementary Fig. 1c).

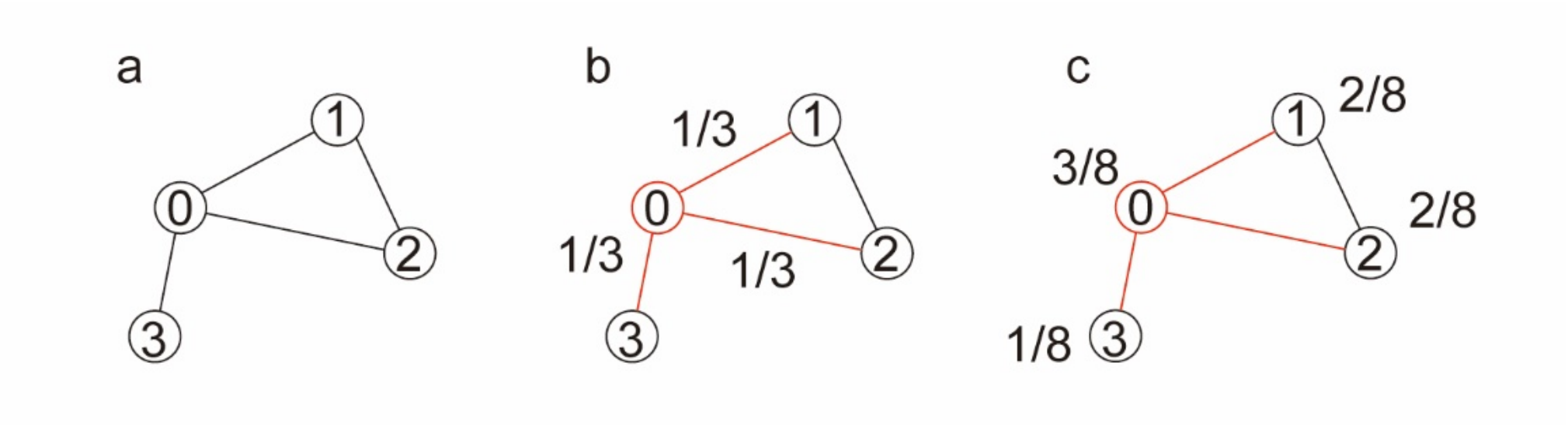

**Supplementary Fig. 1 Graph representation.** Figs. 1a-c. represent three iterations in a graph.

We define a probability vector $\pi$ that corresponds to the stationary distribution of the random walk. At time $t$, $\pi^{t+1} = P \cdot \pi^t = AD^{-1} \cdot \pi^t$, and $\pi(v_i) = \frac{\mathrm{d}_{v_i}}{2m}$. This breaks the periodicity of the random walk and forgets the initial graph information.

Because a deep GCN faces the oversmoothing problem, we first consider a multilayer GCN:

$$H^{(l+1)} = \tilde{P} \cdots \sigma(\tilde{P}\sigma\big(\tilde{P}XW^{(0)}\big)W^{(1)}) \cdots W^{(l)}$$

$W^{(l)}$is a layer-specific trainable weight matrix, $H^{(l)}$is the matrix of activations in the $l$th layer, $H(0) = X$, and $\sigma(\cdot)$ denotes an activation function. First, ignoring $\sigma(\cdot)$, we can describe the matrix as $H^{(K)} = \tilde{P}^K XW, \tilde{P} = \tilde{D}^{-1/2}\tilde{A}\tilde{D}^{-1/2}$, and then expand the calculation; we obtain

$$\tilde{P}^K = \tilde{D}^{-1/2}\tilde{A}\tilde{D}^{-1}\tilde{A}\tilde{D}^{-1} \cdots \tilde{A}\tilde{D}^{-1}\tilde{A}\tilde{D}^{-1/2}$$

$$= \tilde{D}^{-1/2}(\tilde{A}\tilde{D}^{-1})(\tilde{A}\tilde{D}^{-1}) \cdots (\tilde{A}\tilde{D}^{-1})\tilde{A}\tilde{D}^{-1/2} \cdot \tilde{D}^{-1/2} \cdot \tilde{D}^{1/2}$$

$$= \tilde{D}^{-1/2}(\tilde{A}\tilde{D}^{-1})^K\tilde{D}^{1/2}$$

This demonstrates that as the number of layers increases, the nodes in the GCN converge to certain values; this convergence makes the initial information indistinguishable, degrading the performance of the GCN.

**Supplementary Table 1** Overall statistics of the datasets extracted from PDBbind, BindingDB and Binding MOAD. The refined set of PDBbind only contains the measurements with $K_i$ and $K_d$. BindingDB is rich in measured $IC_{50}$ values (more than 500 thousand data, due to the defects of the PDB file and some compounds could not apply graph representation (atoms with more than 6 adjacent nodes), we only obtained 218,615 compound-protein paired data, while the collections of the measured values obtained based on $K_i$ and $K_d$ are significantly smaller (40 thousand $K_i$ measurements and 28 thousand $K_d$ measurements are recorded). In this paper, to construct large datasets from BindingDB, we only select the measured $IC_{50}$ values to generate training data. Note: Although the data of compound-protein pairs are fairly rich in BindingDB, the diversity of proteins remains very low compared with the data in PDBbind [7]. To test the generalization ability of the models, we constructed new datasets from the Binding MOAD database and excluded the complexes that appeared in the datasets (training, validation, and test datasets) constructed from PDBbind. For a fair comparison of the generalization ability, we limit the datasets constructed from Binding MOAD with the measurement of $IC_{50}$ and KIKD to the same number of compounds. Thus, we constructed the dataset using the results of the $IC_{50}$ and KIKD measurements from the “all of Binding MOAD” and “nonredundant MOAD” sets in the Binding MOAD database.

| Database | Measurement | Quantity | Compound Amount | PDB Entries | Max Affinity | Min Affinity |
|---|---|---|---|---|---|---|
| **PDBbind (general)** | **KIKD** | 7,156 | 5,691 | 7,156 | 15.2218 | 0.3979 |
| | **$IC_{50}$** | 5,543 | 5,243 | 5,543 | 11.5229 | 0.4498 |
| **PDBbind (refined)** | **KIKD** | 2768 | 2475 | 2768 | 11.9208 | 2.0 |
| **BindingDB** | **$IC_{50}$** | 218,615 | 183,584 | 2,248 | 11.0458 | 2.3468 |
| **Binding MOAD** | **$IC_{50}$** | 1963 | 1862 | 1952 | 11.5003 | 0.4226 |
| | **KIKD** | 1915 | 1285 | 1884 | 13.9586 | 0.0773 |

**Supplementary Table 2** Parameter settings of FeatNN training on the datasets generated from the PDBbind (both refined and general sets) and BindingDB databases. Note: $FeatNN^{optm}$ follows the same settings.

| parameter name | value |
|---|---|
| Hidden size (in the entire architecture) | 128 |
| Dropout probability | 0.1 |
| Number of attention heads in the deep GCN block | 4 |
| Number of attention heads in the Evo-Updating block | 4 |
| Layers of deep GCN blocks | 6 |
| Layers of Evo-Updating blocks | 2 |
| $\alpha$ in the deep GCN block | 0.2 |
| $\lambda$ in the deep GCN block | 0.5 |
| Maximum number of neighbors for each atom node | 6 |
| DDM word embedding size | 40 |
| Torsion size (both the sine and cosine values of $\Phi$ and $\psi$ on the backbone) | 4 |
| Kernel size in all CNN layers | 11 |
| Padding size in all CNN layers | 5 |
| Stride in all CNN layers | 1 |

**Supplementary Table 3** Model performance comparisons for the compound-clustered group and protein-clustered group. The models are ordered by their perfo

clustered test group in terms of the $R^2$ for $IC_{50}$. FeatNN outperforms the other models by significant margins in all metrics and on both affinity measurements. I shown as the mean value and standard deviation (SD) by 5-fold cross-validations with 10 independent experiments. The mean value (and SD) of each indepen shown in the table.

| Type | Threshold | Model | $R^2$ | | RMSE | | Pearson | | Spear |
|---|---|---|---|---|---|---|---|---|---|
| | | | $IC_{50}$ | KIKD | $IC_{50}$ | KIKD | $IC_{50}$ | KIKD | $IC_{50}$ |
| Compound-Cluster | 0.3 | FeatNN | 0.512(0.022) | 0.487(0.027) | 1.130(0.045) | 1.442(0.046) | 0.724(0.015) | 0.716(0.015) | 0.697(0.018) |
| | | MONN | 0.422(0.033) | 0.416(0.033) | 1.215(0.015) | 1.485(0.054) | 0.682(0.019) | 0.679(0.024) | 0.661(0.02) |
| | | BACPI | 0.318(0.029) | 0.381(0.043) | 1.289(0.027) | 1.507(0.051) | 0.614(0.010) | 0.633(0.024) | 0.592(0.007) |
| | | GATNet | 0.011(0.031) | 0.182(0.032) | 1.986(0.031) | 1.764(0.034) | 0.182(0.044) | 0.441(0.026) | 0.182(0.038) |
| | | GATGCN | 0.139(0.040) | 0.248(0.027) | 1.853(0.044) | 1.692(0.030) | 0.401(0.040) | 0.511(0.021) | 0.582(0.024) |
| | | GCNNet | 0.124(0.055) | 0.193(0.043) | 1.869(0.058) | 1.752(0.047) | 0.374(0.057) | 0.467(0.026) | 0.361(0.054) |
| | | GINConvNet | 0.164(0.059) | 0.216(0.036) | 1.825(0.064) | 1.727(0.040) | 0.480(0.063) | 0.488(0.026) | 0.517(0.071) |
| | | SIGN | -0.108(0.067) | -0.05(0.032) | 1.366(0.042) | 1.493(0.023) | 0(0) | 0.167(0) | 0(0) |
| | 0.4 | FeatNN | 0.442(0.031) | 0.406(0.073) | 1.202(0.047) | 1.540(0.079) | 0.684(0.018) | 0.669(0.040) | 0.669(0.02) |
| | | MONN | 0.385(0.014) | 0.369(0.057) | 1.251(0.025) | 1.534(0.064) | 0.655(0.010) | 0.643(0.030) | 0.631(0.008) |
| | | BACPI | 0.365(0.020) | 0.251(0.020) | 1.378(0.022) | 1.667(0.022) | 0.632(0.009) | 0.575(0.008) | 0.610(0.006) |
| | | GATNet | -0.003(0.039) | 0.182(0.016) | 2.000(0.039) | 1.764(0.017) | 0.173(0.051) | 0.439(0.019) | 0.173(0.053) |
| | | GATGCN | 0.137(0.047) | 0.235(0.015) | 1.855(0.051) | 1.706(0.017) | 0.397(0.040) | 0.503(0.012) | 0.379(0.043) |
| | | GCNNet | 0.081(0.019) | 0.224(0.021) | 1.915(0.019) | 1.718(0.023) | 0.327(0.034) | 0.491(0.021) | 0.313(0.038) |
| | | GINConvNet | 0.231(0.036) | 0.213(0.063) | 1.752(0.040) | 1.729(0.069) | 0.522(0.010) | 0.490(0.037) | 0.560(0.006) |
| | | SIGN | -0.064(0.09) | -0.005(0.003) | 1.37(0.057) | 1.575(0.07) | 0.089(0) | 0.036(0.033) | 0.103(0) |
| | 0.5 | FeatNN | 0.365(0.039) | 0.438(0.036) | 1.281(0.039) | 1.507(0.056) | 0.636(0.028) | 0.685(0.016) | 0.608(0.027) |
| | | MONN | 0.331(0.045) | 0.299(0.061) | 1.306(0.045) | 1.624(0.109) | 0.626(0.028) | 0.611(0.051) | 0.603(0.024) |
| | | BACPI | 0.276(0.017) | 0.264(0.048) | 1.372(0.016) | 1.768(0.056) | 0.563(0.007) | 0.542(0.027) | 0.533(0.005) |
| | | GATNet | 0.013(0.023) | 0.161(0.021) | 1.984(0.023) | 1.786(0.022) | 0.188(0.036) | 0.425(0.011) | 0.194(0.034) |
| | | GATGCN | 0.159(0.030) | 0.239(0.029) | 1.831(0.033) | 1.701(0.032) | 0.424(0.031) | 0.504(0.022) | 0.409(0.021) |
| | | GCNNet | 0.083(0.060) | 0.212(0.008) | 1.912(0.062) | 1.731(0.009) | 0.327(0.062) | 0.479(0.007) | 0.315(0.061) |
| | | GINConvNet | 0.221(0.038) | 0.231(0.026) | 1.763(0.042) | 1.710(0.029) | 0.507(0.051) | 0.500(0.022) | 0.540(0.057) |
| | | SIGN | -0.069(0.034) | -0.077(0.081) | 1.484(0.024) | 1.543(0.084) | 0(0) | 0.060(0.171) | 0(0) |
| | 0.6 | FeatNN | 0.339(0.023) | 0.398(0.043) | 1.295(0.053) | 1.429(0.088) | 0.613(0.021) | 0.660(0.028) | 0.59(0.022) |
| | | MONN | 0.210(0.093) | 0.248(0.057) | 1.399(0.092) | 1.681(0.155) | 0.572(0.030) | 0.559(0.057) | 0.554(0.029) |
| | | BACPI | 0.132(0.033) | 0.278(0.024) | 1.569(0.030) | 1.629(0.027) | 0.482(0.009) | 0.582(0.011) | 0.467(0.005) |
| | | GATNet | 0(0.035) | 0.167(0.034) | 1.998(0.035) | 1.780(0.036) | 0.181(0.035) | 0.423(0.033) | 0.176(0.038) |
| | | GATGCN | 0.141(0.060) | 0.244(0.051) | 1.850(0.064) | 1.695(0.057) | 0.399(0.055) | 0.511(0.036) | 0.382(0.064) |
| | | GCNNet | 0.072(0.054) | 0.210(0.019) | 1.923(0.056) | 1.733(0.021) | 0.302(0.059) | 0.475(0.012) | 0.293(0.06) |
| | | GINConvNet | 0.196(0.042) | 0.198(0.083) | 1.791(0.046) | 1.745(0.087) | 0.488(0.026) | 0.483(0.045) | 0.513(0.031) |
| | | SIGN | -0.181(0.032) | -0.066(0.012) | 1.509(0.021) | 1.428(0.085) | 0.098(0) | 0(0) | 0.115(0) |

| Type | Threshold | Model | $R^2$ | | RMSE | | Pearson | | Spear |
|---|---|---|---|---|---|---|---|---|---|
| | | | $IC_{50}$ | KIKD | $IC_{50}$ | KIKD | $IC_{50}$ | KIKD | $IC_{50}$ |
| Protein-Cluster | 0.3 | FeatNN | 0.285(0.039) | 0.326(0.050) | 1.371(0.068) | 1.647(0.067) | 0.552(0.027) | 0.586(0.036) | 0.538(0.024) |
| | | MONN | 0.247(0.058) | 0.306(0.063) | 1.383(0.046) | 1.642(0.049) | 0.537(0.042) | 0.579(0.044) | 0.515(0.048) |
| | | BACPI | 0.154(0.015) | 0.276(0.034) | 1.446(0.013) | 1.771(0.041) | 0.491(0.009) | 0.558(0.020) | 0.475(0.01) |
| | | GATNet | 0.012(0.015) | 0.161(0.009) | 1.985(0.016) | 1.786(0.010) | 0.169(0.061) | 0.423(0.006) | 0.177(0.062) |
| | | GATGCN | 0.211(0.037) | 0.252(0.015) | 1.774(0.042) | 1.687(0.017) | 0.468(0.034) | 0.519(0.009) | 0.453(0.037) |
| | | GCNNet | 0.062(0.045) | 0.224(0.021) | 1.934(0.046) | 1.718(0.023) | 0.279(0.078) | 0.487(0.017) | 0.276(0.076) |
| | | GINConvNet | 0.234(0.023) | 0.244(0.018) | 1.748(0.027) | 1.696(0.020) | 0.525(0.009) | 0.512(0.013) | 0.562(0.009) |
| | | SIGN | -0.066(0.087) | 0.047(0.125) | 1.464(0.06) | 1.433(0.098) | 0.091(0) | 0.232(0.264) | 0.112(0) |
| | 0.4 | FeatNN | 0.292(0.045) | 0.324(0.029) | 1.364(0.045) | 1.643(0.039) | 0.559(0.035) | 0.586(0.028) | 0.535(0.049) |
| | | MONN | 0.244(0.048) | 0.289(0.027) | 1.399(0.060) | 1.671(0.050) | 0.551(0.035) | 0.568(0.021) | 0.528(0.04) |
| | | BACPI | 0.128(0.026) | 0.279(0.015) | 1.529(0.023) | 1.758(0.019) | 0.476(0.005) | 0.566(0.008) | 0.458(0.007) |
| | | GATNet | 0.014(0.037) | 0.176(0.014) | 1.983(0.037) | 1.770(0.015) | 0.210(0.036) | 0.433(0.017) | 0.215(0.046) |
| | | GATGCN | 0.131(0.056) | 0.240(0.018) | 1.861(0.060) | 1.701(0.020) | 0.397(0.048) | 0.507(0.013) | 0.381(0.052) |
| | | GCNNet | 0.114(0.047) | 0.214(0.045) | 1.880(0.049) | 1.729(0.049) | 0.366(0.046) | 0.478(0.028) | 0.35(0.047) |
| | | GINConvNet | 0.202(0.053) | 0.223(0.065) | 1.784(0.059) | 1.718(0.072) | 0.512(0.029) | 0.497(0.048) | 0.552(0.028) |
| | | SIGN | -0.019(0.029) | -0.015(0.007) | 1.394(0.019) | 1.446(0.005) | 0.143(0) | 0.162(0) | 0.179(0) |
| | 0.5 | FeatNN | 0.283(0.041) | 0.307(0.021) | 1.378(0.045) | 1.659(0.057) | 0.552(0.021) | 0.570(0.020) | 0.538(0.017) |
| | | MONN | 0.249(0.065) | 0.288(0.028) | 1.426(0.027) | 1.662(0.055) | 0.556(0.050) | 0.566(0.020) | 0.536(0.062) |
| | | BACPI | 0.081(0.029) | 0.304(0.016) | 1.464(0.023) | 1.727(0.02) | 0.461(0.008) | 0.575(0.007) | 0.444(0.006) |
| | | GATNet | 0.027(0.042) | 0.174(0.031) | 1.970(0.042) | 1.772(0.033) | 0.192(0.098) | 0.432(0.029) | 0.191(0.1) |
| | | GATGCN | 0.153(0.035) | 0.245(0.022) | 1.838(0.038) | 1.694(0.025) | 0.410(0.035) | 0.508(0.020) | 0.395(0.039) |
| | | GCNNet | 0.096(0.052) | 0.233(0.033) | 1.899(0.055) | 1.708(0.037) | 0.341(0.057) | 0.493(0.035) | 0.331(0.063) |
| | | GINConvNet | 0.212(0.044) | 0.201(0.044) | 1.772(0.049) | 1.743(0.048) | 0.493(0.051) | 0.480(0.044) | 0.529(0.059) |
| | | SIGN | -0.033(0.025) | -0.058(0.059) | 1.389(0.048) | 1.515(0.041) | 0.156(0) | 0(0) | 0.186(0) |
| | 0.6 | FeatNN | 0.285(0.032) | 0.343(0.050) | 1.366(0.054) | 1.640(0.045) | 0.555(0.015) | 0.598(0.032) | 0.532(0.023) |
| | | MONN | 0.204(0.059) | 0.288(0.043) | 1.440(0.045) | 1.676(0.053) | 0.514(0.030) | 0.570(0.030) | 0.500(0.03) |
| | | BACPI | 0.012(0.036) | 0.286(0.021) | 1.575(0.029) | 1.700(0.025) | 0.412(0.014) | 0.558(0.013) | 0.385(0.015) |
| | | GATNet | 0.024(0.033) | 0.184(0.023) | 1.973(0.034) | 1.761(0.025) | 0.220(0.040) | 0.444(0.020) | 0.225(0.038) |

| | | | | | | | | |
|---|---|---|---|---|---|---|---|---|
| | GATGCN | 0.174(0.035) | 0.225(0.040) | 1.815(0.038) | 1.716(0.045) | 0.435(0.027) | 0.499(0.018) | 0.418(0.031) |
| | GCNNet | 0.090(0.041) | 0.232(0.013) | 1.905(0.042) | 1.710(0.014) | 0.341(0.029) | 0.491(0.012) | 0.328(0.034) |
| | GINConvNet | 0.186(0.057) | 0.226(0.070) | 1.801(0.061) | 1.714(0.077) | 0.485(0.058) | 0.503(0.036) | 0.527(0.061) |
| | SIGN | -0.033(0.079) | -0.015(0.013) | 1.409(0.055) | 1.436(0.009) | 0.074(0.166) | 0.084(0) | 0.088(0.198) |

**Supplementary Table 4** Comparison of the performances of FeatNN on the datasets generated from the general set and from refined set of PDBbind with th protein-clustered strategy. The results of each group were obtained with 5 independent experiments by 5-fold cross-validation strategy. Details are provided in

| FeatNN | Threshold | RMSE | | Pearson | | Spearman | | |
|---|---|---|---|---|---|---|---|---|
| Type | | refined | general | refined | general | refined | general | refined |
| | 0.3 | 1.38(0.071) | 1.442(0.046) | 0.735(0.029) | 0.716(0.015) | 0.729(0.032) | 0.714(0.019) | 0.512(0.05 |
| Compound- | 0.4 | 1.469(0.063) | 1.54(0.079) | 0.698(0.029) | 0.669(0.04) | 0.700(0.03) | 0.677(0.041) | 0.448(0.05 |
| Clustered | 0.5 | 1.448(0.09) | 1.507(0.056) | 0.699(0.061) | 0.685(0.016) | 0.690(0.064) | 0.674(0.023) | 0.440(0.11 |
| | 0.6 | 1.442(0.148) | 1.429(0.088) | 0.672(0.019) | 0.66(0.028) | 0.636(0.029) | 0.634(0.022) | 0.421(0.03 |
| Protein- | 0.3 | 1.642(0.104) | 1.647(0.067) | 0.558(0.125) | 0.586(0.036) | 0.552(0.127) | 0.577(0.039) | 0.278(0.18 |
| Clustered | 0.4 | 1.617(0.062) | 1.643(0.039) | 0.578(0.143) | 0.586(0.028) | 0.579(0.143) | 0.572(0.018) | 0.305(0.20 |
| | 0.5 | 1.615(0.111) | 1.659(0.057) | 0.579(0.066) | 0.570(0.02) | 0.577(0.081) | 0.562(0.017) | 0.305(0.07 |
| | 0.6 | 1.654(0.052) | 1.64(0.045) | 0.549(0.073) | 0.598(0.032) | 0.553(0.082) | 0.583(0.027) | 0.277(0.07 |

**Supplementary Table 5** Comparison of the performances of the SOTA baseline (MONN) on the datasets generated from the general set and refined set of PD clustered and protein-clustered strategy. The results of each group were obtained with 5 independent experiments by 5-fold cross-validation strategy. Details are Fig. 5.

| MONN | Threshold | RMSE | | Pearson | | Spearman | | |
|---|---|---|---|---|---|---|---|---|
| Type | | refined | general | refined | general | refined | general | refined |
| | 0.3 | 1.438(0.075) | 1.485(0.054) | 0.716(0.021) | 0.679(0.024) | 0.71(0.022) | 0.679(0.031) | 0.481(0.024 |
| Compound- | 0.4 | 1.514(0.172) | 1.534(0.064) | 0.684(0.021) | 0.643(0.03) | 0.683(0.028) | 0.641(0.039) | 0.391(0.084 |
| Clustered | 0.5 | 1.516(0.083) | 1.624(0.109) | 0.668(0.029) | 0.611(0.051) | 0.664(0.032) | 0.605(0.054) | 0.403(0.053 |
| | 0.6 | 1.496(0.132) | 1.681(0.155) | 0.638(0.055) | 0.559(0.057) | 0.617(0.051) | 0.544(0.032) | 0.36(0.094 |
| Protein- | 0.3 | 1.702(0.075) | 1.642(0.049) | 0.539(0.079) | 0.579(0.044) | 0.537(0.089) | 0.572(0.038) | 0.236(0.112 |
| Clustered | 0.4 | 1.651(0.072) | 1.671(0.05) | 0.546(0.069) | 0.568(0.021) | 0.544(0.066) | 0.561(0.021) | 0.251(0.098 |
| | 0.5 | 1.646(0.089) | 1.662(0.055) | 0.552(0.069) | 0.566(0.02) | 0.557(0.079) | 0.559(0.023) | 0.281(0.08 |
| | 0.6 | 1.68(0.123) | 1.676(0.053) | 0.502(0.058) | 0.57(0.03) | 0.500(0.054) | 0.558(0.037) | 0.176(0.075 |

**Supplementary Table 6** Comparison of FeatNN and the SOTA baseline (MONN) with regard to the generalization ability on the datasets generated from the g

PDBbind. The generalization abilities of $FeatNN^{refine}$ and the SOTA $baseline^{refine}$ decrease compared with the corresponding methods trained on the genera because the amount of the data affects the training process. The results of each group were tested on the dataset constructed from Binding MOAD with at lea

Considering that the refined set only contains the measurement of $K_i$ and $K_d$, we use the $FeatNN^{general}$ trained on the KIKD dataset constructed from PDBbind as th the test dataset constructed from Binding MOAD in this part is based on the measurement of $K_i$ and $K_d$.

| **Model** | **RMSE** | | **Pearson** | | **Spearman** | | **$R^2$** | |
|---|---|---|---|---|---|---|---|---|
| | **general** | **refined** | **general** | **refined** | **general** | **refined** | **general** | **refin** |
| **FeatNN** | 1.656(0.033) | 1.925(0.046) | 0.647(0.017) | 0.47(0.024) | 0.656(0.02) | 0.465(0.023) | 0.359(0.025) | 0.133(0 |
| **SOTA Baseline** | 1.668(0.082) | 2.042(0.072) | 0.612(0.052) | 0.378(0.025) | 0.592(0.055) | 0.358(0.019) | 0.348(0.067) | 0.024( |

**Supplementary Table 7** Pretraining and Fine-tuning Results of FeatNN. This process was applied to the datasets constructed from the general set of PDBbin results. FeatNN$^{general}$ is also trained on the general set of PDBbind with the measurement of $IC_{50}$. The results of each group were obtained from 10 independ cross-validation strategy.

| Model | RMSE | Pearson | Spearman | $R^2$ |
|---|---|---|---|---|
| **FeatNN$^{general}$** | 1.130(0.045) | 0.724(0.015) | 0.697(0.018) | 0.512(0.022) |
| **FeatNN$^{optm}$** | 1.094(0.006) | 0.738(0.003) | 0.712(0.004) | 0.540(0.005) |

**Supplementary Table 8** Comparison of FeatNN$^{general}$, FeatNN$^{optm}$ and the SOTA baseline (MONN which is trained on the dataset constructed from the general s to the generalization ability. Because FeatNN$^{optm}$ is pretrained on BindingDB only with $IC_{50}$ measurements, all of these models are tested on the datasets constr using the $IC_{50}$ measurement results. Thus, both the FeatNN$^{general}$ and SOTA baseline$^{general}$ are trained on the general set of PDBbind using the measured $IC_{50}$ v group were obtained from at least 10 independent experiments.

| Model | RMSE | Pearson | Spearman | $R^2$ |
|---|---|---|---|---|
| **SOTA Baseline$^{general}$** | 1.339(0.044) | 0.657(0.014) | 0.621(0.016) | 0.385(0.041) |
| **FeatNN$^{general}$** | 1.267(0.036) | 0.687(0.019) | 0.660(0.018) | 0.449(0.032) |
| **FeatNN$^{optm}$** | 1.238(0.019) | 0.701(0.011) | 0.683(0.008) | 0.475(0.016) |

**Supplementary Table 9** Ablation study for module deletion in FeatNN. "Entire FeatNN" refers to the full proposed FeatNN model. Here, "Only Sequence" or "Only Structure" indicate only the protein sequence information or structure information being used when representing the features of protein by protein extractor. "Without Interact Mat" indicates the ablation of the compound-protein interactive matrix in the affinity learning module, which could help FeatNN to learn the interaction information between compound and protein possibly. The performances are sorted by the $R^2$ values of the respective variant models. The results of each group were obtained from 10 independent experiments by 5-fold cross-validation strategy. The mean value (and SD) of each independent experimental group is shown in the table.

| Name | $R^2$ | RMSE | Pearson | Spearman |
|---|---|---|---|---|
| Entire FeatNN | **0.512(0.022)** | **1.130(0.045)** | **0.724(0.015)** | **0.697(0.018)** |
| Without Torsion Info | 0.443(0.045) | 1.196(0.025) | 0.692(0.021) | 0.672(0.027) |
| Without MasterNode | 0.388(0.048) | 1.255(0.032) | 0.653(0.028) | 0.636(0.026) |
| Without Evo-Updating | 0.352(0.055) | 1.292(0.024) | 0.640(0.029) | 0.624(0.027) |
| Without Deep GCN | 0.326(0.055) | 1.302(0.042) | 0.611(0.024) | 0.591(0.036) |
| Only Sequence | 0.325(0.089) | 1.325(0.105) | 0.619(0.043) | 0.592(0.05) |
| Without Interact Mat | 0.317(0.078) | 1.299(0.093) | 0.584(0.071) | 0.56(0.077) |
| Only Structure | 0.157(0.058) | 1.477(0.034) | 0.480(0.022) | 0.451(0.021) |

**Supplementary Table 10** Targeting SARS-CoV-2 3C-like protease with the conformation constructed from the PDB file with PDB-id of 7CWC, we applied FeatNN and MONN (SOTA baseline) to predict the listed 28 validated bioactive compounds to test the CPAs prediction precision of FeatNN. Each result was obtained by the average of 15 independent experiments. Real affinity values were collected from published papers and are listed in the references.

| Compound Name | Real Affinity Value | FeatNN Prediction Value | MONN Prediction Value |
|---|---|---|---|
| Darunavir [8] | 4.442 | 5.467(0.282) | 5.955(0.285) |
| Cobicistat [9] | 7.495 | 6.346(0.557) | 6.457(0.583) |
| Ritonavir[10] | 4.863 | 5.430(0.326) | 5.469(0.219) |
| Tipranavir [11] | 4.875 | 5.074(0.091) | 5.733(0.264) |
| Ivermectin [12] | 5.699 | 6.279(0.405) | 7.821(0.397) |
| REMDESIVIR [11] | 4.943 | 4.388(0.274) | 4.443(0.113) |
| PF-07321332 [13] | 7.638 | 6.816(0.348) | 5.872(0.551) |
| PF-00835231 [14] | 8.398 | 6.785(0.254) | 5.947(1.406) |
| Lufotrelvir [15] | 8.097 | 5.775(0.253) | 6.049(0.572) |
| ML188 [16] | 5.824 | 5.011(0.219) | 4.709(0.236) |
| FB2001 [17] | 6.276 | 5.473(0.751) | 5.995(0.515) |
| Dalcetrapib [18] | 4.752 | 4.622(0.127) | 4.875(0.421) |
| EGCG Octaacetate [19] | 4.857 | 5.715(0.517) | 5.571(0.378) |
| Ellagic acid [19] | 4.928 | 5.810(0.511) | 5.571(0.683) |
| Curcumin [19] | 4.924 | 4.821(0.193) | 5.040(0.468) |
| Resveratrol [19] | 4.772 | 5.307(0.248) | 4.408(0.931) |
| Quercetin [19] | 4.631 | 5.750(0.291) | 5.422(0.319) |
| Chloroquine [20] | 5.567 | 4.910(0.155) | 4.805(0.210) |
| Lopinavir [11] | 5.040 | 4.954(0.172) | 4.927(0.283) |
| Azithromycin [11] | 5.674 | 5.614(0.519) | 5.753(0.510) |
| N4-Hydroxycytidine [11] | 6.523 | 5.921(0.658) | 6.023(0.707) |
| Molnupiravir [17] | 6.523 | 5.447(0.825) | 4.622(0.874) |
| GC-373 [15] | 6.456 | 6.681(0.277) | 6.549(0.465) |
| PF-07304814 [15] | 8.097 | 5.775(0.253) | 6.049(0.572) |
| Nirmatrelvir [21] | 7.796 | 6.816(0.348) | 5.872(0.551) |
| Boceprevir [11, 15] | 5.384 | 5.307(0.268) | 5.614(0.775) |
| Calpeptin [15] | 4.971 | 4.548(0.309) | 4.885(0.283) |
| Telaprevir [15] | 4.940 | 5.933(0.417) | 5.544(0.682) |

**Supplementary Table 11** Targeting Akt-1 protease with the conformation constructed from the PDB file with PDB-id of 3O96, we applied FeatNN and MONN (SOTA baseline) to predict the listed 10 validated bioactive compounds to test the CPA prediction precision of FeatNN. Each result was obtained by the average of 15 independent experiments. Real affinity values were collected from published papers and are listed in the references.

| Compound Name | Real Affinity Value | FeatNN Prediction Value | MONN Prediction Value |
|---|---|---|---|
| Capivasertib [22] | 9.046 | 6.963(0.108) | 6.763(0.809) |
| Ipatasertib [23] | 8.456 | 6.308(0.031) | 6.213(0.703) |
| GSK690693 [24] | 8.699 | 6.623(0.260) | 7.065(0.356) |
| Miransertib [25] | 8.569 | 6.815(0.874) | 6.29(0.262) |
| BAY1125976 [26] | 8.284 | 6.521(0.632) | 6.813(0.572) |
| AT7867 [27] | 7.495 | 6.311(0.788) | 6.322(0.073) |
| AT13148 [28] | 7.420 | 6.234(0.963) | 5.266(0.087) |
| Akti-1/2 [29] | 7.237 | 5.569(0.213) | 5.949(0.122) |
| Uprosertib [30] | 6.745 | 6.840(0.680) | 6.634(0.671) |
| Oridonin [31] | 5.076 | 5.302(0.433) | 5.743(0.403) |

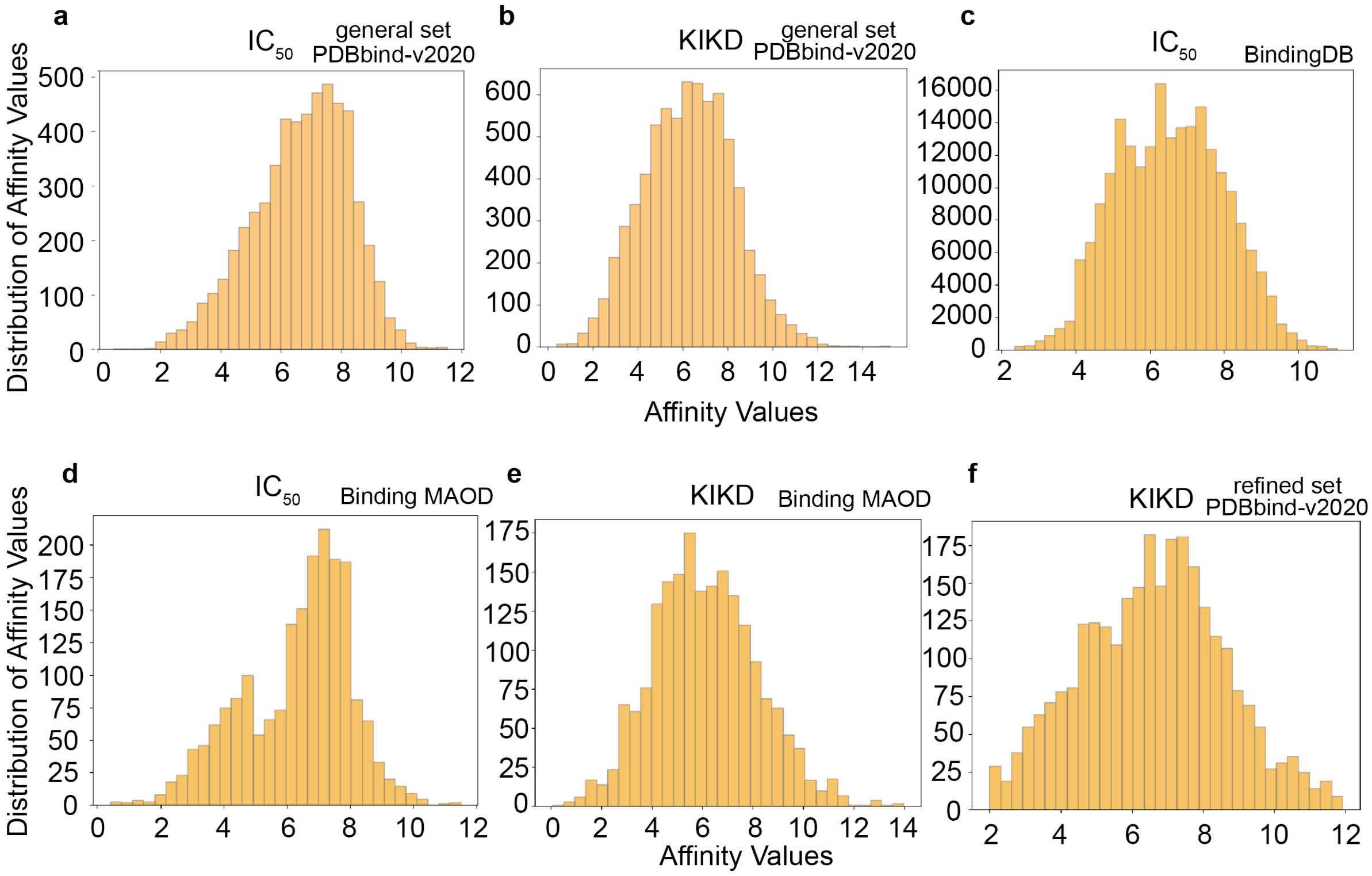

**Supplementary Fig. 2** The overall distributions of the affinity values in the PDBbind-v2020 dataset (**a.** $IC_{50}$ and **b.** KIKD, general set) and **c.** BindingDB dataset ($IC_{50}$). For a fair comparison of the generalization ability, we limit the datasets constructed from Binding MOAD with the measurements of **d.** $IC_{50}$ and **e.** KIKD to the same amount of data. Thus, we constructed the dataset with $IC_{50}$ and KIKD measurements from the "all of Binding MOAD" and "nonredundant MOAD" sets. **f.** shows the affinity value distribution on the refined set of PDBbind-v2020 that only contains the measurement of KIKD. All of these datasets produce approximately normal distributions with their values.

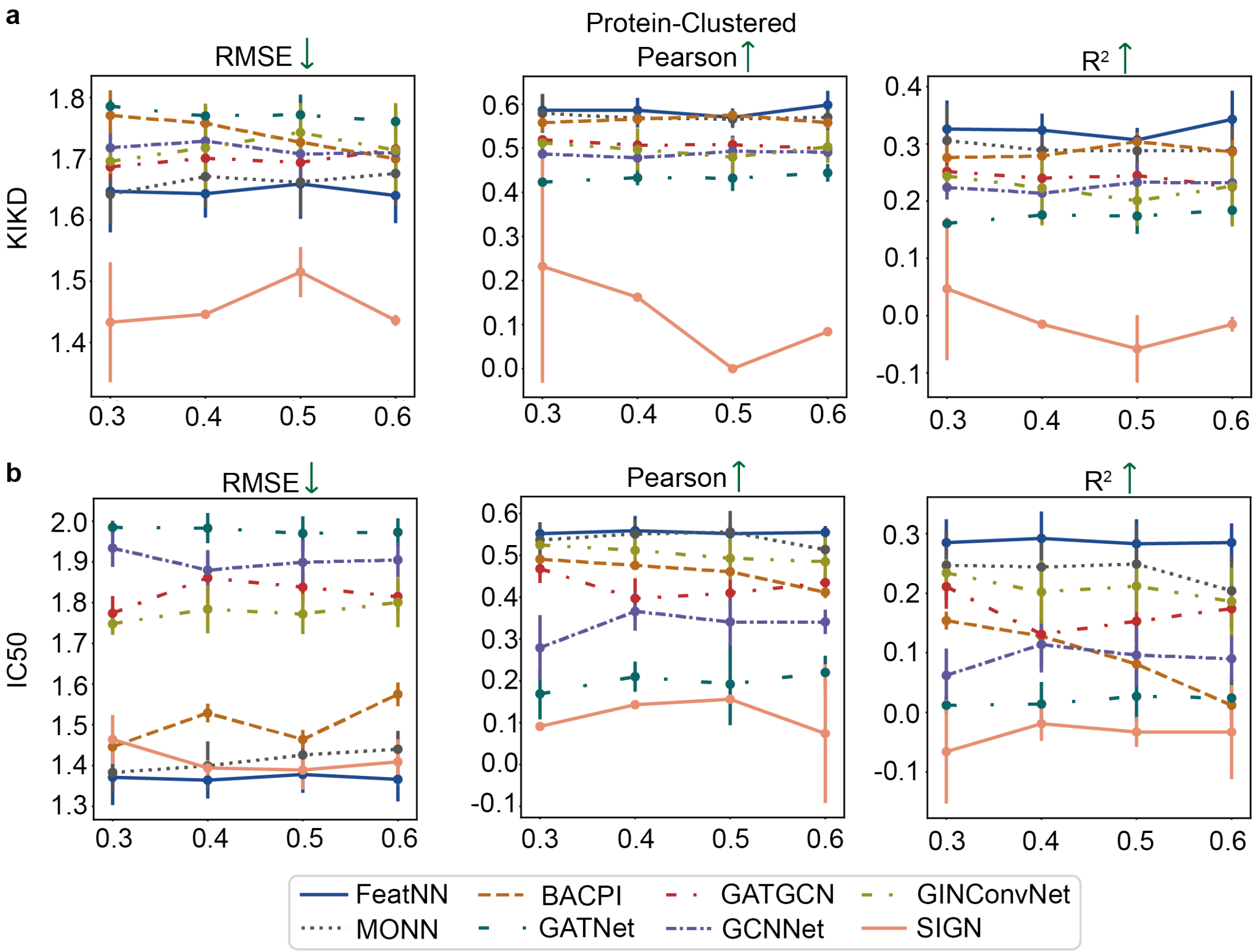

**Supplementary Fig. 3** Evaluation performances on the datasets generated from PDBbind with the protein-cluster strategy. **a.** Performance evaluated on the dataset generated from PDBbind with KIKD measurements. **b.** Performance evaluated on the dataset generated from PDBbind with $IC_{50}$ measurements. Performance results are plotted as the mean values and standard deviations (SD) by 5-fold cross-validation with 10 independent experiments. Each point represents the mean of an independent experimental group, with error bars indicating SD. Note: the results present here were slightly different from the results reported by the original literature [32], possibly because we use PDBbind-v2020 as our benchmark database instead of PDBbind-v2016 used in their study. In addition, considering the biology means behind the data, we split the dataset into two parts ("IC50" and "KIKD" [33]) instead of simply mixing the affinity measured with "$IC_{50}$", "$K_i$", and "$K_d$" together in their study. Moreover, we applied compound-cluster and protein-cluster strategies in our study to avoid data leakage caused by the biology-correlated knowledge (similarity structure or sequence in protein or compound). Thus, the results here may differ from the results in their article [32].

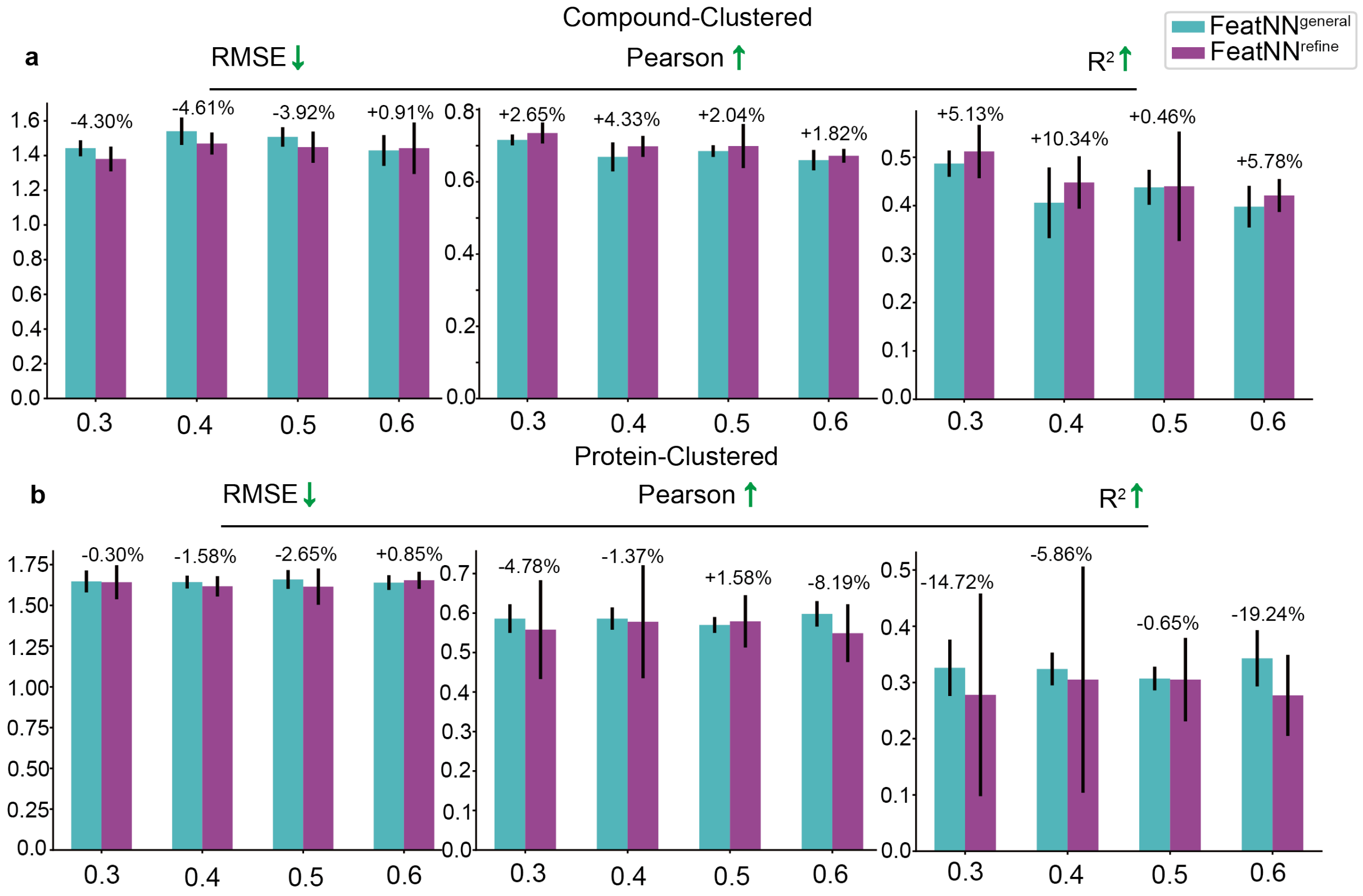

**Supplementary Fig. 4** Comparison of the performances of FeatNN on the datasets generated from the general set and refined set of PDBbind with the compound-clustered and protein-clustered strategy. **a.** Based on the compound-clustered method, $FeatNN^{refine}$ shows improved performance compared with $FeatNN^{general}$, possibly because high-quality structural information is introduced into the training process. **b.** However, the performance of $FeatNN^{refine}$ based on the protein-clustered method is much worse than that of $FeatNN^{general}$. The performance of $FeatNN^{refine}$ declined strongly, particularly at the threshold of 0.6 (which means that less similar proteins will appear during the training process). This result may be obtained because in the training process, both the amount of data and the diversity of protein information are more important than data quality [7]. The detailed data can be found in Supplementary Table 4. Note: the text on each group bar indicates the difference between the performance of the model trained on the refined dataset and the performance of the same model trained on the general dataset. $FeatNN^{refine}$ indicates the FeatNN trained and tested on the datasets generated from the refined set of PDBbind. $FeatNN^{general}$ indicates the FeatNN trained and tested on the datasets generated from the general set of PDBbind. Performance results are plotted as the mean values and standard deviations (SD) obtained by 5-fold cross-validation with 5

independent experiments. Each bar represents the mean of an experimental group, with error bars indicating the SD.

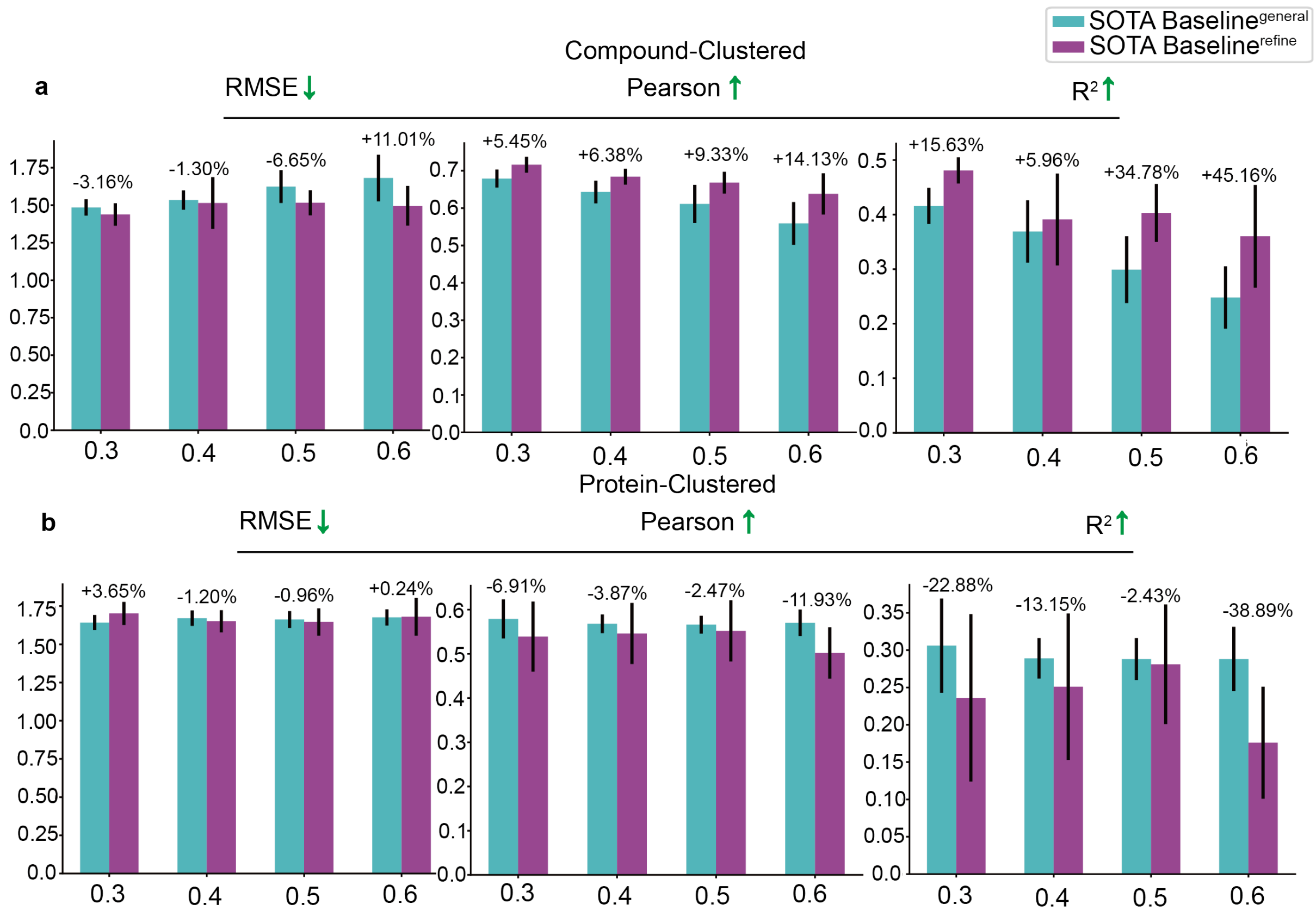

**Supplementary Fig. 5** Comparison of the performances of the SOTA baseline (MONN) on datasets generated from the general set and refined set of PDBbind with the compound-clustered and protein-clustered strategy. **a.** he performances of the SOTA baseline$^{refine}$ based on the compound-clustered method is more or less improved with the SOTA baseline$^{general}$, which is similar to the FeatNN$^{refine}$ results in Supplementary Fig. 4a. **b.** The performances of the SOTA baseline$^{refine}$ based on the protein-clustered method are much worse compared with the SOTA baseline$^{general}$. Additionally, the performance of the SOTA baseline$^{refine}$' declined significantly at the threshold of 0.6, supporting the hypothesis and result shown in Supplementary Fig. 4b. The detailed data can be found in Supplementary Table 5. Note: The text on each group bar indicates the difference between the performance of the model trained on the refined dataset and the performance of the same model trained on the general dataset. SOTA Baseline$^{refine}$ indicates the SOTA baseline trained and tested on the datasets generated from the refined set of PDBbind. SOTA Baseline$^{general}$ indicates the SOTA baseline trained and tested on the datasets generated from the general set of PDBbind. Performance results are plotted as the mean values and standard deviations (SD) obtained by 5-fold cross-validation

with 5 independent experiments. Each bar represents the mean of an experimental group, with error bars indicating the SD.

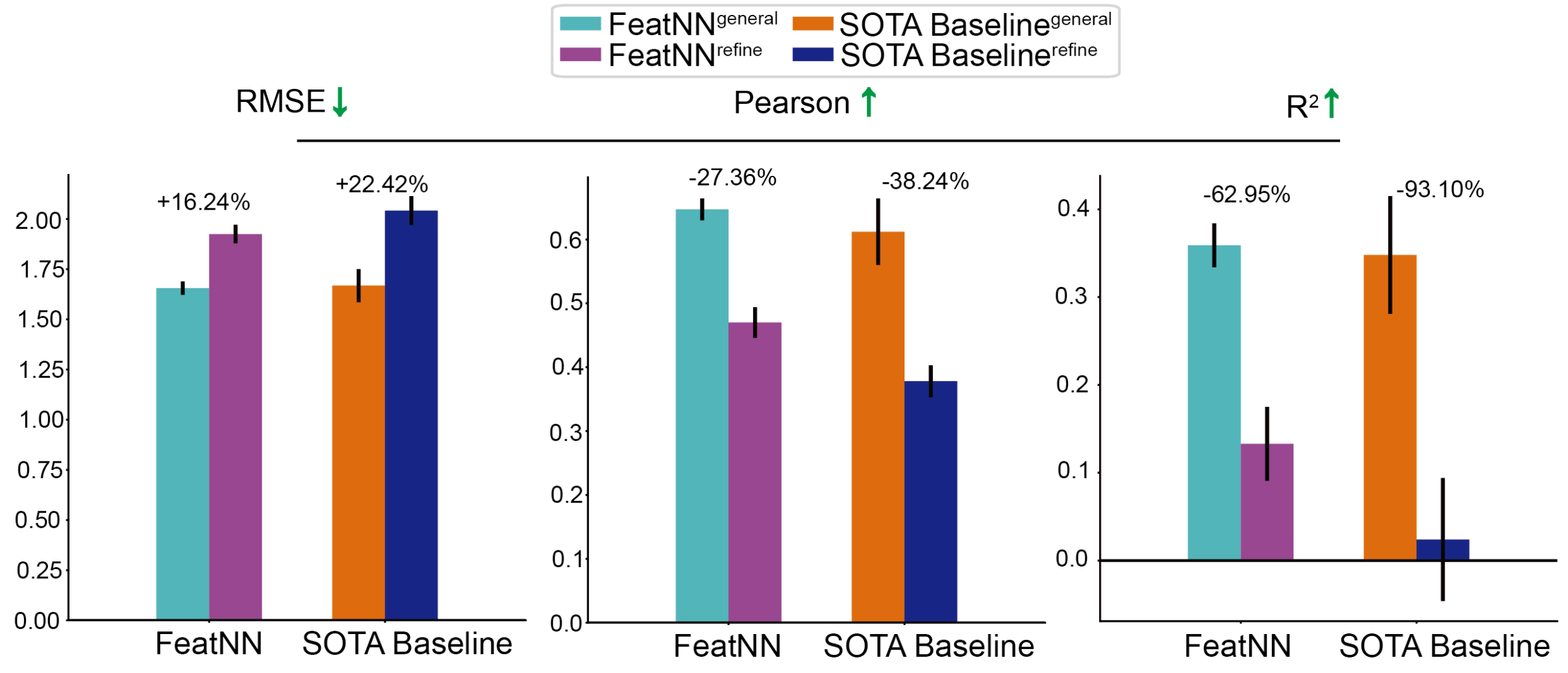

**Supplementary Fig. 6** Comparison of the generalization performance of FeatNN$^{general}$ versus FeatNN$^{refine}$ and SOTA baseline$^{general}$ versus SOTA baseline$^{refine}$ on the dataset generated from the Binding MOAD database. The detailed data can be found in Supplementary Table 6. Note: the text on each group bar indicates the difference between the performance of the model trained on the refined dataset and the performance of the same model trained on the general dataset. FeatNN$^{refine}$ and SOTA Baseline$^{refine}$ indicate that these two models were trained on the datasets generated from the refined set of PDBbind. FeatNN$^{general}$ and SOTA Baseline$^{general}$ indicate that these two models were trained on the datasets generated from the general set of PDBbind. Performance results are plotted as the mean values and standard deviations (SD). Each bar represents the mean of an experimental group, with error bars indicating the SD.

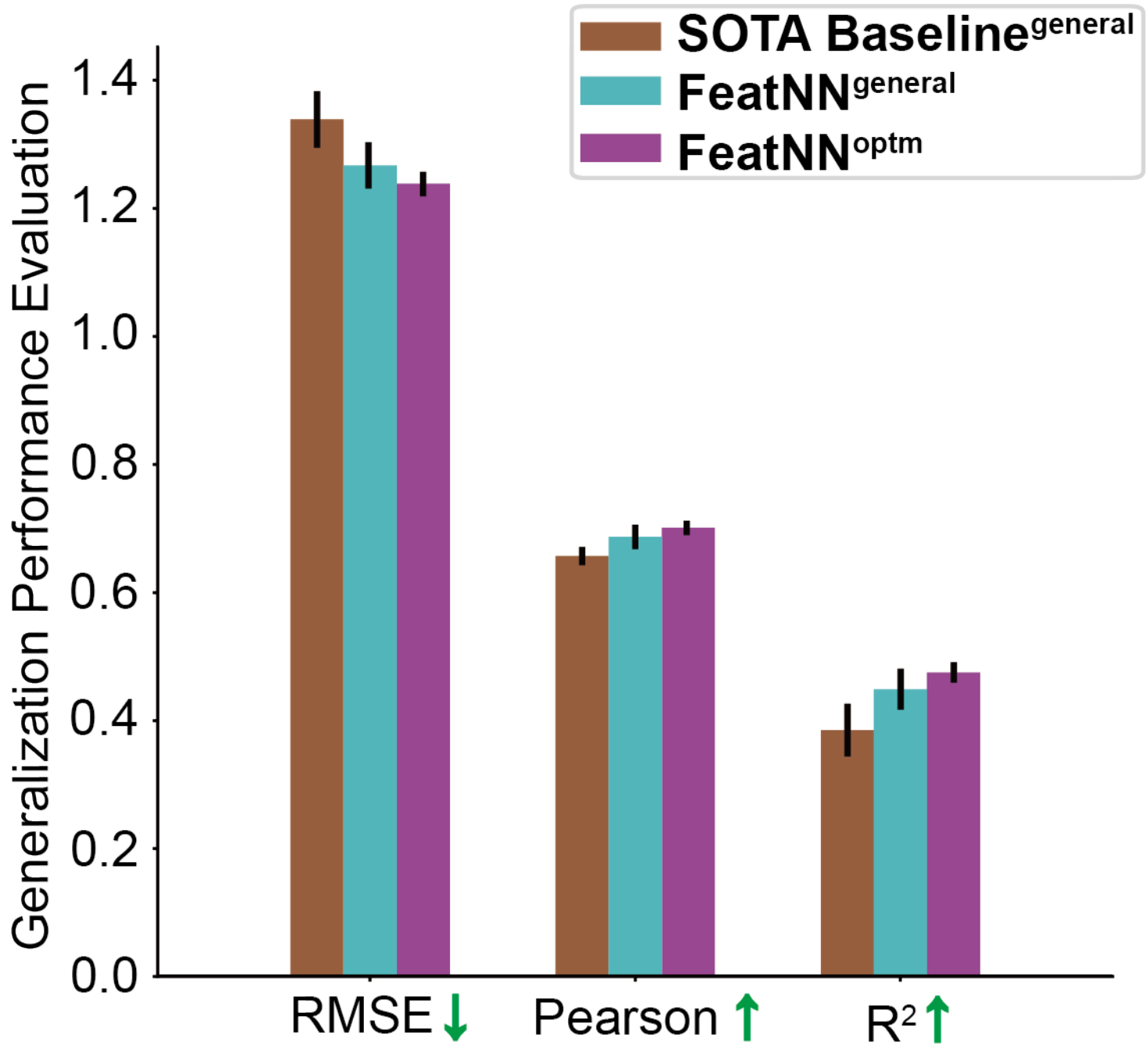

**Supplementary Fig. 7** Comparison of the generalization performances of FeatNN$^{general}$, FeatNN$^{optm}$ and SOTA baseline$^{general}$ (MONN) on the dataset generated from the Binding MOAD database. Generalization of the Pearson and $R^2$ of FeatNN$^{general}$ are improved by 4.57% and 16.62% compared with the SOTA baseline$^{general}$. FeatNN$^{optm}$ is further improved by 2.04% and 5.79% in Pearson and $R^2$ compared with FeatNN$^{general}$. Detailed data can be found in Supplementary Table 8.

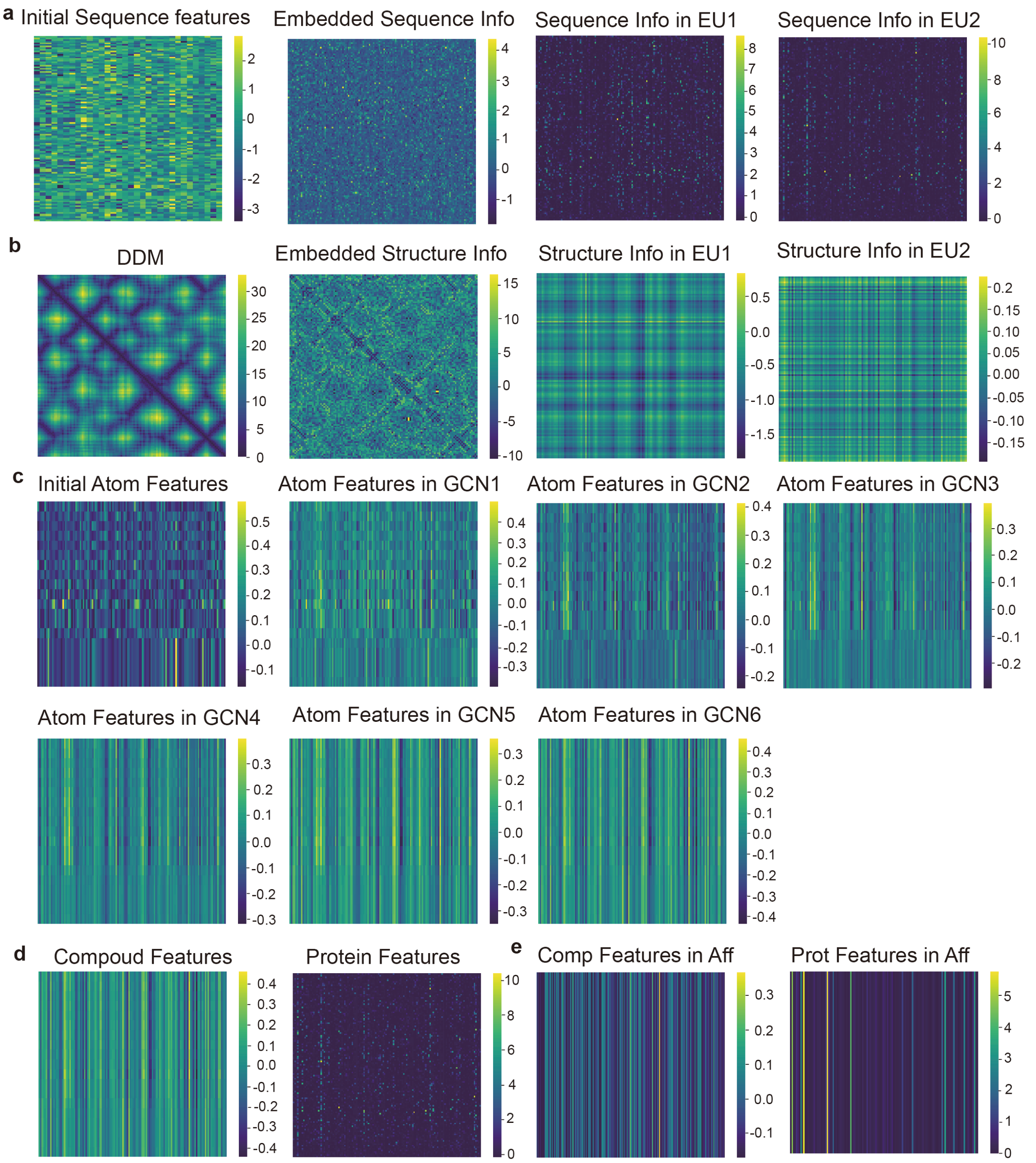

**Supplementary Fig. 8 Visualization of original features in FeatNN. a.** Sequence information in the Prot-Aggregation module and Evo-Updating module. **b.** Structure information in the Prot-Aggregation module and Evo-Updating module. **c.** Atom features in the deep GCN. **d.** Protein and compound features extracted by the protein extractor and compound extractor. **e.** Compound and protein feature interactions in the affinity learning module. Abbrev. Info: information. DDM: Discrete Distance Matrix. EU1: Evo-Updating of Layer 1. EU2: Evo-Updating of Layer 2. GCN1: GCN block of Layer 1. GCN2: GCN block of Layer 2. GCN3: GCN block of Layer 3. GCN4: GCN block of Layer 4. GCN5: GCN block of Layer 5. GCN6: GCN block of Layer 6. Aff: affinity learning module.

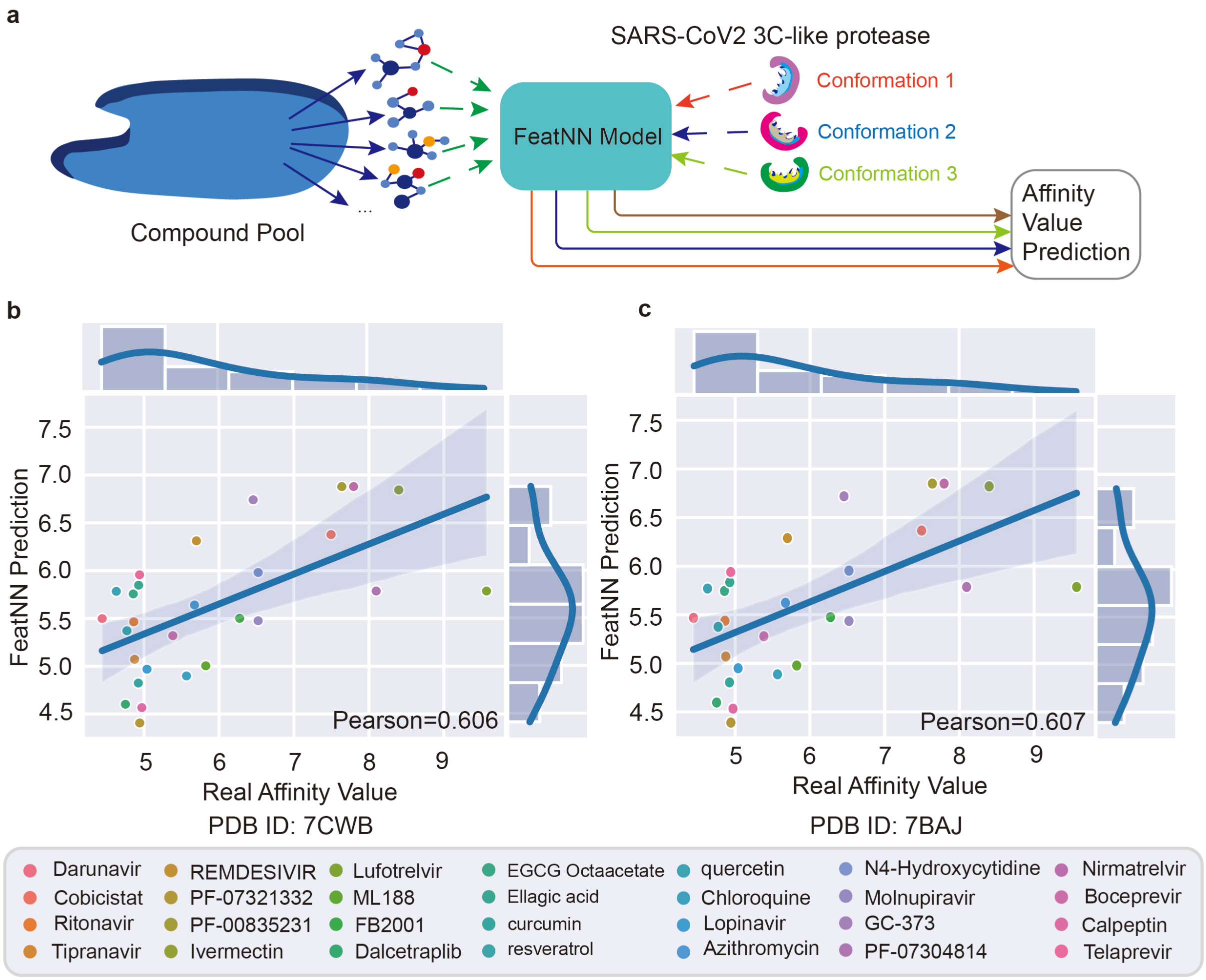

**Supplementary Fig. 9 Affinity prediction results obtained based on receptors (SARS-CoV-2 3C-like protease) with different protein conformations**. **a.** We applied FeatNN to predict the binding affinity for 28 validation compounds and different conformations of the same target protein (SARS-CoV-2 3C-like protease, PDB-ids: 7CWC (Fig. 6b), 7CWB, 7BAJ). **b.** Affinity prediction results of 28 validation bioactive compounds (Supplementary Table 10) by FeatNN based on the conformation of 7CWB in the PDB file. **c.** Affinity prediction results of 28 validation bioactive compounds by FeatNN based on the conformation of 7BAJ in the PDB file. Each point was obtained by the average of 15 independent experiments. Note: All protein conformations were selected based on the ligand-free structure. In addition, the affinity prediction results among different protein conformations did not show significant differences.

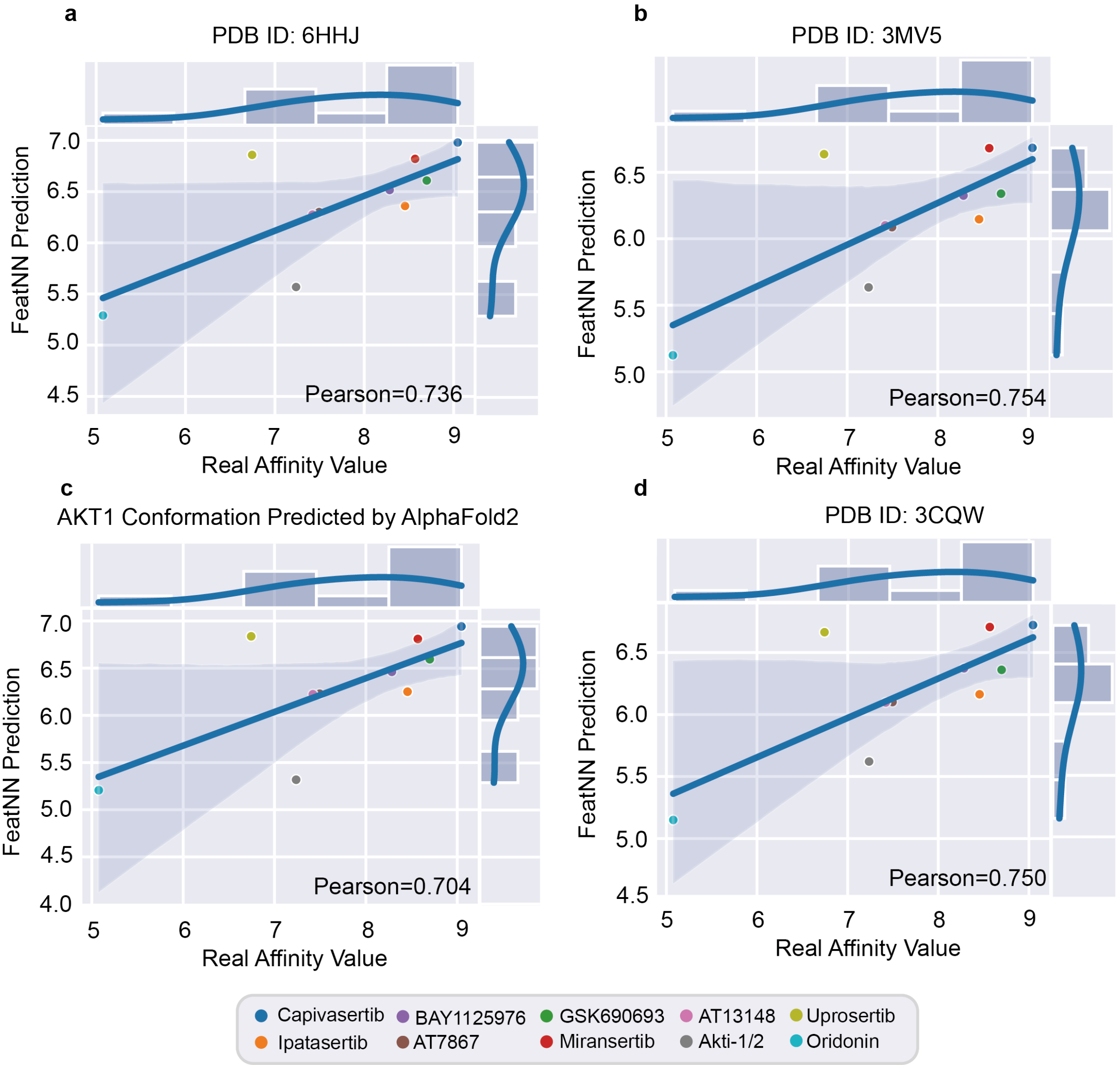

**Supplementary Fig. 10 Affinity prediction results obtained based on receptor Akt-1 with different protein conformations.** We applied FeatNN to predict the binding affinity for 10 validation compounds (Supplementary Table 11) and different conformations of the same target protein (Akt-1, PDB-ids: 3O96 (Fig. 6d), **a.** 6HHJ, **b.** 3MV5, **d.** 3CQW and **c.** The conformation predicted by AlphaFold2 [36]). Note: We did not find the ligand-free structure in the Protein Data Bank. All protein conformations that we selected to bind with small molecules. We obtained the ligand-free structure from the prediction of AlphaFold2. In addition, the affinity prediction results among different protein conformations did not show significant differences.

### 1.2. Influence of the number of layers with a deep graph convolution block and Evo-Updating block and the convergence rates of different models

The RMSE, $R^2$, and Pearson correlation metrics are utilized to evaluate the performance of FeatNN in predicting binding affinities on the IC50 dataset generated from PDBbind. For both panels, FeatNN is evaluated under 5-fold cross-validation settings with a clustering threshold of 0.3, and the layers of the Evo-Updating block are fixed as 2. The means and SDs of the metrics over five cross-validations are shown in Supplementary Figs. 11a-c. It is clear that FeatNN performance is gradually optimized as the number of GCN layers increases (from 1 to 6 layers). The FeatNN with a deep GCN block outperforms the same model without the deep GCN block, emphasizing the importance of addressing the oversmoothing problem in the traditional GCN.

For Supplementary Fig. 11d, FeatNN is evaluated under 5-fold cross-validation settings with a clustering threshold of 0.3, and the layers of the deep GCN block are fixed as 6. The performances of the FeatNN with different numbers of layers of the Evo-Updating block are shown in Supplementary Fig. 11d.

For Supplementary Fig. 11e, we test FeatNN, MONN, BACPI, and GraphDTA on the $IC_{50}$ dataset and evaluate them under 5-fold cross-validation settings with a clustering threshold of 0.3. We use the RMSE in each epoch to represent the convergence rate. The convergence rates of different modes are given below (Supplementary Fig. 11e).

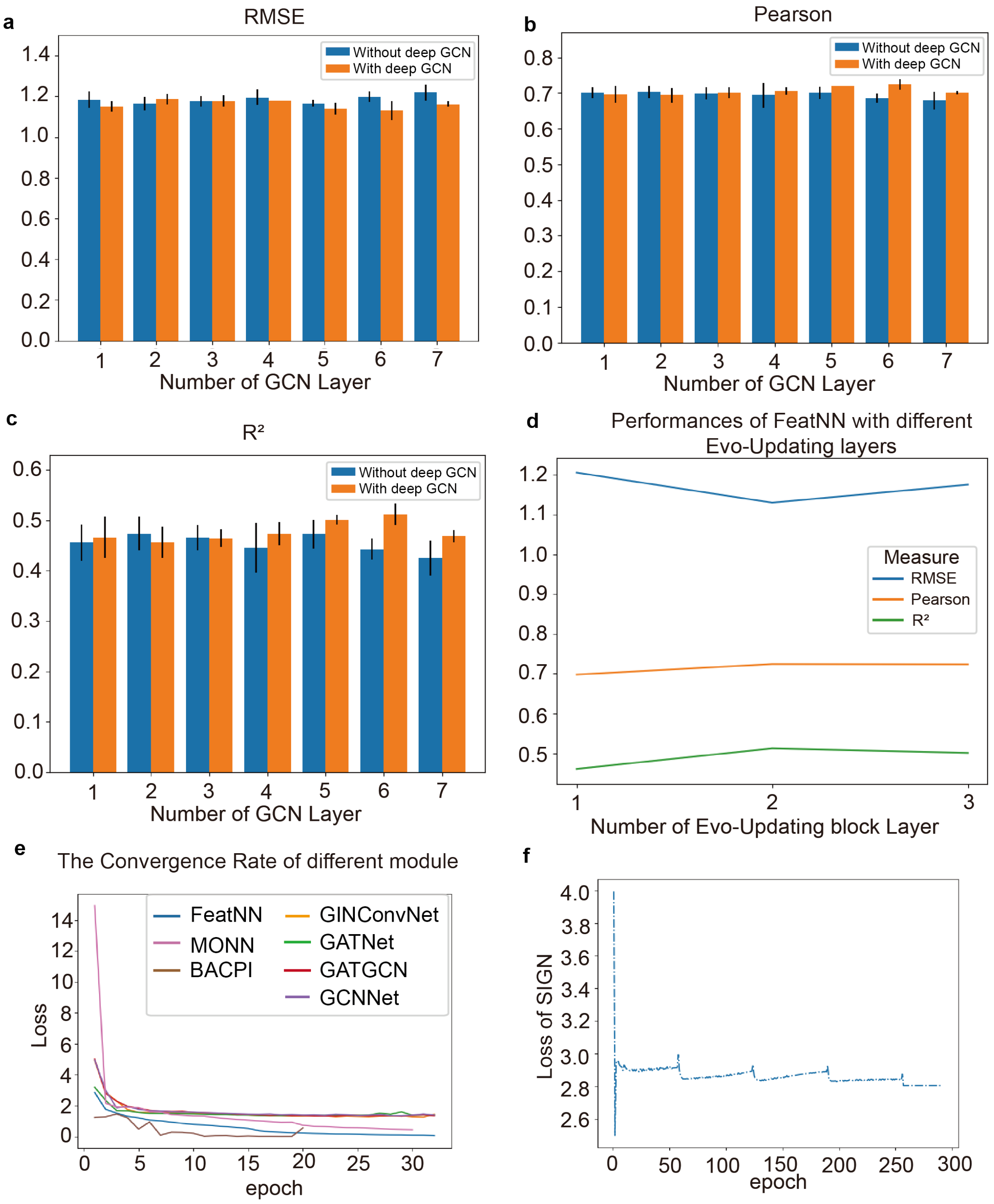

**Supplementary Fig. 11 Influence of the number of layers with a deep graph convolution block and Evo-Updating block and the convergence rates of different models. a-c.** With the deepening of the GCN module layers (from 1 layer to 6 layers), the RMSE, Pearson and $R^2$ performance metrics of CPA prediction improve. The performance metrics of FeatNN with the deep GCN block are superior to those of FeatNN without a deep GCN block. **d.** Performance metrics of FeatNNs with different numbers of Evo-Updating block layers. **e.**

During the training process with 32 epochs, the convergence rate of FeatNN is compared with those of the baseline models. **f.** Since 300 epochs were used for SIGN , and the value of the initial loss exceeds 1e6, we remove the outliers and separately present the result of SIGN here.

# 2. Supplementary Architecture

## 2.1. Notation Definitions

*Linear* $(\cdot)$ indicates a fully connected linear layer without an activation function. $matmul(\cdot)$ represents the multiplication operation between two tensors. $DimentionReshape(\cdot)$ indicates the dimension reshaping operation. $Embedding(\cdot)$ indicates the embedding layer based on the word embedding strategy. $concat(\cdot)$ indicates the concatenation operation between two tensors. $LayerNorm(\cdot)$ indicates the layer normalization operation on a specific channel with learnable per-channel gains and biases. $COMBINE(\cdot)$ indicates the aggregation operation based on the message passing mechanism. $dropout(\cdot)$ is the dropout regulation method. $tanh(\cdot)$, $sigmoid(\cdot)$, $Softmax(\cdot)$ and $gelu(\cdot)$ serve as the activation functions. For the definition of the calculation process, we use $\odot$ for the elementwise product and $\oplus$ for the outer sum.

## 2.2. Block I: Compound Extractor

The deep GCN and multihead attention representation are illustrated in the compound extractor module (Supplementary Fig. 13). In the graph network, the compound information is extracted using graph representation, in which the main nodes (each atom in the compound) and the master node (the node sum of all atoms in the compound) are employed to aggregate the local and global information of the compound, respectively. A deep graph convolution unit (Supplementary Fig. 12) and a multihead attention representation strategy are used to update the main node information. The gate warp strategy interactively regulates and updates the information between the main nodes and the master node. A gated recurrent unit (GRU) is responsible for aggregating the multilayer information in the compound extractor module and updating the features of the main nodes and the master node.

In the GCN, the message passing unit gathers the information of a node's neighbors and passes it to that node for local feature updating (Fig. 1a, Supplementary Fig. 12). Here, we apply a master node to maintain the global features for nodes over long distances. This helps to mitigate the oversmoothing problem in the GCN (for a detailed explanation of the oversmoothing problem, please refer to Supplementary Note 1.1). Moreover, by applying the master node, the number of layers in the GCN can be deepened to better extract the features of compounds, thus contributing to the multihead attention representation (Supplementary Fig. 14). Finally, the local information and global information representations of compounds are jointly input into the affinity prediction module with the protein features learned from the protein extractor module, ultimately benefiting the CPA prediction process.

### 2.2.1. Algorithm 1: Deep GCN Block

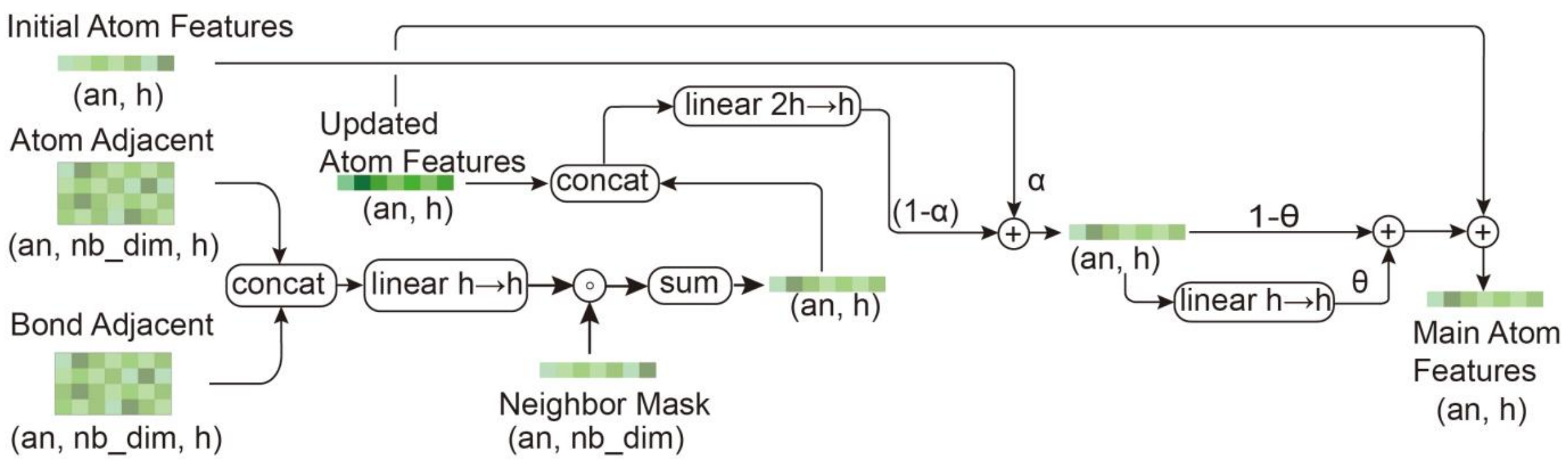

**Supplementary Fig. 12 Deep GCN block.** Atom features are combined by a message passing mechanism in a deep GCN block.

**Define:** $F^{vertex}_{N_{atom},h}$ indicates the features of the atoms in the compound, and $F^{edge}_{N_{atom},nbs}$ indicates the features of the bonds in the compound. $Adj^{atom}_{N_{atom},nbs}$ and $Adj^{bond}_{N_{atom},nbs}$ are the adjacency matrices of atoms and bonds, respectively, which are used to aggregate adjacent vertex and bond information into each atom. $F^{h0}_{N_{atom},h}$ denotes the initial features of atoms that are vital for addressing the oversmoothing problem. $theta$ and $alpha$ are hyperparameters of the residual connection in the deep GCN block.

**def**

$GraphCovNN(\{F^{vertex}_{N_{atom},h}\},\{F^{edge}_{N_{atom},nbs}\},\{Adj^{atom}_{N_{atom},nbs}\},\{Adj^{bond}_{N_{atom},nbs}\},\{F^{h0}_{N_{atom},h}\},\{theta\},\{alpha\})$:

$ver_{neighbor} = COMBINE\ (F^{vertex}_{N_{atom},h}, Adj^{atom}_{N_{atom},nbs})$

$edge_{neighbor} = COMBINE(F^{edge}_{N_{atom},nbs}, Adj^{bond}_{N_{atom},nbs})$

$con_{neighbor} = concat_{neighbor}(ver_{neighbor}, edge_{neighbor})$

$$neighbor_label \;=\; gelu(Linear(con_{neighbor}))$$

$$hi \;=\; Linear(concat_h(F^{vertex}_{N_{atom},h}, neighbor_label))$$

$$support \;=\; (1 - alpha) \odot hi + alpha \odot F^{h0}_{N_{atom},h}$$

$$output \;=\; theta \odot Linear(support) + (1 - theta) \odot support$$

**return**$\rightarrow\{F^{output}_{N_{atom},h}\}$

### 2.2.2. Algorithm 2: Compound Extractor

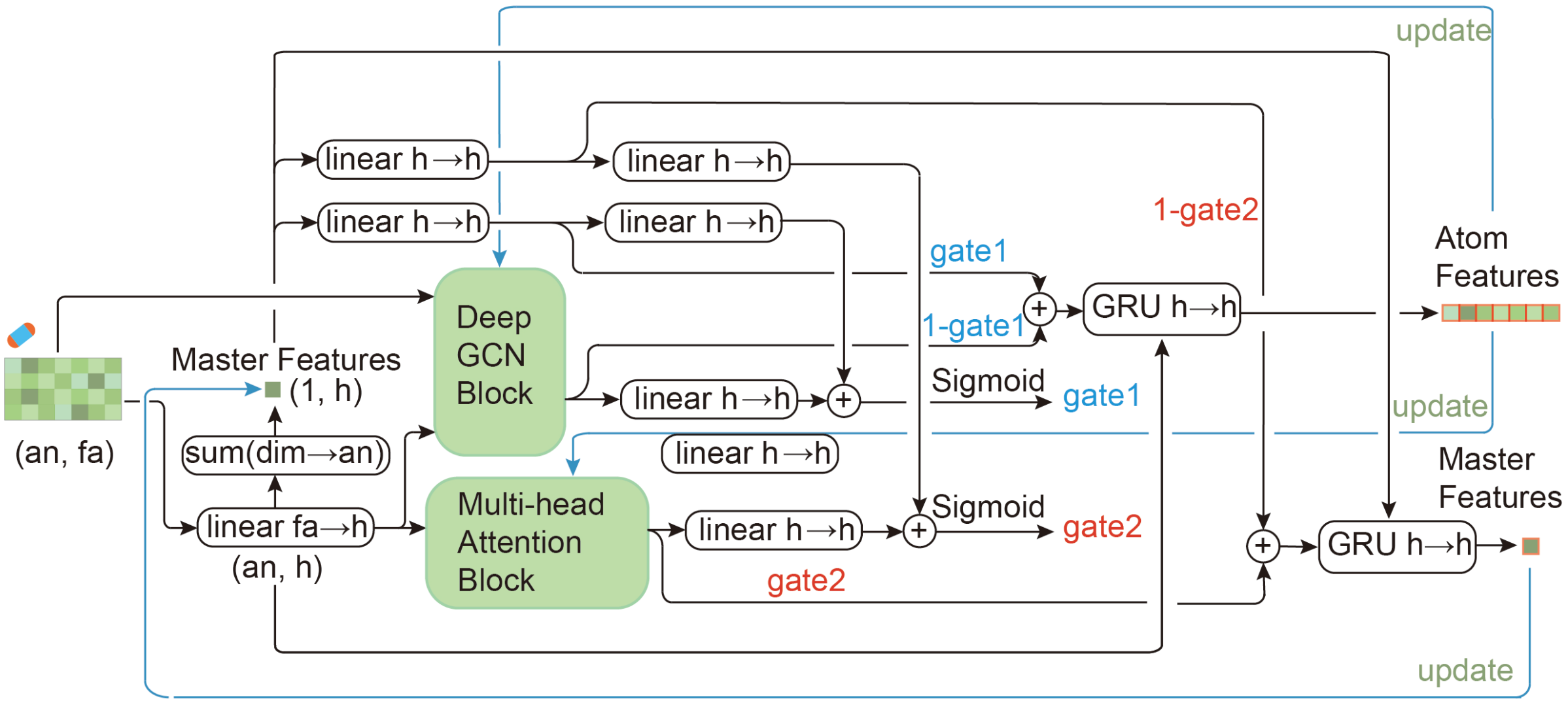

**Supplementary Fig. 13 Outlines of the compound extractor.** The deep GCN block and multihead attention block function form the core of the compound extractor.

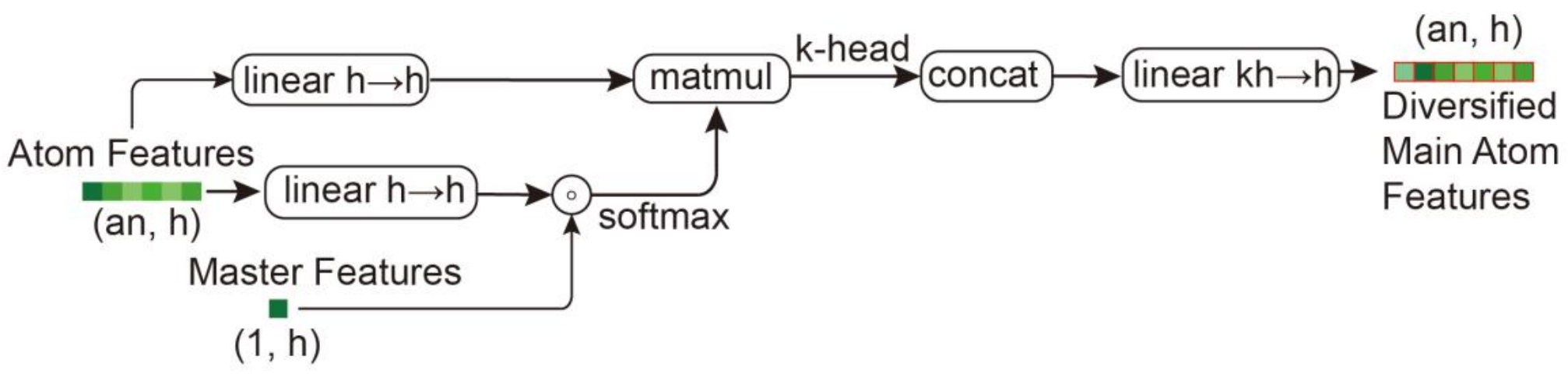

**Supplementary Fig. 14 Multihead attention block in the compound extractor.** A multihead attention block is applied to enhance the diversification of atom features.

**Define:** The definitions of $F^{vertex}_{N_{atom},h}$, $F^{edge}_{N_{atom},nbs}$, $F^{edge}_{N_{atom},nbs}$, $Adj^{atom}_{N_{atom},nbs}$ and $Adj^{bond}_{N_{atom},nbs}$ are the same as those in Algorithm 1. In particular, ${F^{master}_{1,h}}_{l_c}$ and ${F^{atom}_{N_{atom},h}}_{l_c}$ indicate the master features (the sum over all atom features in the compound) and the atom features extracted from the GCN's $l_c th$ layer. ${F^{master}_{1,h}}_{l_0}$ and ${F^{atom}_{N_{atom},h}}_{l_0}$ indicate the initial states of the master and atom features. $mask^{vertex}_{N_{atom}}$ indicates the mask matrix of the vertex in the compound graph.

**def** $CompExtractor(\{F^{vertex}_{N_{atom},h}\},\{F^{edge}_{N_{atom},nbs}\},\{Adj^{atom}_{N_{atom},nbs}\},\{Adj^{bond}_{N_{atom},nbs}\},\{mask^{vertex}_{N_{atom}}\})$:

${F^{atom}_{N_{atom},h}}_{l_0} = gelu(Linear(F^{vertex}_{N_{atom},h}))$

$F^{h0}_{N_{atom},h} = {F^{atom}_{N_{atom},h}}_{l_0}$

${F^{master}_{1,h}}_{l_0} = sum({F^{atom}_{N_{atom},h}}_{l_0} \odot mask^{vertex}_{N_{atom}})$

**for** $l_c \in [l_0, \dots . N_{Comp}]$:

**for**$k \in [0, \dots . head_num]$:

$main_vertex = tanh(Linear({F^{atom}_{N_{atom},h}}_{l_c-1}))$

$vertex = Linear(main_vertex \odot {F^{master}_{1,h}}_{l_c-1})$

$attention_score = softmax(vertex + mask^{vertex}_{N_{atom}})$

$k_head_atom_to_master = bmm(attention_score, Linear({F^{atom}_{N_{atom},h}}_{l_c-1}))$

**if** k == 0:

$m_atom_to_master = k_head_atom_to_master$

**else**:

$m_atom_to_master = concat(m_atom_to_master, k_head_atom_to_master)$

**end if**

$atom_to_master = tanh(Linear(m_atom_to_master))$

$atom_feat = dropout({F^{atom}_{N_{atom},h}}_{l_c-1})$

$vert_{agg}$

$= GraphCovNN(atom_feat, F^{edge}_{N_{atom},nbs}, Adj^{atom}_{N_{atom},nbs}, Adj^{bond}_{N_{atom},nbs}, F^{h0}_{N_{atom},h}, theta, alpha)$

$master_to_atom = gelu(Linear({F^{master}_{1,h}}_{l_c-1}))$

$master_agg = gelu(Linear({F^{master}_{1,h}}_{l_c-1}))$

$gate_{atom} = sigmoid(Linear(vert_{agg}) + Linear(master_to_atom))$

$updated_atom = (1 - gate_{atom}) \odot vert_{agg} + gate_{atom} \odot master_to_atom$

${F^{atom}_{N_{atom},h}}_{l_c} = GRU(updated_atom, {F^{atom}_{N_{atom},h}}_{l_c-1})$

$gate_{master} = sigmoid(Linear(master_self) + Linear(atom_to_master))$

$updated_master = (1 - gate_{master}) \odot master_agg + gate_{master} \odot atom_to_maste$

${F^{master}_{1,h}}_{l_c} = GRU(updated_master, {F^{master}_{1,h}}_{l_c-1})$

**end for**

**end for**

**return**→ $\{{F^{atom}_{N_{atom},h}}_{N_{Comp}}\}, \{{F^{master}_{1,h}}_{N_{Comp}}\}$

## 2.3. Block II: Protein Extractor Module

Most importantly, the direct introduction of the 3D structures of proteins may drastically increase the computational costs of our model. The continuous Euclidean distance information between protein residues in the traditional distance matrix is difficult to discriminate within a small scope. In this study, the protein distance matrix is discretely encoded, and its continuous values are divided into 40 mapping intervals that conform to a normal distribution in statistics. Between 3.25 Å and 50.75 Å, the distance matrix is mapped to 38 intervals with equal distances and widths (1.25 Å per unit). Two additional intervals are added to store any larger distances (when the distances between residues are greater than 50.75 Å) and smaller distances (when the distances between residues are less than 3.25 Å). Therefore, the computational cost is greatly reduced. Furthermore, the sequence information and torsion angle information of the protein are introduced in the protein extractor module, and the DDM and protein residue sequence information are further characterized.

Many traditional networks can only update one type of data source at a time, while a multimodal mechanism can learn more comprehensive information from a variety of data sources. In contrast with previous work, we innovatively aggregate the sequence and structure features of the proteins with the protein aggregation unit (Prot-Aggregation, Supplementary Fig. 17). The torsion matrix is aggregated into sequence features through the linear mapping and Hadamard product operation in the protein aggregation unit. A mechanism employed by the evolutionary updating block (Evo-Updating) can interactively update these two properties. The Prot-Aggregation block and Evo-Updating block jointly construct the backbone of the protein extractor module (Supplementary Fig. 15). The DDM updates sequence features by summation over its columns (Supplementary Fig. 18).

Message communication from the evolving DDM to the sequence features in the Evo-Updating unit (Supplementary Fig. 18) is enabled by an enormous amount of matrix multiplications that serves as the core of the protein encoder module (Supplementary Fig. 16). The embedded distance matrix (embedded DM) is transformed into distance vectors that possess the same shapes as the sequence features through column sum and row sum operations. A merging matrix is constructed by multiplying the embedded sequence features with the distance vectors through a batch-dot-product operation, and this matrix is then added to the features of the embedded DM. The sequence features are finally renovated by the attention mechanism and gate unit updating methods. These sequence features are then evolutionarily projected to structure information through the outer sum operation and gate unit updating method. Such an intricate network architecture satisfies the requirement of multimodal pattern feature extraction, ensuring that the overall Evo-Updating unit can fully mix information regarding sequence and structure features and is sufficient for accurate affinity prediction.

### 2.3.1. Algorithm 3: Protein Extractor

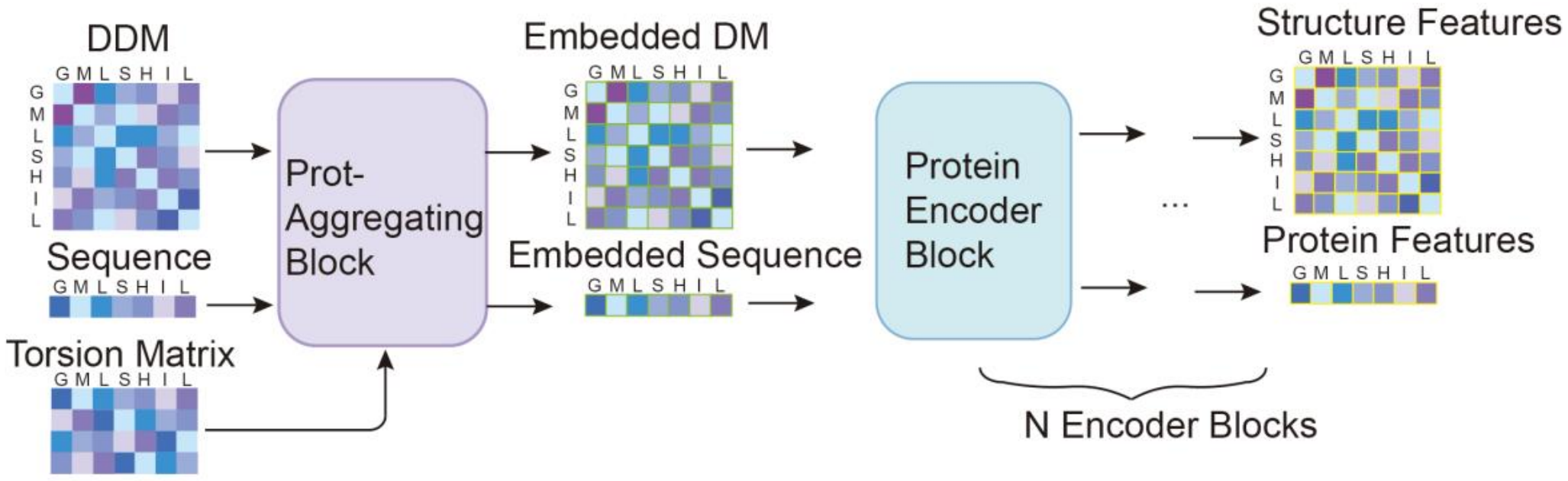

**Supplementary Fig. 15 Overview of the protein extractor.** The protein extractor consists of the Prot-Aggregation block and protein encoder block.

**Define:** $Seq_{init}$, $DDM_{init}$ and $TorMat_{init}$ indicate the initial information of the protein residue sequence, *DDM* and torsion matrix, respectively. $mask_{N_{res}}^{Seq}$ and $mask_{N_{res},N_{res}}^{DDM}$ are the mask matrices of the protein residue sequence, and $DDM$.

**def** *ProtExtractor*($\{Seq_{init}\}$,$\{DDM_{init}\}$,$\{TorMat_{init}\}$,$\{mask_{N_{res},N_{res}}^{DDM}\}$,$\{mask_{N_{res}}^{Seq}\}$):

$F_{N_{res},h_{init}}^{seq}, F_{N_{res},e_{init}}^{DDM} \leftarrow ProtAggregation(Seq_{init}, DDM_{init}, TorMat_{init}, mask_{N_{res},N_{res}}^{DDM}, mask_{N_{res}}^{Seq})$

$F_{N_{res},h}^{seq}, F_{N_{res},e}^{DDM} \leftarrow ProtEncoder(F_{N_{res},h}^{seq}, F_{N_{res},e}^{DDM}, mask_{N_{res}}^{Seq}, mask_{N_{res},N_{res}}^{DDM})$

**return** $\rightarrow$ $\{F_{N_{res},h}^{seq}\}, \{F_{N_{res},e}^{DDM}\}$

### 2.3.2. Algorithm 4: Protein Encoder

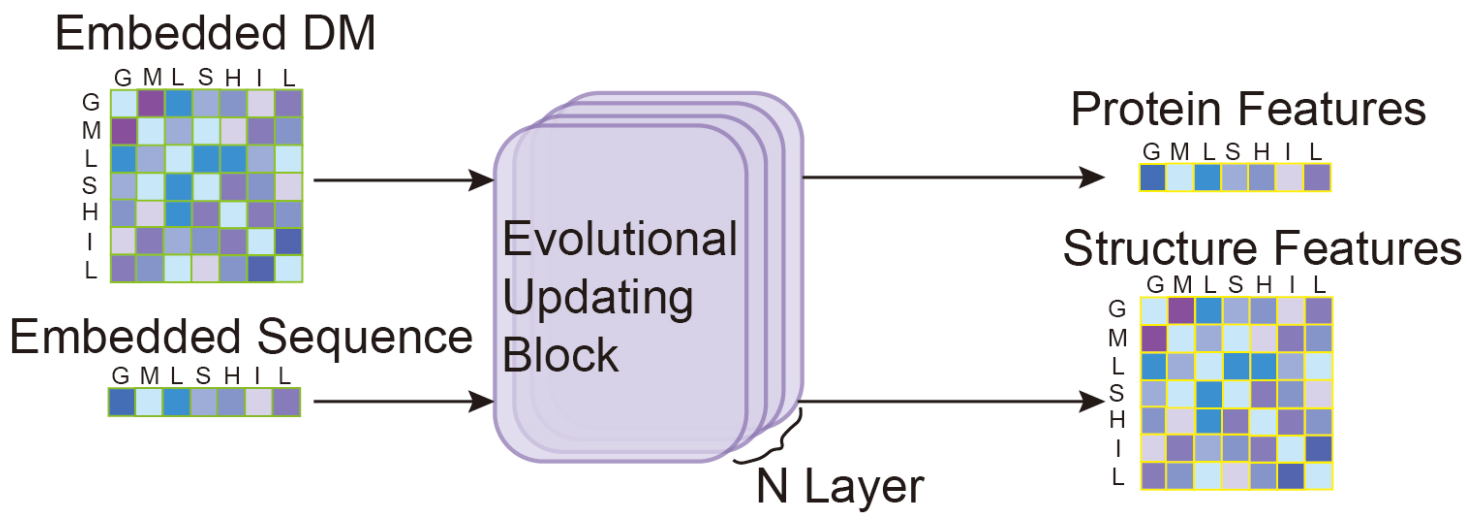

**Supplementary Fig. 16 Architecture of the protein encoder.** With the Evo-Updating block, the protein encoder interactively updates the sequence and structure features.

**Define:** $F_{N_{res},h}^{seq}$ and $F_{N_{res},e}^{DDM}$ are the features of the residue sequence and *DDM* embedded by the Prot-Aggregation Algorithm, and $mask_{N_{res}}^{Seq}$ and $mask_{N_{res},N_{res}}^{DDM}$ are the mask matrices of the protein residue sequence and $DDM$, respectively. ${F_{N_{res},h}^{seq}}_{l_{Encoder}}$ and ${F_{N_{res},e}^{DDM}}_{l_{Encoder}}$ indicate the DDM and sequence features extracted from the $l_{Encoder}th$ layer of the protein encoder.

**def** *ProtEncoder*($\{F_{N_{res},h}^{seq}\}$,$\{F_{N_{res},e}^{DDM}\}$, $\{mask_{N_{res}}^{Seq}\}$, $\{mask_{N_{res},N_{res}}^{DDM}\}$):

**for all** $l_{Encoder} \in [init, 1,2, \ldots, N_{Prot}]$ do:

$${Seq_{N_{res},h}^{final}}_{l_{Encoder}}, {DDM_{N_{res},e}^{final}}_{l_{Encoder}} \leftarrow EvoUpdating({F_{N_{res},h}^{seq}}_{l_{Encoder}-1}, {F_{N_{res},e}^{DDM}}_{l_{Encoder}-1}, mask_{N_{res}}^{Seq}, mask_{N_{res},N_{res}}^{DDM})$$

**end for**

**return** $\rightarrow \{{Seq_{N_{res},h}^{final}}_{N_{Prot}}\}, \{{DDM_{N_{res},e}^{final}}_{N_{Prot}}\}$

### 2.3.3. Algorithm 5: Prot-Aggregation

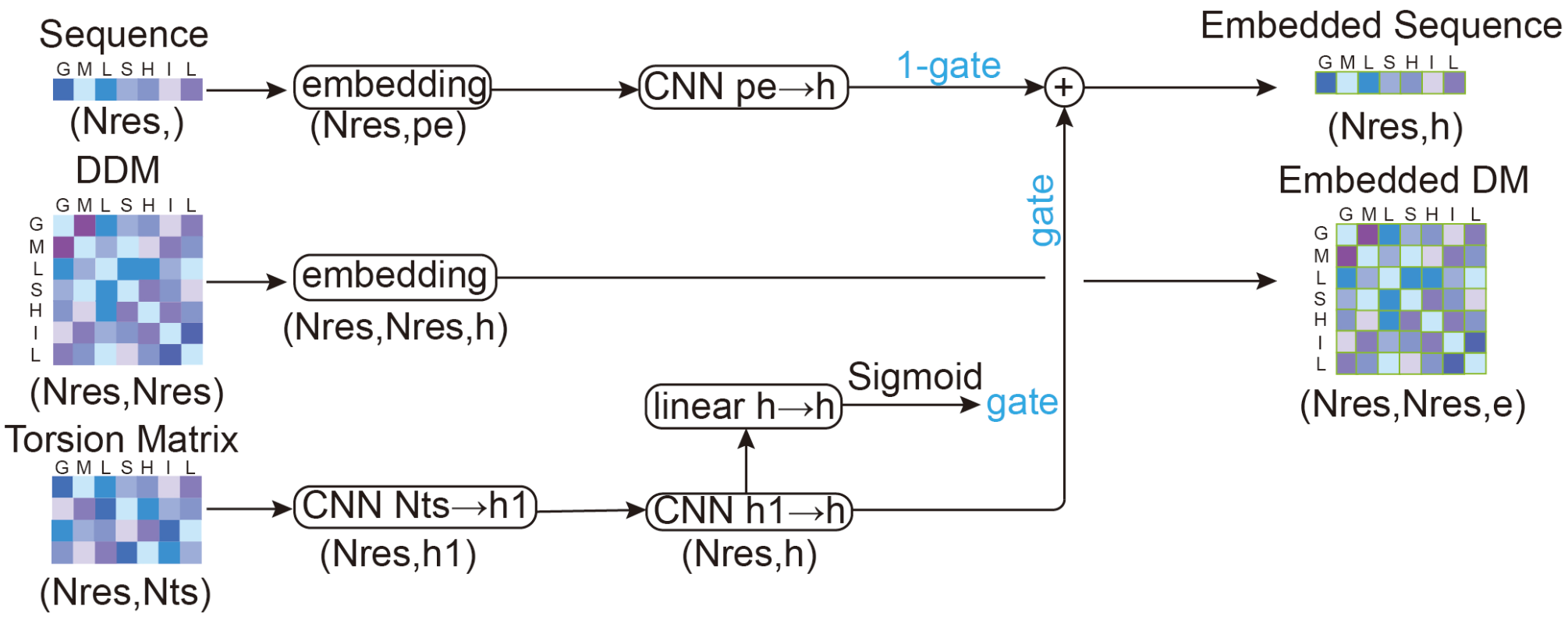

**Supplementary Fig. 17 Prot-aggregation block.** Based on the input raw protein data, the sequence features and structure features are embedded and aggregated in the Prot-Aggregation block.

**Define:** The definitions of $Seq_{init}$, $DDM_{init}$, $TorMat_{init}$, $mask_{N_{res}}^{Seq}$ and $mask_{N_{res},N_{res}}^{DDM}$ are the same as those in Algorithm 3.

**def** $ProtAggregation(\{Seq_{init}\},\{DDM_{init}\},\{TorMat_{init}\},\{mask^{DDM}_{N_{res},N_{res}}\},\{mask^{Seq}_{N_{res}}\})$:

$$seq_embed = Embedding(Seq_{init})$$

$$seq_features = CNN(seq_embed \odot mask^{Seq}_{N_{res}})$$

$$torsion_embed = CNN_{Nts \to h1}(TorMat_{init})$$

$$torsion_vector = CNN_{h1 \to h}(torsion_embed)$$

$$gate = Sigmoid(Linear(torsion_vector))$$

$$Embed^{Seq}_{N_{res},h} \leftarrow gate \odot torsion_vector + (1 - gate) \odot seq_features$$

$$Embed^{DM}_{N_{res},N_{res},e} = Embedding(DDM_{init}) \odot mask^{DDM}_{N_{res},N_{res}}$$

**return** $\to \{Embed^{DM}_{N_{res},N_{res},e}\},\{Embed^{Seq}_{N_{res},h}\}$

### 2.3.4. Algorithm 6: Evo-Updating

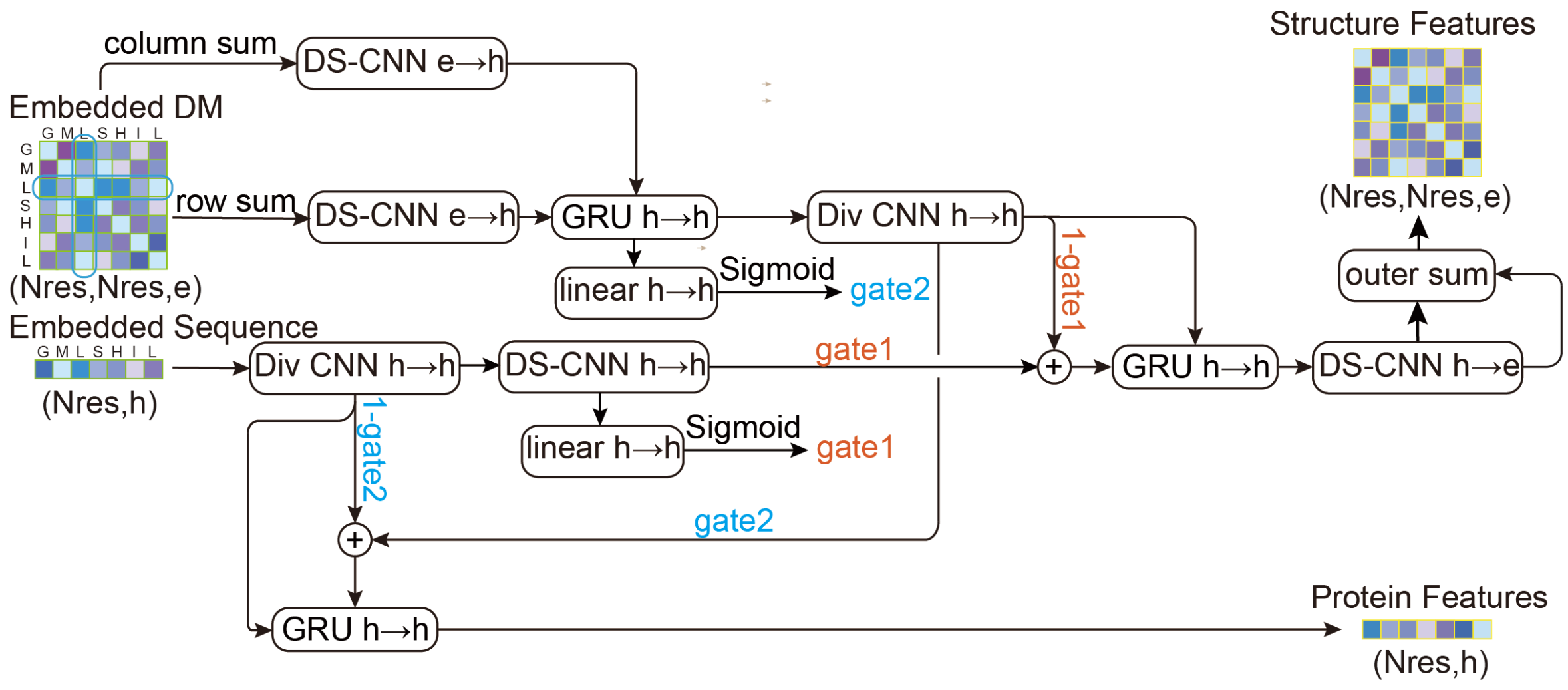

**Supplementary Fig. 18 The protein Evo-Updating block.** The protein residue sequence and structure features are coevolutionarily updated through the Evo-Updating block, so that the sequence features include structure information and forcing the structure features to contain sequence information. Abbrev: DS-CNN: Depthwise Separable Convolution Neural Network, Div CNN: Diversification Convolution neural network.

**Define:** $F^{seq}_{N_{res},h}$ and $F^{DDM}_{N_{res},e}$ are the features of the residue sequence and *DDM* embedded by the Prot-Aggregation algorithm. $mask^{Seq}_{N_{res}}$ and $mask^{DDM}_{N_{res},N_{res}}$ are the mask matrices of the protein residue sequence and $DDM$, respectively.

**def** *EvoUpdating*($\{F^{seq}_{N_{res},h}\},\{F^{DDM}_{N_{res},e}\},\{mask^{Seq}_{N_{res}}\},\{mask^{DDM}_{N_{res},N_{res}}\}$):

$$PairKey1 = DeepSparseCNN(RowSum(F^{DDM}_{N_{res},e}))$$

$$PairKey2 = DeepSparseCNN(ColumnSum(F^{DDM}_{N_{res},e}))$$

$$PairKey2 = DeepSparseCNN(ColumnSum(F^{DDM}_{N_{res},e}))$$

$$MixKey = GRU(PairKey1, PairKey2)$$

$$Struct_Features = DivCNN(MixKey)$$

$$Seq_Features = DivCNN(F^{seq}_{N_{res},h})$$

$Seq2Struct = DeepSparseCNN(F^{seq}_{N_{res},h})$

$$SeqGate = Sigmoid(Linear(Seq2Struct))$$

$StructGate = Sigmoid(Linear(MixKey))$

$Seq2Struct_Vector = SeqGate \odot Seq2Struct + (1 - SeqGate) \odot Struct_Features$

$Struct_Vector = GRU(Seq2Struct_Vector, Struct_Features)$

$$Struct2Seq_Mapping = DeepSparseCNN(Struct_Vector)$$

${F^{DistMat}_{N_{res},e}}_{output} \leftarrow Struct2Seq_Mapping \oplus Struct2Seq_Mapping$

$Seq_Vector = StructGate \odot Struct_Features + (1 - StructGate) \odot Seq_Features$

${F^{sequence}_{N_{res},h}}_{output} = GRU(Seq_Vector,\ Seq_Features) \odot mask^{Seq}_{N_{res}}$

**return** $\rightarrow \{{F^{sequence}_{N_{res},h}}_{output}\},\{{F^{DistMat}_{N_{res},e}}_{output}\}$

### 2.3.5. Algorithm 7: Div CNN

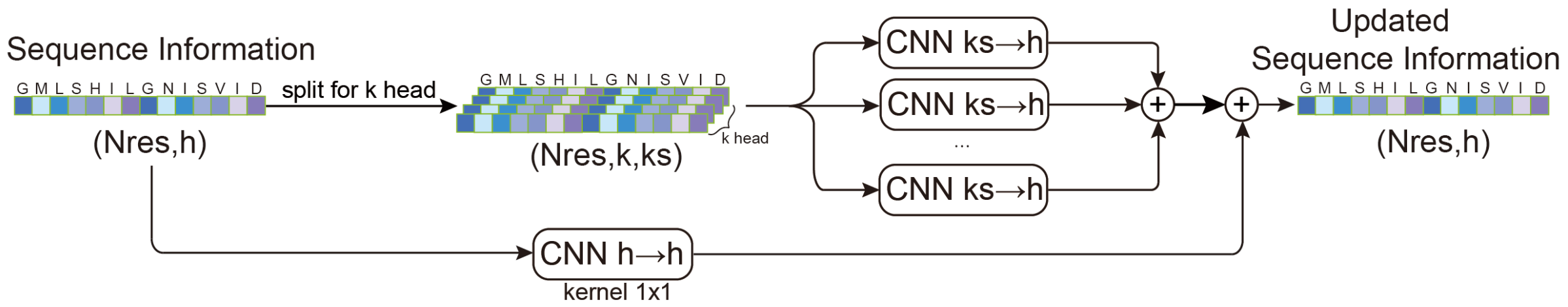

**Supplementary Fig. 19 The diversification convolution neural network (Div CNN) block.** Div CNN is used to enhance the diversification of structure features and sequence features with the multihead mechanism.

**def** *DivCNN* (*x*):

$x0 = CNN_{kernel=1}(x)$

# x (seq, hidden_size) → $x_{k-head}$(seq, k, head_size)

# where k is the number of heads, and head_size = hidden_size/k

$x_{k-head} = TransposeForScores(x)$

$x_{total} = \sum_{n=k-head} CNN_{\text{head_size}\rightarrow\text{hidden_size}}(x_{k-head}) + x0$

**return** → $\{x_{total}\}$

### 2.3.6. Algorithm 8: TransposeForScores

**def** *TransposeForScores({input}):*

# *input dimension:*$(N_{res}/N_{atom}, h)$

# output dimension:$(k, N_{res}/N_{atom}, ks)$

$output = DimentionReshape(input)$

**return** → $\{output\}$

## 2.4. Block III: Affinity Learning Module

Based on an end-to-end architecture, the protein and compound features extracted from the upstream model (included in the protein extractor and compound extractor) are fed into the affinity learning module (Supplementary Fig. 20. The mapping information between proteins and compounds is constructed as a pairwise matrix through the matrix multiplication operation to achieve feature aggregation between proteins and compounds, enabling the fitting and learning of the potential interaction information between the proteins and the compounds. Finally, the CPA predictions are given.

### 2.4.1. Algorithm 9: Affinity Prediction

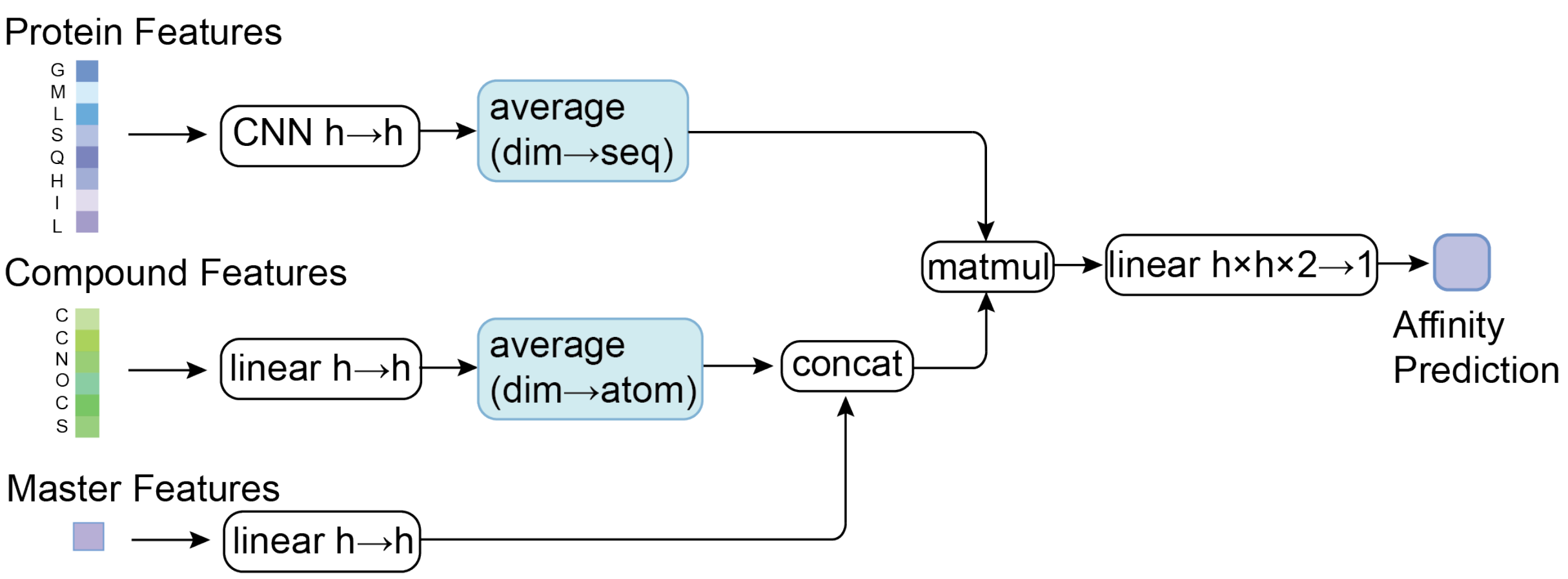

**Supplementary Fig. 20 Architecture of the affinity learning module.**

**Define** $F_{N_{atom},h}^{Compound}$ and $F_{1,h}^{Master}$ as the compound atom features and master features extracted from the compound extractor algorithm, respectively. $F_{N_{res},h}^{Protein}$ denotes the protein features extracted from the protein extractor algorithm that contain both the sequence and structure information of the protein. $mask_{N_{res}}^{Seq}$ and $mask_{N_{atom}}^{vertex}$ are the mask matrices of the protein residue sequence and the vertex in the compound graph.

**def** *AffinityPrediction*($\{F_{N_{atom},h}^{Compound}\}, \{F_{N_{res},h}^{Protein}\}, \{F_{1,h}^{Master}\}, \{mask_{N_{res}}^{Seq}\}, \{mask_{N_{atom}}^{vertex}\}$):

# Inputs projections

$$feature_{init}^{Compound} = Linear(F_{N_{atom},h}^{Compound})$$

$$feature_{init}^{Protein} = CNN(F_{N_{res},h}^{Protein})$$

$$feature^{Master\ Node} = Linear(F_{1,h}^{Master})$$

$$feature_{final}^{Compound} \leftarrow (\sum_{a=N_{atom}} feature_{init}^{Compound}{}_a \odot mask_a^{vertex}) / \sum_{a=N_{atom}} mask_{N_{atom}}^{vertex}$$

$$feature_{final}^{Protein} \leftarrow (\sum_{s=N_{res}} feature_{init}^{Protein}{}_s \odot mask_s^{Seq}) / \sum_{s=N_{res}} mask_{N_{res}}^{Seq}$$

$$feature_{mixture}^{Compound} = concat_h(feature_{final}^{Compound}, feature^{Master\ Node})$$

# Output projection

$$Affinity_{Prediction} = Linear(matmul(feature_{mixture}^{Compound}, feature_{final}^{Protein}))$$

**return**→ $\{Affinity_{Prediction}\}$

## 3. Supplementary Methods

### 3.1 Calculation details for the discrete distance matrix and torsion angle matrix

To calculate the distance between the residues and construct the discrete distance matrix (DDM) in each protein, we followed the steps listed below:

1. The 3D position of each residue in a protein was represented by the amino acids' beta carbon position for all amino acids except glycine because glycine does not have beta carbon; therefore, for glycine the position of its alpha carbon is used instead;
2. Then, based on the represented 3D position of each residue in a protein, we calculated the Euclidean distance between each residue to construct a distance matrix ($N_{res} \times N_{res}$, where $N_{res}$ is the residue number in the protein) of the protein;
3. The distance between every two residues was discretized into 40 bins: the number tokens from 1 to 38 represent 38 bins of equal width between 3.25 Å and 50.75 Å, 0 represents distances smaller than 3.25 Å, and 39 represents distances larger than 50.75 Å.

Ultimately, a discrete distance matrix with lower storage and calculation requirements was constructed.

The torsion angle matrix was calculated through the following steps:

1. We first calculated the ψ (the torsion between Cα-C) and Φ (the torsion between N-Cα) angles in each residue;
2. Then, sine and cosine functions were applied to encode the torsion angles of ψ and Φ to accurately represent the torsion information of each protein;

Ultimately, a torsion angle matrix with the dimensions of $N_{res} \times 4$ was constructed.

### 3.2 Parameter Settings of the FeatNN

In this work, for fast and convenient calculation, we utilized 6 layers of deep GCN blocks and 2 layers of Evo-Updating blocks. The hidden size in the entire architecture was set as 128. The number of attention heads in the deep GCN blocks and the Evo-Updating blocks was 4. The detailed parameter settings can be found in Supplementary Table 2.

### 3.3 Details of Dataset Construction from PDBbind, BindingDB and Binding MOAD

The datasets constructed from the general set of PDBbind contain the CPA values with the $K_i$, $K_d$, and $IC_{50}$ measurements between drugs and proteins, while datasets constructed from the refined set of PDBbind only contain the CPA values with the measurements of $K_i$ and $K_d$. BindingDB is rich in $IC_{50}$ measured values (more than 500 thousand data), while the collections of the measurement values based on $K_i$ and $K_d$ are significantly smaller (40 thousand $K_i$ measurements and 28 thousand $K_d$ measurements were recorded). In this paper, to construct large datasets from BindingDB, we only selected the measured $IC_{50}$ values to generate training data. To test the generalization ability of the models, we constructed new datasets from the Binding MOAD database and excluded the complexes that appeared in the datasets (training, validation, and test datasets) constructed from PDBbind. For a fair comparison of the generalization ability, we limit the datasets constructed from Binding MOAD with the measurement of $IC_{50}$ and KIKD ($K_i$ and $K_d$) to the same amount of data. Thus, we constructed the dataset with $IC_{50}$ and KIKD measurements from the "all of Binding MOAD" and "nonredundant MOAD" sets in the Binding MOAD database.

### 3.4 Molecular Similarity Calculation

Molecular structures were represented by 1024-dimensional binary Morgan fingerprints with radii of 2, while the Tanimoto coefficient was utilized to measure molecular similarities. Finally, the compounds in the dataset, according to similarity thresholds from 0.3 to 0.6 (with a step of 0.1), and similar compounds were grouped into the same dataset (training, valid or test set) to prevent data leakage.

### 3.5 Homologous Protein Calculation

The homology between proteins was quantified by multisequence alignment (MSA) methods, and based on the thresholds from 0.3 to 0.6 (with a step of 0.1), the obtained similarity scores were applied to divide the homologous proteins into the same subset to ensure that similar proteins did not appear in the same dataset (training, valid or test set), which is similar to the method in Note 3.3.

### 3.6 Generalization evaluation on Binding MOAD

All generalization testing processes are evaluated on the dataset constructed from Binding MOAD. We selected all models trained on the refined set and general set of PDBbind-v2020 to investigate the differences in the generalization ability (Supplementary Fig. 6) when trained with different amounts or qualities of data. We further tested all of the generalization abilities of FeatNN$^{optm}$ (Supplementary Fig. 7) to prove the effectiveness of the pretraining strategy that utilized the high- and low-quality data from PDBbind and BindingDB, respectively.

### 3.7 Details of Optimization with a Pretraining Strategy

FeatNN was pretrained for 32 epochs on the datasets generated by BindingDB and used as the initial fine-tuning model. Then, we froze the parameters of the compound extractor and trained (fine-tuned) for 30 epochs on the protein extractor and affinity learning module with the training dataset generated with the measurement of $IC_{50}$ (because the BindingDB dataset that we constructed here only has the affinity values calculated from the $IC_{50}$ data) based on the general set of PDBbind-v2020.

### 3.8 Details of the Ablation Experiment

We used the $IC_{50}$ dataset constructed from PDBbind's general set to generate the training datasets for the module ablation experiment. The other dataset generation steps and model parameter settings were the same as those used to train FeatNN on the benchmark datasets generated from the general set of PDBbind in the main text.

In the model architecture modification step, we directly deleted or replaced the module to be ablated with a simple linear layer. Finally, the RMSE, Pearson coefficient and $R^2$ were selected to compare and evaluate the comprehensive performance of these models.

## 4. Full Algorithm Details

The pseudocodes for each module are available in the supplementary methods.

### Notations for the Operations Between Vectors and Matrix

The definitions of operations and variables are listed as follows. We use $\oplus$ for the outer sum, $\odot$ for the elementwise product, namely, the Hadamard product, $\sigma(\cdot)$ for the sigmoid activation function $\sigma(x) = 1/(1 + e^{-x})$, $tanh(\cdot)$ for the tanh activation function $tanh(x) = (e^x - e^{-x})/(e^x + e^{-x})$, and $f(\cdot)$ for the Gaussian error linear unit (GELU) activation function $GELU(x) = 0.5x(1 + erf(\frac{x}{\sqrt{2}}))$, where $erf(\cdot)$ serves as the Gaussian error function, $erf(x) = \frac{2}{\sqrt{\pi}}\int_0^x e^{-t^2}dt$, and $softmax(x_i)$ is used for the softmax function $exp(x_i)/\sum_i exp(x_i)$.

### 4.1. Compound Extractor Module

In this study, a graph representation of a compound is utilized to describe the specific correlation between its atom features and bond features. Given a graph representation $\{V, E\}$, vertices and edges are used to represent atom and bond features in the compound, respectively. More specifically,

$\{V\} = \{\text{element name,aromatic type,vertex degree, "atom valence"}\}$, in which the features are encoded by a one-hot-encoding strategy and then are concatenated into an all-one vector as $\{F_i^{atom} \in R^{\mathrm{h}}\}_{i=1}^{N_a}$ for each atom. Similarly, $\{E\} = \{\text{"bond type", "shape"}\}$ is also applied, obtaining the embedded bond feature vector as $\{F_j^{bond} \in R^{\mathrm{h}}\}_{j=1}^{N_b}$, where $i = 1,2, \ldots, N_a$, $j = 1,2, \ldots, N_b$, $h$ is the dimensionality of the hidden size, $N_a$ is the number of compound atoms, and $N_b$ is the number of protein residues. Original atom features are defined as $F^0 \in R^{N_a \times \mathrm{h}}$, and master node features are defined as summaries of atom features, that is, $F^{master} = \sum_{i=1}^{N_a} F_i^{atom}$. Considering that there are $l_c$ graph convolution layers where $l_c = 1,2, \ldots, l_{comp}$ and $k_c$ attention heads where $k_c = 1,2, \ldots, k_{comp}$, $l_{comp}$ is the total number of graph convolution layers, and $k_{comp}$ is the total number of compound feature attention heads. The atom features, bond features, and master features in the $l_c th$ layer are defined as $F_{atom}^{l_c}$, $F_{bond}^{l_c}$ and $F_{master}^{l_c}$, respectively, and the variables $V$ with $k_c$ heads are defined as $V^{k_c}$. For example, $F_{atom}^{l_c,k_c}$ represents the atom features in the $l_c th$ layer of the GCN with $k_c$ heads. For a detailed description, see Supplementary Session 2.2.

#### 4.1.1 Multihead Attention Block

Main vertex (atom) features are obtained with a multihead attention mechanism and the elementwise product operation. The main vertex features are updated as $v_{comain}^{l_c}$ in each layer:

$$v_{tmain}^{l_c,k_c} = softmax(W_{vm}^{l_c,k_c}(tanh(W_{vmain}^{l_c,k_c}F_{atom}^{l_c-1}) \odot F_{master}^{l_c-1})) \otimes W_{ms}^{l_c,k_c}F_{atom}^{l_c-1}$$

$$v_{main}^{l_c} = tanh(W_{cmat}^{l_c}[v_{tmain}^{l_c,k_c}]_{k_c})$$

$$v_{comain}^{l_c} = dropout(F_{atom}^{l_c-1})$$

where $W_{vmain}^{l_c,k_c} \in R^{\mathrm{h}\times\mathrm{h}}$, $W_{vm}^{l_c,k_c} \in R^{\mathrm{h}\times\mathrm{h}}$, $W_{ms}^{l_c,k_c} \in R^{\mathrm{h}\times\mathrm{h}}$, and $[\cdot]_{k_c}$ indicates the integration of the information from multihead attention. A detailed description and the pseudocode are provided in Supplementary Section 2.2.2.

#### 4.1.2 Deep GCN

The atom features are sequentially updated using a message passing unit and a graph warp unit at each iteration of the GCN.

$$mt_{main}^{l_c} = W_{lu2}^{l_c}[v_{comain}^{l_c}, \sum_{v_k \in Neighbor(v_i)} f(W_{ln}^{l_c}[v_{comain}^{l_c}, F_{bond}^{l_c}]_{h+h+bn})]_{h+h}$$

where $bn$ is the shape or size of bond neighbors, $[\cdot,\cdot]_m$ indicates the concatenation operation on different dimensions, $W_{lu2}^{l_c} \in R^{2\mathrm{h}\times\mathrm{h}}$, and $W_{ln}^{l_c} \in R^{\mathrm{h}+(h+bn)}$.

To avoid the oversmoothing problem in the graph convolution process, we use the initial vertex features $F^0$ as the identity information and the residual connection pathway:

$$r^{l_c} = (1-\alpha)mt_{main}^{l_c} + \alpha F^0$$

$$v_{comp}^{l_c} = \theta W_{fu}^{l_c} r^{l_c} + (1-\theta) r^{l_c}$$

where $W_{fu}^{l_c} \in R^{\mathrm{h}\times\mathrm{h}}$, and both $\alpha$ and $\theta$ are hyperparameters.

Next, $ht^{l_c}_{master} = W^{l_c}_{mas}F^{l_c-1}_{master}$, $ht^{l_c}_{mas2m} = W^{l_c}_{mas2m}F^{l_c-1}_{master}$ is defined. Both the main vertex and master node features can be mutually updated through the graph warp unit and $GRU$ layers, and $GRUs$ are used to determine the proportions of the main vertex and master node features updated at layer $l_c$.

$$g^{l_c}_{main} = \sigma(W^{l_c}_{zm1}v^{l_c}_{comp} + W^{l_c}_{zm2}ht^{l_c}_{mas2m})$$

$$ht^{l_c}_{main} = g^{l_c}_{main}ht^{l_c}_{mas2m} + (1 - g^{l_c}_{main})v^{l_c}_{comp}$$

$$v^{l_c}_{comain} = GRU_{main}(ht^{l_c}_{main}, v^{l_c-1}_{comain})$$

With the same process, the master node features are also updated as $v^{l_c}_{comaster}$ in each graph convolution layer, that is,

$$g^{l_c}_{master} = \sigma(W^{l_c}_{zs1}ht^{l_c}_{master} + W^{l_c}_{zs2}v^{l_c}_{main})$$

$$ht^{l_c}_{master} = g^{l_c}_{master}v^{l_c}_{main} + (1 - g^{l_c}_{master})ht^{l_c}_{master}$$

$$v^{l_c}_{comaster} = GRU_{master}(ht^{l_c}_{master}, v^{l_c-1}_{comaster})$$

where $W^{l_c}_{mas2m} \in R^{\mathrm{h}\times\mathrm{h}}$, $W^{l_c}_{mas} \in R^{\mathrm{h}\times\mathrm{h}}$, $W^{l_c}_{zs1} \in R^{\mathrm{h}\times\mathrm{h}}$, and $W^{l_c}_{zs2} \in R^{\mathrm{h}\times\mathrm{h}}$.

After the iterations of the deep GCN block, the final features of the main vertex and master features are obtained as $v^{l_{comp}}_{comain}$ and $v^{l_{comp}}_{comaster}$ that are defined above as $F^{fatom}$ $and$ $F^{fmaster}$. A detailed description and pseudocode are provided in Supplementary Section 2.2.1.

### 4.2. Protein Extractor Module

#### 4.2.1 Protein Aggregation Module

Sequence and distance features are embedded through a word embedding strategy, and torsion features are embedded through a linear layer. The protein embedding module takes sequence features $\{F^{seq}_n \in R^h\}^{N_{res}}_{n=1}$, the DDM $\{\boldsymbol{F}^{DDM}_n \in R^e\}^{N_{res}\times N_{res}}_{n=1}$ and the torsion matrix $\{\boldsymbol{F}^{TM}_n \in R^h\}^{N_{res}}_{n=1}$ of proteins as input data. In addition, $N_{ts}$ is the initial torsion dimension, $h$ is the hidden size, $e$ is the embedding size, $k$ is the number of attention heads and $k_h$ is the hidden size of the attention heads, where $k_h = h/k$. A linear layer is used to update these features, that is,

$$l_n^{seq} = CNN_{e\to h}^{seq} F_n^{seq}$$

$$\boldsymbol{E}_{i,j}^{DDM} = I_n F_n^{DDM}$$

$$l_n^{TM} = CNN_{e\to h}^{tor2} f\left(CNN_{N_{ts}\to e}^{tor1} \boldsymbol{F}_n^{TM}\right)$$

where $I_n$ is the identity matrix, $N_{res}$ is the length of the amino acid in each protein, $N_{ts}$ is the initial dimensionality of the torsion size, $h$ is the hidden size, $\mathrm{m}$ is the kernel size $e$ is the embedding size, and $pe$ is the preembedding size. All $CNN_{e\to h}^{seq}$, $CNN_{N_{ts}\to e}^{tor1}$ and $CNN_{e\to h}^{tor2}$ retain the width and height of the input matrix but change the feature dimensions with specific kernel sizes and padding sizes.

The aggregation of the protein sequence, distance and torsion features together is a novel strategy for use prior to the extraction of protein features.

$$torgate = \sigma(W_{gt} l_n^{TM})$$

$$\boldsymbol{E}_n^{seq} = torgate \odot l_n^{TM} + (1 - torgate) \odot l_n^{seq}$$

where $W_{gt} \in R^{h\times h}$.

Ultimately, the embedded sequence vector $\{E_n^{seq} \in R^{\mathrm{h}}\}_{n=1}^{N_{res}}$ and embedded DDM $\{\boldsymbol{E}_{i,j}^{DDM} \in R^{\mathrm{e}}\}_{i=1,j=1}^{N_{res}\times N_{res}}$ are obtained from the protein embedding block. A detailed description and pseudocode are provided in Supplementary Section 2.3.3.

### 4.2.2 Evo-Updating Module

We use an evolutionary updating strategy to update the sequence and structure features in the Evo-Updating model block by combining the information derived from the protein embedding block.

The sequence features $\{E_n^{seq} \in R^{\mathrm{h}}\}_{n=1}^{N_{res}}$ and structure features $\{\boldsymbol{E}_{i,j}^{DDM} \in R^{\mathrm{e}}\}_{i=1,j=1}^{N_{res} \times N_{res}}$ are embedded via the protein aggregation algorithm, and $i, j$ are the row and column of the embedded DDM, respectively. We define the input features in layer $l_p$ as $F_{seq}^{l_p-1}$ and $\boldsymbol{DDM}_{i,j}^{l_p-1}$, where $l_p = 1,2, \dots, L_p$.

$$key_{mix}^{l_p} = GRU(CNN_{e \to h}^{row^{l_p}} \sum_{i=1}^{n} \boldsymbol{DDM}_{i,j}^{l_p-1}, CNN_{e \to h}^{column^{l_p}} \sum_{j=1}^{n} \boldsymbol{DDM}_{i,j}^{l_p-1})$$

$$key_{mix}^{l_p,kn} = TransposeForScores(key_{mix}^{l_p})$$

$$struct^{l_p} = \sum_{M=1}^{k} CNN_{k_h \to h}^{sn}{}^{l_p,M} key_{mix}^{l_p,M} + CNN_{h \to h}^{init_res^{l_p}} key_{mix}^{l_p}$$

Protein features are first processed through the gated recurrent unit (GRU) cell with row and column pooling features of $\boldsymbol{DDM}_{i,j}^{l_p-1}$, while all $CNN_{e \to h}^{row^{l_p}}$, $CNN_{e \to h}^{column^{l_p}}$, $CNN_{kn \to h}^{sn}{}^{l_p,kn}$ and $CNN_{h \to h}^{init_res^{l_p}}$ models retain the width and height of the input matrix but change the feature dimensions with specific kernel sizes and padding sizes. In particular, $TransposeForScores()$ is an algorithm described in Supplementary Section 2.3.6. We use the outer sum operation to update and map the information derived from the sequence and use multihead attention to learn the diversified correlation of $\boldsymbol{DDM}_{i,j}^{l_p-1}$:

$$prot_vec_{seq}^{l_p} = f(\sum_{sn=1}^{k} CNN_{k_h \to h}^{zwei}{}^{l_p,sn} F_{seq}^{l_p-1,kn} + CNN_{h \to h}^{zwei_res^{l_p}} F_{seq}^{l_p-1})$$

$$seq_{initial}^{l_p} = CNN_{h \to h}^{dz1^{l_p}} prot_vec_{seq}^{l_p}$$

$$seq2struct^{l_p} = f(CNN_{h \to h}^{dz2^{l_p}} seq_{initial}^{l_p})$$

We use a gate mechanism to gather more useful information from the input features and aggregate the sequence features onto the structure features, and the GRU cell is used to aggregate both updated and initial structure features,

$$g_{\mathrm{seq2str}}^{l_p} = \sigma(W_{seq2str}^{l_p} seq2struct^{l_p})$$

$$g_{\mathrm{str2seq}}^{l_p} = \sigma(W_{str2seq}^{l_p} key_{mix}^{l_p})$$

$$v_{struct}^{l_p} = g_{\mathrm{seq2str}}^{l_p} \odot seq2struct^{l_p} + (1 - g_{\mathrm{seq2str}}^{l_p}) \odot struct^{l_p}$$

$$v_{sequence}^{l_p} = g_{\mathrm{str2seq}}^{l_p} \odot struct^{l_p} + (1 - g_{\mathrm{str2seq}}^{l_p}) \odot seq_{initial}^{l_p}$$

$$p_{struct}^{l_p} = CNN_{h \to e}^{map^{l_p}} f(GRU(v_{struct}^{l_p}, struct^{l_p}))$$

$$p_{sequence}^{l_p} = f(GRU(v_{sequence}^{l_p}, seq_{initial}^{l_p}))$$

where $W_{seq2str}^{l_p} \in R^{h \times h}$ $and$ $W_{str2seq}^{l_p} \in R^{h \times h}$ and $CNN_{k_h \to h}^{zwei^{l_p,sn}}$, $CNN_{h \to h}^{zwei_res^{l_p}}$, $CNN_{h \to h}^{dz1^{l_p}}$ $CNN_{h \to h}^{dz2^{l_p}}$ and $CNN_{h \to e}^{map^{l_p}}$ retain the width and height of the input matrix but change the feature dimensions with specific kernel sizes and padding sizes.

We aggregate the features with the outer sum (to create a symmetric matrix with a highly correlated DDM) and the gate. The updated features of the sequence and DDM in the $l_p th$ layer of the Evo-Updating block are given as $F_{seq}^{l_p}$ and $\boldsymbol{DDM}_{i,j}^{l_p}$, respectively,

$$F_{seq}^{l_p} = I_n p_{sequence}^{l_p}$$

$$\boldsymbol{DDM}^{l_p} = p_{struct}^{l_p} \oplus p_{struct}^{l_p}$$

where $I_n$ is the identity matrix and $\oplus$ is as the outer sum operation.

After calculating $L_p$ iterations of the protein encoder, we obtain the final feature representations $\{F_{seq,i}^{l_p} \in \mathrm{R}^{\mathrm{h}}\}_{i=1}^{N_{res}}$ and $\{\boldsymbol{DDM}_{i,j}^{l_p} \in \mathrm{R}^{\mathrm{e}}\}_{i=1,j=1}^{N_{res} \times N_{res}}$. A detailed description and pseudocode are provided in Supplementary Section 2.3.4. All specific information can be found in Supplementary Section 2.3.

**4.3. Affinity Learning Module**

The affinity learning module integrates the mutual information between the compounds and proteins during noncovalent interaction affinity prediction. Suppose we are given atom features $\{F_i^{fatom} \in R^{\mathrm{h}}\}_{i=1}^{N_a}$ and master node features $\{F^{fmaster} \in R^{\mathrm{h}}\}$ from the compound extractor, as well as protein features $\{F_i^{seq} \in \mathrm{R}^{\mathrm{h}}\}_{i=1}^{N_{res}}$ extracted from the protein extractor. In particular, both the compound and protein features are separately transformed into a compatible space by single linear layers, that is, $F_{atom}^{Comp} = f(W_{atom}F_i^{fatom})$ and $F_{seq}^{Prot} = f(CNN_{h\to h}^{Fseq}F_i^{seq})$, where $i = 1,2,..,N_a$, $j = 1,2,..,N_{res}$, $W_{atom} \in \mathrm{R}^{\mathrm{h}\times\mathrm{h}}$, and $CNN_{h\to h}^{Fseq}$ retains the width and height of the input matrix but changes the feature dimensions with specific kernel sizes and padding sizes.

The protein and compound features are eventually calculated after the $l_{aff}th$ iteration. Prior to performing affinity prediction, the feature aggregation operation between the master node features and main graph features should be considered with the help of a summation operation, that is,

$$C_{final} = \sum_{a=N_a} F_{atom\ a}^{Comp} / N_a$$

$$C_{aggre} = [C_{final}, F^{fmaster}]_{\mathrm{h+h}}$$

The same operations are also utilized for the protein features, that is,

$$P_{final} = \sum_{s=N_{res}} F_{seq\ s}^{Prot} / N_{res}$$

where $[\cdot,\cdot]_m$ indicates the concatenation operation between the hidden sizes of the main vertex features and master node features.

Finally, with a single linear mapping layer, the affinity value is calculated by vectors $C_{aggre}$ and $P_{final}$, that is,

$$affinity = W_{aff}(f(C_{aggre}P_{final}))$$

where $W_{aff} \in \mathrm{R}^{2\mathrm{h}^2\times 1}$.

A detailed description and pseudocode are provided in Supplementary Section 2.4.

### 4.4. Quantification and Statistical Analysis

Evaluation Metrics

We use eight metrics that are commonly used for this problem to evaluate the prediction performance of our model. These metrics are defined as follows.

The $R^2$ score, RMSE, and Pearson coefficient are often used in regression analysis. They describe the distance between the predicted values and true values. The higher the values of R2 and the Pearson coefficient are, the closer the model prediction results are to the real values. The smaller the RMSE value is, the smaller the error in the prediction value, that is, the higher the accuracy.

The RMSE is the standardized value of the MSE that is typically used as the training loss in machine learning studies. It is defined as follows:

$$RMSE(y, \hat{y}) = \sqrt{\frac{1}{n}\sum_{i=1}^{n}(y_i - \hat{y_i})^2}$$

The $R^2$ score is a dimensionless score describing the effectiveness of the model. It compares the output prediction to a random guess according to the average of the true values:

$$\begin{aligned} R^2(y, \hat{y}) &= 1 - \frac{SS_{residual}}{SS_{total}} \\ &= 1 - \frac{\sum_i (y_i - \hat{y_i})^2}{\sum_i (y_i - \overline{y_i})^2} \end{aligned}$$

We use a coefficient that can describe the correlation between the predicted values and true values: namely the Pearson product-moment correlation coefficient.

The Pearson correlation coefficient describes the linear correlation between two values and is defined as:

$$Pearson(y,\hat{y}) = \frac{Cov(y,\hat{y})}{\sigma_y \sigma_{\hat{y}}} = \frac{\sum_i (y_i - \overline{y})(\widehat{y_i} - \overline{\widehat{y_i}})}{\sqrt{\sum_i (y_i - \overline{y_i})^2}\sqrt{\sum_i (\widehat{y_i} - \overline{\widehat{y_i}})^2}}$$

where $y_i$ are the prediction values, and $\widehat{y_i}$ are the true values in the dataset, $i$ =1,2...,n, where n is the total amount of the dataset.

In this paper, Pearson was selected to evaluate the accuracy of CPA prediction when predicting the affinity of 28 bioactive small molecules binding to SARS-CoV-2 3C-like protease, and the calculation and statistical method are consistent with the above description.

## References:

1. Kip F TN, Welling M. Semi-Supervised Classification with Graph Convolutional Networks, arXiv 2016.
2. Collins EM, Raghavachari K. A Fragmentation-Based Graph Embedding Framework for QM/ML, J Phys Chem A 2021;125:6872-6880.
3. Wang X, Li Z, Jiang M et al. Molecule Property Prediction Based on Spatial Graph Embedding, J Chem Inf Model 2019;59:3817-3828.
4. Li Q, Han Z, Wu XM. Deeper Insights into Graph Convolutional Networks for Semi-Supervised Learning, AAAI 2018:3538-3545.
5. He K, Zhang X, Ren S et al. Deep Residual Learning for Image Recognition, IEEE 2016.
6. Chen M, Wei Z, Huang Z et al. Simple and Deep Graph Convolutional Networks, PMLR 119 2020:1725-1735.
7. Freschlin CR, Fahlberg SA, Romero PA. Machine learning to navigate fitness landscapes for protein engineering, Curr Opin Biotechnol 2022;75:102713.
8. Costanzo M, De Giglio MAR, Roviello GN. SARS-CoV-2: Recent Reports on Antiviral Therapies Based on Lopinavir/Ritonavir, Darunavir/Umifenovir, Hydroxychloroquine, Remdesivir, Favipiravir and other Drugs for the Treatment of the New Coronavirus, Curr Med Chem 2020;27:4536-4541.
9. Chen J, Xia L, Liu L et al. Antiviral Activity and Safety of Darunavir/Cobicistat for the Treatment of COVID-19, Open Forum Infect Dis 2020;7:ofaa241.
10. Mahdi M, Motyan JA, Szojka ZI et al. Analysis of the efficacy of HIV protease inhibitors against SARS-CoV-2's main protease, Virol J 2020;17:190.
11. Xiang R, Yu Z, Wang Y et al. Recent advances in developing small-molecule inhibitors against SARS-CoV-2, Acta Pharm Sin B 2022;12:1591-1623.
12. Lopez-Medina E, Lopez P, Hurtado IC et al. Effect of Ivermectin on Time to Resolution of Symptoms Among Adults With Mild COVID-19: A Randomized Clinical Trial, JAMA 2021;325:1426-1435.
13. Li J, Lin C, Zhou X et al. Structural basis of main proteases of coronavirus bound to drug candidate PF-07321332, bioRxiv 2021.
14. Hoffman RL, Kania RS, Brothers MA et al. Discovery of Ketone-Based Covalent Inhibitors of Coronavirus 3CL Proteases for the Potential Therapeutic Treatment of COVID-19, J Med Chem 2020;63:12725-12747.
15. Vankadara S, Wong YX, Liu B et al. A head-to-head comparison of the inhibitory activities of 15 peptidomimetic SARS-CoV-2 3CLpro inhibitors, Bioorg Med Chem Lett 2021;48:128263.
16. Hosseini-Zare MS, Thilagavathi R, Selvam C. Targeting severe acute respiratory syndrome-coronavirus (SARS-CoV-1) with structurally diverse inhibitors: a comprehensive review, RSC Adv 2020;10:28287-28299.
17. Zhao L, Li S, Zhong W. Mechanism of Action of Small-Molecule Agents in Ongoing Clinical Trials for SARS-CoV-2: A Review, Front Pharmacol 2022;13:840639.
18. Niesor EJ, Boivin G, Rheaume E et al. Inhibition of the 3CL Protease and SARS-CoV-2 Replication by Dalcetrapib, ACS Omega 2021;6:16584-16591.

19. Bahun M, Jukic M, Oblak D et al. Inhibition of the SARS-CoV-2 3CL(pro) main protease by plant polyphenols, Food Chem 2022;373:131594.
20. Shiyong Fan DX, Yanming Wang, Lianqi Liu, Xinbo Zhou, and Wu Zhong. Research progress on repositioning drugs and specific therapeutic drugs for SARS-CoV-2, Future Medicinal Chemistry 2020;12:1565-1578.
21. Rosales R, McGovern BL, Rodriguez ML et al. Nirmatrelvir, Molnupiravir, and Remdesivir maintain potent in vitro activity against the SARS-CoV-2 Omicron variant, bioRxiv 2022.
22. Andrikopoulou A, Chatzinikolaou S, Panourgias E et al. "The emerging role of capivasertib in breast cancer", Breast 2022;63:157-167.
23. Weisner J, Landel I, Reintjes C et al. Preclinical Efficacy of Covalent-Allosteric AKT Inhibitor Borussertib in Combination with Trametinib in KRAS-Mutant Pancreatic and Colorectal Cancer, Cancer Res 2019;79:2367-2378.
24. Rhodes N, Heerding DA, Duckett DR et al. Characterization of an Akt kinase inhibitor with potent pharmacodynamic and antitumor activity, Cancer Res 2008;68:2366-2374.
25. Nandan D, Zhang N, Yu Y et al. Miransertib (ARQ 092), an orally-available, selective Akt inhibitor is effective against Leishmania, PLoS One 2018;13:e0206920.
26. Politz O, Siegel F, Barfacker L et al. BAY 1125976, a selective allosteric AKT1/2 inhibitor, exhibits high efficacy on AKT signaling-dependent tumor growth in mouse models, Int J Cancer 2017;140:449-459.
27. Grimshaw KM, Hunter LJ, Yap TA et al. AT7867 is a potent and oral inhibitor of AKT and p70 S6 kinase that induces pharmacodynamic changes and inhibits human tumor xenograft growth, Mol Cancer Ther 2010;9:1100-1110.
28. McLeod R, Kumar R, Papadatos-Pastos D et al. First-in-Human Study of AT13148, a Dual ROCK-AKT Inhibitor in Patients with Solid Tumors, Clin Cancer Res 2020;26:4777-4784.
29. Wu WI, Voegtli WC, Sturgis HL et al. Crystal structure of human AKT1 with an allosteric inhibitor reveals a new mode of kinase inhibition, PLoS One 2010;5:e12913.
30. Iksen, Pothongsrisit S, Pongrakhananon V. Targeting the PI3K/AKT/mTOR Signaling Pathway in Lung Cancer: An Update Regarding Potential Drugs and Natural Products, Molecules 2021;26.
31. Song M, Liu X, Liu K et al. Targeting AKT with Oridonin Inhibits Growth of Esophageal Squamous Cell Carcinoma In Vitro and Patient-Derived Xenografts In Vivo, Mol Cancer Ther 2018;17:1540-1553.
32. Li S, Zhou J, Xu T et al. Structure-aware Interactive Graph Neural Networks for the Prediction of Protein-Ligand Binding Affinity. Proceedings of the 27th ACM SIGKDD Conference on Knowledge Discovery & Data Mining. 2021, 975-985.
33. Li S, Wan F, Shu H et al. MONN: A Multi-objective Neural Network for Predicting Compound-Protein Interactions and Affinities, Cell Systems 2020;10:308-322.e311.
34. Chicco D, Warrens MJ, Jurman G. The coefficient of determination R-squared is more informative than SMAPE, MAE, MAPE, MSE and RMSE in regression analysis evaluation, PeerJ Comput Sci 2021;7:e623.

35. Tjur T. Coefficients of Determination in Logistic Regression Models—A New Proposal: The Coefficient of Discrimination, The American Statistician 2009;63:366-372.
36. Jumper J, Evans R, Pritzel A et al. Highly accurate protein structure prediction with AlphaFold, Nature 2021;596:583-589.